%% file: paper.tex
\documentclass[12pt]{article}
\usepackage{epsfig}
\usepackage{amsmath}
\usepackage{hhline}
\usepackage{amssymb}
\usepackage{times}

\newlength{\dinwidth}
\newlength{\dinmargin}
\begin{document} 
\newcommand{\photoproduction}{$\gamma p$}
\newcommand{\ptmiss} {$P_{T}^{\rm miss}$}
\newcommand{\epz}    {$E{\rm-}p_z$}
\newcommand{\vap}    {$V_{ap}/V_p$}
\newcommand{\Zero}   {\mbox{$Z^{\circ}$}}
\newcommand{\Ftwo}   {\mbox{$\tilde{F}_2$}}
\newcommand{\Ftwoz}  {\mbox{$\tilde{F}_{2,3}$}}
\newcommand{\Fz}     {\mbox{$\tilde{F}_3$}}
\newcommand{\FL}     {\mbox{$\tilde{F}_{_{L}}$}}
\newcommand{\Fem}    {\mbox{$F_2$}}
\newcommand{\Fgam}   {\mbox{$F_2^{\gamma}$}}
\newcommand{\Fint}   {\mbox{$F_2^{\gamma Z}$}}
\newcommand{\Fwk}    {\mbox{$F_2^{Z}$}}
\newcommand{\Fzint}  {\mbox{$F_3^{\gamma Z}$}}
\newcommand{\Fzwk}   {\mbox{$F_3^{Z}$}}
\newcommand{\Gev}    {\mbox{${\rm GeV}$}}
\newcommand{\Gevv}   {\mbox{${\rm GeV}^2$}}
\newcommand{\QQ}     {\mbox{${Q^2}$}}
\newcommand{\gapprox}{\raisebox{-0.7ex}{$\stackrel {\textstyle>}{\sim}$}}
\newcommand{\lapprox}{\raisebox{-0.7ex}{$\stackrel {\textstyle<}{\sim}$}}
\newcommand{\htab}{\rule[-1mm]{0mm}{6mm}}
\newcommand{\vtab}{\rule[-1mm]{0mm}{4mm}}

\hyphenation{brems-strahl-ung}

\begin{titlepage}

\noindent
\begin{flushleft}
DESY-99-107  \hfill  ISSN 0418-9833 \\
August 1999
\end{flushleft}

\vspace*{4cm}

\begin{center}
\begin{Large}
{\bf Measurement of Neutral
and Charged Current \\ Cross-Sections 
in Positron-Proton Collisions \\
at Large Momentum Transfer}

\vspace*{2.0cm}

H1 Collaboration \\
\vspace*{1.5cm}
\end{Large}

\end{center}

\vspace{1.5cm}

\begin{abstract}
\noindent The inclusive single and double differential cross-sections
for neutral
and charged current processes with four-momentum transfer 
squared $Q^2$ between $150$ and $30\,000$~\Gevv\ and with Bjorken $x$
between $0.0032$ and $0.65$ are measured in $e^+ p$ collisions. The
data were taken with the H1 detector at HERA between 1994 and 1997,
and they correspond to an integrated luminosity of 
$35.6 \ {\rm pb}^{-1}$. The $Q^2$ evolution of the parton 
densities of the proton is tested, 
yielding no significant deviation from the
prediction of perturbative QCD.
The proton structure function $F_2(x,Q^2)$ is determined.
An extraction of the $u$ and $d$ quark distributions
at high $x$ is presented.
At high $Q^2$ electroweak effects of the heavy 
bosons $Z^\circ$ and $W$ are 
observed and found to be consistent with Standard Model expectation.
\vspace*{3.cm}
\end{abstract}
\begin{center}
{\it Submitted to the European Journal of Physics}
\end{center}

\end{titlepage}

\include{h1auts}

\newpage

\section{Introduction}

The deep-inelastic scattering (DIS) of leptons off nucleons has played
a fundamental role in understanding the structure of matter and in the
foundation of the Standard Model as the theory of strong and electroweak
interactions. 
The first DIS measurements revealed
the existence of partons in the proton~\cite{slac} and opened the way
to the development of Quantum Chromodynamics (QCD) as the theory of
strong interactions. The establishment of electroweak theory followed
the observation of neutral current neutrino
scattering~\cite{lagarigue}. Subsequent (fixed target) DIS
experiments~\cite{fixtarg,bcdms,nmc} have helped to constrain the
electroweak parameters of the Standard Model and the partonic
structure of the proton.

At HERA, the first electron-proton ($ep$) collider ever built,
the study of DIS has been further pursued since 1992.
There are two
contributions to DIS, both of which can be measured at HERA, neutral
current (NC) interactions, \mbox{$ep \rightarrow eX$}, and charged
current (CC) interactions, \mbox{$ep \rightarrow \nu X$}. In the
Standard Model a photon ($\gamma$) or a $Z^{\circ}$ boson is
exchanged in a NC interaction, and a $W^{\pm}$ boson is exchanged in a CC
interaction.
DIS can be described 
by the
four-momentum transfer squared $Q^2$, Bjorken $x$ 
and inelasticity $y$ defined as
\begin{equation}
Q^2 = -q^2 \equiv -(k-k')^2
\hspace*{2cm}
x = \frac{Q^2} {2 p \cdot q}
\hspace*{2cm}
y = \frac{p \cdot q} {p \cdot k} \, ,
\end{equation}
with $k (k^{'})$ and $p$ being the four-momentum of the incident (scattered) lepton and
proton. 
The centre-of-mass energy $\sqrt{s}$ of the $ep$ interaction is given
by $s \equiv (p+k)^2 = Q^2/xy$ when neglecting
the proton and positron masses. 

The kinematic range of DIS measurements is extended 
to $Q^2=30\,000$~\Gevv\ at high $x$ with this analysis.
The fixed target experiments covered the kinematic plane up to 
$Q^2=250$~\Gevv\ and down to $x \approx 10^{-2}$.
Previous results by the HERA experiments, H1 and ZEUS, 
extended to higher values of $Q^2=5000$~\Gevv\ and to lower values of 
$x \approx\ 10^{-5} $ at low $Q^2$~\cite{H194,ZEUS94}. 
The extensions in kinematic domain
are made possible at HERA by the positron
and proton beam energies 
of $E_e = 27.6$ \Gev\ and $E_p=820$ \Gev\ and consequently large 
 $\sqrt{s} \approx 300$ \Gev. 

For NC interactions at low $x$, 
the measurements of the proton structure function $F_2$ 
revealed~\cite{H192-93,ZEUS92-93} a strong rise with decreasing $x$,
which can be understood within perturbative QCD in the form of 
Next-to-Leading Order (NLO) DGLAP~\cite{bb.dglap} evolution equations. 
The kinematic reach in $x$ at high $Q^2$ allowed
cross-sections 
which are directly related to the valence quark distributions of the 
proton to be measured, albeit with limited precision. 
For CC interactions measurements of $e^-p$ and $e^+p$
single differential cross-sections extended the results obtained
in fixed target neutrino and antineutrino scattering to higher 
$Q^2$~\cite{H1CC94abc,ZEUSCC94ab}.
The measurements
were used to determine the $W$ propagator mass 
$M_{W}$.
 
In this paper measurements of the NC and CC cross-sections at high
$Q^2$ are presented. The results are obtained using $e^+p$ data taken
between 1994 and 1997. 
The integrated luminosity of \mbox{$35.6 \ {\rm pb}^{-1}$} is more than a
factor of $10$ greater than for previously published measurements of
NC and CC cross-sections by H1~\cite{H194,H1CC94abc}. The increase in
integrated luminosity enables both the influence of the $Z^{\circ}$
boson in NC interactions and the helicity structure of the CC
interaction to be tested in the high $Q^2$ domain. The behaviour of
the NC and CC cross-sections at the highest $Q^2$ is of particular
interest following the observation by H1 and
ZEUS~\cite{h1hiq296,zeushiq296} of an excess over Standard Model
expectation of NC events at $Q^2$ greater than $15\,000$~\Gevv\ using
the $e^+p$ data taken between 1994 and 1996. 
A detailed analysis of the significance of this excess using the
complete $e^+p$ data set used here is presented in~\cite{yves}.
Recently measurements of the NC and CC 
cross-sections at high $Q^2$ have been
reported by the ZEUS experiment~\cite{zeushiq2nc,zeushiq2cc}.
 
This paper consists of five sections. In section 2 the experimental
technique used for the measurements is presented. In section 3 the
procedures used for the cross-section measurement, and the QCD
analysis which is used subsequently to interpret the data, are
described. In section 4 the cross-section measurements and their
interpretation are presented. The paper is summarized in section 5.

\section{Experimental Technique}

\subsection{Kinematic Reconstruction}
\label{sec:kinematics}

The measurement of the differential DIS cross-sections 
relies on the precise
determination of the kinematic variables of each event.
Different reconstruction methods are used for CC and NC interactions.

For CC interactions the kinematic variables
can only be reconstructed using the 
hadronic final state because the neutrino ($\nu$) is not detected. 
To characterize the hadronic 
final state, it is convenient to introduce the quantity
$\Sigma$, the transverse momentum $P_{T,h}$, and the inclusive hadronic
angle $\gamma_h$ defined by
\begin{equation}
 \Sigma=\sum_i{(E_i-p_{z,i})}
 \hspace*{1.cm}
 P_{T,h}=\sqrt{(\sum_i{p_{x,i}})^2+(\sum_i{p_{y,i}})^2}
 \hspace*{1.cm}
\tan{\frac{\gamma_h}{2}} = \frac{\Sigma}{P_{T,h}} \, .
\end{equation}
Here $E_i$ and $p_{z,i}$ are the energy and longitudinal momentum component
of a particle
 $i$, and $p_{x,i}$, $p_{y,i}$ are its momentum components in the 
 plane orthogonal to the $z$-axis\footnote{The forward direction and the 
positive $z$-axis are defined at HERA as the proton beam direction.}. 
The summation is over all hadronic final
state particles, whose rest masses are neglected\footnote{
The $p_{x,h}$ and $p_{y,h}$ components
of the hadronic transverse momentum vector $\vec{P}_{T,h}$
are defined using the same summation over $p_{x,i}$ and $p_{y,i}$ 
respectively.}. 
The kinematic variables are then obtained from~\cite{jb}
\begin{equation}
 y_{h} = \frac{\Sigma}{ 2 \ E_e }
 \hspace*{2cm}
 Q^2_{h} = \frac{P_{T,h}^2}{ 1-y_{h}}
 \hspace*{2cm}
 x_h=\frac{Q^2_h} {s \ y_h} \, .
\end{equation}
This ``hadron method'' ($h$ method)
gives moderate precision in the reconstruction of the kinematic
variables because of particle losses in the beam-pipe 
and because of fluctuations of the detector response to
hadronic final state particles.
It is thus used only for the CC interactions. 

For NC interactions different methods of determining the kinematic variables
are possible since there is
redundant information from the simultaneous reconstruction 
of the scattered positron and of the hadronic final state.
The choice of the method determines the corrections due to resolution and
radiative effects, and the size of the systematic errors.
In the ``electron method'' ($e$ method)
the energy $E_e^{\prime}$ and the polar angle
 $\theta_e$ 
of the scattered positron 
are used to determine the variables
\begin{equation}
y_e =
 1- \frac {E_e^{'} (1-\cos{\theta_e})} {2 \ E_e}
\hspace*{1cm}
 Q^2_{e} = \frac{P_{T,e}^2}{ 1-y_e} 
\hspace*{1.5cm} x_e = \frac{Q^2_e}{s \ y_e}
\end{equation}
with $ P_{T,e} = E^{\prime}_e \sin{\theta_e}$.
The $e$ method has excellent resolution in
$Q^2$ and in $x$ at large $y$. The $\Sigma$ method~\cite{bb3}
makes use of the positron and the hadronic
final state variables. It has a better resolution in $x$ at low $y$
and is less sensitive to radiative effects since 
\begin{equation}
\label{eq:sigma} 
y_{\Sigma} = \frac{{\Sigma}} {E-p_z} \hspace*{1cm} {\mbox{with}}
\hspace*{1cm}
E-p_z = \Sigma + E^{\prime}_e(1-\cos{\theta_e})
\end{equation}
does not depend on the energy of the incoming positron.
A combination of the
$e$ and $\Sigma$ methods, the $e\Sigma$ method~\cite{bb3}, 
is thus used to optimize the kinematic reconstruction
in the NC measurement; $Q^2$ is taken from the $e$ method 
and $x$ from the $\Sigma$ method. 
Both these variables display good resolution in the complete
kinematic range and the radiative corrections remain small
compared to those of the $e$ method. 
The $e\Sigma$ formulae are
\begin{equation}
y_{e\Sigma} = \frac{2 E_e}{E-p_z} y_{\Sigma}
\hspace*{1.5cm}
Q^2_{e\Sigma} = \frac{P_{T,e}^2}{ 1-y_e} 
\hspace*{1.5cm}
x_{e\Sigma} = \frac{P_{T,e}^2}{s \
y_{\Sigma} (1-y_{\Sigma}) } \, .
\label{kinematics1}
\end{equation}

\subsection{Detector and Trigger}

The H1 detector~\cite{h1detec} is a nearly hermetic 
multi-purpose apparatus built to investigate $ep$ scattering.
The high $Q^2$ cross-section measurements reported here rely
primarily on the tracking system,
on the Liquid Argon (LAr) calorimeter, on the luminosity detectors, 
and to a lesser extent on the backward calorimeter.
These components are described briefly below.
 
The tracking system includes the central and
forward tracking chambers. 
These detectors are placed around the beam-pipe at $z$ positions
between $-1.5$ and $2.5$ ${\rm m}$. A superconducting solenoid, which
surrounds both the tracking system and the LAr
calorimeter, provides a uniform magnetic field of $1.15~{\rm T}$. 
The central jet chamber (CJC) consists of two concentric drift chambers
covering a polar angular range from $25^{\circ}$ to $155^{\circ}$. 
Particles crossing the CJC are measured with a transverse 
momentum resolution of
${\delta P_T}/{P_T}< 0.01 \cdot P_T/\Gev$ .
To improve the 
determination of the $z$ coordinate of
the tracks, two polygonal drift chambers with wires perpendicular
to the $z$-axis placed at radii of $18$ (CIZ) and $47~{\rm cm}$ (COZ) 
are used. The forward tracking detector measures
charged particles emitted in an angular range from 
$7^{\circ}$ to $25^{\circ}$. 
It is used in this analysis to determine the interaction vertex
for events with no track in the CJC.
 
The
LAr calorimeter~\cite{LARC}, which surrounds the tracking system
in the central and forward regions, covers an angular
region between $4^{\circ}$ and $154^{\circ}$. It is divided in 8
wheels along the $z$-axis, which are themselves subdivided in $\varphi$ in 
up to 8 modules, separated by small regions with inactive material
(``$z$-cracks'' and ``$\varphi$-cracks'' respectively).
The calorimeter consists of an electromagnetic section
with lead absorber plates and a hadronic section with
stainless steel absorber plates. 
Both sections are highly segmented in the transverse and longitudinal
directions with about $44\,000$ cells in total.
The electromagnetic part has a depth between
$20$ and $30$ radiation lengths ($X_{\circ}$). 
The total depth of the calorimeter
varies between $4.5$ and $8$ interaction lengths ($\lambda_I$). 
The systematic uncertainty of 
the electromagnetic (hadronic) energy scale of the LAr 
calorimeter is between $0.7$ and $3\%$ ($2\%$) (section~\ref{calib}).

In the backward region a lead/scintillating fibre calorimeter 
(SPACAL)~\cite{spacal} was 
installed in 1995 to replace the previous lead/scintillator 
electromagnetic calorimeter (BEMC).
The new calorimeter has both an electromagnetic and a hadronic section
with a total depth of about $2\lambda_I$, compared with 
the $1\lambda_I$ depth of the BEMC. 
Together with the LAr calorimeter, its angular acceptance 
($154^{\circ}< \theta< 177.8^{\circ}$) 
makes possible complete coverage 
for the detection of the hadronic final state in the H1 apparatus
apart from the remnants of the proton.
The uncertainty in the measurement of hadronic energy 
in the SPACAL is $7\%$, compared with $15\%$ for the BEMC 
which was operational when the 1994 data were taken. 
The influence of the backward calorimeter on the analysis 
presented here is small.

The LAr and backward calorimeters are surrounded by
the Instrumented Iron~\cite{h1detec} which is
used for muon identification and for the measurement of
hadronic energy leaking from the other calorimeters. In this analysis
it is used to reject muon induced background.

The $ep$ luminosity is determined by comparison 
of the QED cross-section for
the bremsstrahlung reaction $ep \rightarrow ep\gamma$ 
with the measured event rate. 
The photon is detected in a calorimeter
(photon ``tagger'') close to the beam-pipe which is situated at a large
distance from the main detector ($z=-103~\rm{m}$).
The precision of the luminosity determination is $1.5\%$~\cite{lumipap}.

An electron tagger is placed at $z=-33~\rm{m}$ adjacent 
to the beam-pipe. It is used to check the
luminosity measurement and to provide information on $ep \rightarrow
eX$ events at very low $Q^2$ (photoproduction) where the positron
scatters through a small angle ($\pi-\theta_e<5$ ${\rm mrad}$).

The ``trigger'' for the high $Q^2$ events uses mainly information from 
the LAr calorimeter. In NC events the positron initiates 
a trigger ``tower''~\cite{h1detec,LARC}
of electromagnetic energy which points towards the event vertex.
Above the threshold energy of the analysis ($11~\Gev $)
the trigger efficiency is $\ge 99.5\%$. 
For CC events the missing transverse momentum \ptmiss ,
determined from the vector sum of the calorimeter
towers\footnote{For CC events this scalar quantity \ptmiss\ is
equal to $P_{T,h}$.}, is used as the trigger.
During 1997 data taking, a trigger which used track information
supplemented the \ptmiss\ trigger in events with small missing
transverse momentum and large angles of the hadrons. The combined
efficiency of these triggers~\cite{beate} for the CC analysis reaches 
$98\%$ for a missing transverse momentum \ptmiss\ above
$25~\Gev $, and is about $50\%$ at the minimum \ptmiss\ of the
analysis ($12~\Gev$).

\subsection{Event Simulation}

In order to determine acceptance corrections and background
contributions for the DIS cross-section measurements, 
the detector response to events produced
by various Monte Carlo generation programs is simulated in detail
using a program based on GEANT~\cite{GEANT}. 
These simulated events are then subject 
to the same reconstruction and analysis chain as the real data.

DIS processes are generated using the DJANGO~\cite{django} program
which is based on HERACLES~\cite{heracles} for the
electroweak interaction and on LEPTO~\cite{lepto}, using the colour
dipole model as implemented in ARIADNE~\cite{cdm} to generate the QCD
dynamics. The JETSET~\cite{jetset} program is used for the hadron
fragmentation.
The implementation of HERACLES in DJANGO includes the real
bremsstrahlung from the positron and the effects of vacuum
polarization~\cite{jegerlehner}. The simulated events are produced
with the MRSH~\cite{mrsh} parton distributions, reweighted to the H1
NLO QCD Fit described in section~\ref{QCDA} which then gives a better
description of the data.

The NC (CC) analysis makes use of a sample of simulated events
corresponding to an integrated luminosity of 
about $3$ (75) times that of the data.
At $Q^2>1000$~\Gevv\ and $x>0.3$ additional samples of
simulated events are included, which amount to between $10$ and $1000$
times the integrated luminosity of the data.

The main background contribution to NC and CC processes is due
to photoproduction (\photoproduction ) events.
These are simulated using the PYTHIA~\cite{pythia} generator with GRV
leading order parton distribution functions for the proton and
photon~\cite{ggrv}. 
This background is described in detail in section~\ref{epinduced}.

Further potential background contributions 
resulting from the following $ep$ processes have been simulated:
\ 1) elastic and inelastic QED Compton events can fake NC processes
and are generated by the
COMPTON~\cite{epcompt} program; 
\, 2) elastic and inelastic $\gamma \gamma$ processes 
producing pairs of leptons ($l$), 
$ep \rightarrow e p \ l^+ \ l^- (e X \ l^+ \ l^-)$, 
are generated using the LPAIR~\cite{lpair} program
(processes with $l^\pm=e^\pm$ can contribute to the NC sample,
while processes with $l^\pm=\mu^\pm$ are more likely to
contribute in the CC sample);
\, 3) prompt photon production with $\gamma-e$ misidentification can fake
NC events and is generated as a dedicated PYTHIA sample;
\, 4) real production of heavy gauge
bosons, $ep \rightarrow eXW^\pm (eXZ)$, 
followed by leptonic decays of the $W$ or $Z$
is generated using the EPVEC~\cite{epvec} program.
These processes were found to produce only a small (\lapprox 1\%)
contamination in the measured $(x,Q^2)$ domain. They have been taken into
account and will not be discussed henceforth.

\subsection{Event Selection}
\label{sec:select}

For CC events the selection is based on the observation of large
\ptmiss , which is assumed to be the transverse
momentum $p_T^{\nu}$ carried by the outgoing neutrino.
For NC events it is based on the identification of a scattered
positron with large $P_{T,e}$. For both CC and NC events an
event vertex, which is reconstructed using central or forward tracks,
is required to be within $\pm 35\ {\rm cm}$ of its nominal position.
Fiducial (NC) and kinematic cuts (CC and NC) are then applied. The
reconstruction of the kinematic variables for each selection follows
the methods described in section~\ref{sec:kinematics}, and uses the
measurements of the positron and the hadronic final state
which are described in section~\ref{calib}.

CC events are selected as follows:
\begin{itemize} 
\item the \ptmiss\ is required to be greater
than $12$ \Gev;
\item the inelasticity $y_h$ is required to be in the 
range $0.03$ to $0.85$ to 
restrict the measurement to a region where the kinematic reconstruction 
is precise; 
\item the ratio $V_{ap}/V_{p}$ is required
to be less than $0.15$ to reject photoproduction background; 
$V_p$ and $V_{ap}$ are respectively the transverse energy 
flow parallel and antiparallel to
$\vec{P}_{T,h}$; they are determined
from the transverse momentum vectors $\vec{P}_{T,i}$ of all the
particles $i$ which belong to the hadronic final state according to
\begin{equation}
V_{p} = \phantom{-}\sum_{i} \frac{\vec{P}_{T,h}
 \cdot \vec{P}_{T,i}}{P_{T,h}} 
 \hspace*{1cm}
\mbox{for}
 \hspace*{1cm}
\vec{P}_{T,h} \cdot \vec{P}_{T,i} > 0
\label{eq:vp}
\end{equation}
\begin{equation}
V_{ap} = -\sum_{i} \frac{\vec{P}_{T,h}
 \cdot \vec{P}_{T,i}}{P_{T,h}} 
 \hspace*{1cm}
\mbox{for}
 \hspace*{1cm}
\vec{P}_{T,h} \cdot \vec{P}_{T,i} < 0 \, .
\label{eq:vap} 
\end{equation}
\end{itemize}

To identify the positron in NC events, the presence of 
a compact and isolated electromagnetic cluster of energy 
in the LAr calorimeter is required~\cite{bruel}. 
For $\theta_e > 35^{\circ}$ the positron candidate 
is validated only if it is
associated with a track having a distance of
closest approach to the cluster of less 
than $12 \ {\rm cm}$ (section~\ref{thecalib}).
Fiducial cuts are applied to ensure the
quality of the positron reconstruction (section~\ref{ecalib}).
Events are Êalso not included if $\theta_e \ \gapprox \ 153^{\circ}$,
because then the electromagnetic shower of the scattered positron is
not fully contained in the LAr calorimeter. The measurements are
thus restricted to $Q^2 \ge 150$~\Gevv . NC events with such an
identified positron are required to satisfy the following 
cuts:
\begin{itemize}
\item a cluster energy $E_e^{\prime}$ greater than $11 \ \Gev$;
\item an inelasticity $y_e$ lower than $0.9$;
\item a longitudinal momentum balance verifying
 \epz\ (eq.~\ref{eq:sigma}) greater than $35 \ \Gev $.
\end{itemize}
These requirements minimize the size of radiative corrections applied in the
analysis (section~\ref{theory}) and reduce the background due to
photoproduction.

After all the different steps of the event selection, 
and after the additional requirements described below to reject events
from background processes, 
the DIS data sample comprises about $75\,000$ NC events and $700$ CC 
events.

\subsection{Background Rejection}

The CC and NC event samples which result from the selection procedures 
described in the previous section contain both non-$ep$
background, arising from particles produced in proton-nucleus 
interactions and from cosmic rays, and $ep$-induced background. 

\subsubsection{Non-\boldmath{$ep$} Background}

Events resulting from processes other than $ep$ collisions originate
from cosmic rays, from ``beam-halo muons'' of the proton beam which
interact in the detector material and cause electromagnetic showers,
and from protons interacting with residual particles in the beam-pipe
(beam-gas events) or with the beam-pipe itself (beam-wall events).

A large fraction of these background events are removed by requiring
that the event time $T_0$, determined from the drift times of hits
from tracks in the CJC, is coincident with the collision time at the
$ep$ interaction region. In the CC analysis the background is further
reduced by using in addition the $T_0$ determined from the rise
times of signals in the LAr calorimeter.

The remaining background is found to be negligible in the NC sample.
Further reductions are necessary in the CC analysis in which the
background mainly originates from random coincidences between soft
photoproduction events and cosmic rays or beam-halo muons. The
majority of these events are rejected by means of topological requirements
following reconstruction of a cosmic ray or a beam-halo muon using
information from the LAr and SPACAL calorimeters, the CJC and the
Instrumented Iron~\cite{H1CC94abc,beate,chab,negri}.

The inefficiency in the CC selection introduced by these requirements is
determined in two different ways. A visual scan of rejected events
yields an overall inefficiency for the CC selection of $5 \pm 2\%$.
This result is consistent with the inefficiency obtained when using NC
events in which the presence of the scattered positron is ignored.
The residual
contamination of non-$ep$ induced background events in the CC 
sample is determined to be $3.7\%$ by visual scanning. 
These events are then rejected so that the remaining uncertainty in the CC 
sample from non-$ep$ background is below $0.5 \%$.

\subsubsection{\boldmath{$ep$}-induced Background}
\label{epinduced}

The main $ep$-induced background in the CC sample originates from
\photoproduction \ events and from NC events in which the scattered
positron is not identified. Mismeasurement of energies and limited
geometrical acceptance can in both cases lead to events which are not
balanced in transverse momentum.

\begin{figure}[b]
\setlength{\unitlength}{1 mm}
\begin{center}
\begin{picture}(150,70)(0,0)
\put(-12,-10){\epsfig{file=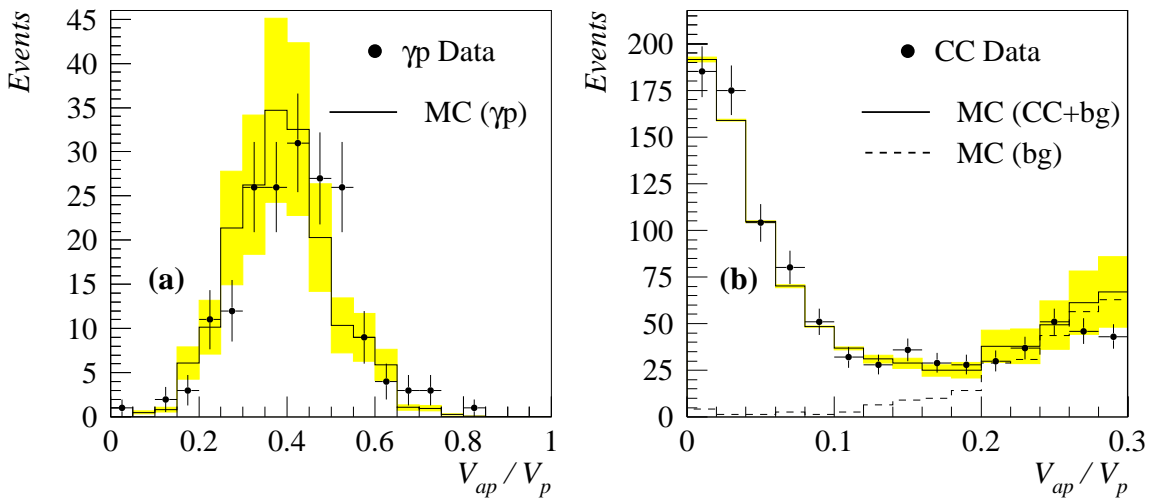,height=9.cm}}
\put(20,25){\bf (a)}
\put(96,25){\bf (b)}
\end{picture}
\end{center}
\caption
{\sl (a) Distribution of $V_{ap}/V_{p}$ 
for tagged \photoproduction\ events passing the CC selection except for
the $V_{ap}/V_{p}$ cut.
The data (points) are compared to the Monte Carlo (MC) simulation
(histogram) of the \photoproduction\ background. 
(b) Distribution of $V_{ap}/V_{p}$ for the CC event sample.
The data (points) are compared to the simulation (histogram)
which includes the CC and the background (${\rm bg} \equiv$ 
\photoproduction\ + NC) events.
A cut $V_{ap}/V_{p}<0.15$ is applied in the CC selection.
The simulation is normalized to the integrated $ep$ luminosity.
The shaded error bands represent the systematic 
uncertainty of the background simulation.
}
\label{fig:sap}
\end{figure}

In CC events the energy flow is concentrated in the hemisphere opposite 
to the transverse momentum of the scattered neutrino, resulting in a low
value for $V_{ap}/V_{p}$
(eqs.~\ref{eq:vp},~\ref{eq:vap}) while in \photoproduction\ 
and NC background events it is more isotropic,
giving values of $V_{ap}/V_{p}$ close to $0.5$. 
This is seen in fig.~\ref{fig:sap}a where the $V_{ap}/V_p$
distribution is shown for \photoproduction \ events 
for which the scattered positron is measured 
in the electron tagger
and which pass the CC selection, apart
from the cut on $V_{ap}/V_p$.
The observed distribution is
well described in shape and
normalization by the \photoproduction \
simulation. 
An error of $\pm 30\%$ on the simulation of the
photoproduction background is shown on the figure.

The same distribution, shown in fig.~\ref{fig:sap}b for all events
which pass the CC selection apart from the cut on $V_{ap}/V_{p}$,
is well described by the simulation
of CC and background (${\rm bg} \equiv$ \photoproduction \ + NC)
events.
The cut of $V_{ap}/V_p < 0.15$, applied in the CC selection, rejects a
large fraction of this background. According to the simulation, about
$70$ ($95$)\% of the CC events with $12<$ \ptmiss\ $<15 \ \Gev $
(\ptmiss\ $>25 \ \Gev $) survive this cut. To evaluate the systematic
uncertainty in the CC selection efficiency which is introduced by this
requirement, the cut value of $V_{ap}/V_{p}$ is varied between $0.13$
and $0.17$ in the simulation while keeping the value fixed for the data.
A variation in the efficiency with which CC events are retained of
$5$ ($2$)\% at low (high) \ptmiss , averaged over $y_h$, is then observed.

In the CC analysis residual background due to NC interactions is
rejected by removing events with
only one track with azimuthal angle opposite to
the hadronic final state ($|\varphi_{track}-\varphi_{h}|>160^{\circ}$).
The azimuthal angle $\varphi_{h}$ of the hadronic final
state is defined by $\tan \varphi_{h}= p_{y,h}$/$p_{x,h}$. Events
with large \ptmiss\ and isolated high momentum leptons
observed recently~\cite{isolated} are removed in this
analysis by applying the selection procedure for such events which is
used there.
The additional inefficiency introduced into the CC selection due to
these requirements is less than $1\%$.
The remaining contamination due to $ep$-induced background
is evaluated from the simulation and
statistically subtracted from the data.
The background corresponds to 
about $10\%$ at the lowest $Q^2$ values and less than $2\%$ 
for $Q^2 > 1000$~\Gevv .

In the NC analysis after all selection cuts,
the only significant background 
is due to events from 
photoproduction processes, in which the scattered positron
escapes the detector along the beam-pipe and
one of the particles of the hadronic final state is misidentified as the
scattered positron.
As in the CC case this background, determined from the 
simulation, is controlled using the sub-sample of about $10\%$ of 
the $\gamma p$ events in which the scattered positron
is detected in the electron tagger, and is subtracted from the data.
It amounts to less than $1\%$ in the total sample and to at most $5\%$ 
in the highest $y$ bins at $Q^2 < 1000$~\Gevv .

\subsection{ Detector Alignment and Calibration}
\label{calib}

At high $Q^2$ the scattered positron and the hadronic final
state are predominantly measured with the LAr calorimeter. 
From test beam data the initial electromagnetic 
and hadronic energy scales were
established with an uncertainty
of about $3\%$ for electrons and pions of energy between
$4$ and $205 \ \Gev\ $~\cite{e-beam,ref.testbeam}. 
These energy scales were verified {\it in situ} at HERA using
the 1994 data~\cite{H194}.
The tracking detectors are used wherever possible to 
improve these measurements
by making use of their good angular precision for the scattered positron, 
and by making use 
of their good precision in both angle and momentum measurement for 
the determination of the hadronic final state energy. This section is
concerned with the
method used to reconstruct the positron angle, 
the absolute calibration of the positron energy, 
and the relative hadronic energy scale between data and
simulation. 
More details of the energy calibration procedures can be found
in~\cite{gayler}.

\subsubsection{Positron Angle Measurement}
\label{thecalib}

The polar angle of the scattered positron is determined using the central
tracking detectors when its track is reconstructed using hits 
in the $3$ central chambers CJC, CIZ and COZ. 
When the positron's track is less well constrained, the angle
is determined from the position of the positron energy cluster in the LAr 
calorimeter and the vertex reconstructed using tracks from charged particles
in the event.

By minimizing the spatial discrepancy between the positron track and the 
location of the calorimeter cluster, 
the alignment of the tracking detectors relative 
to the LAr calorimeter was established to within $1 \ {\rm mm}$ in the
$x,y$ and $z$ directions.

Following this alignment the precision of the angle
measurement with the tracking detectors 
\begin{figure}[htb] 
\begin{center} 
\begin{picture}(100,70)(0,0)
\setlength{\unitlength}{1 mm}
\put(-32,-8){\epsfig{file=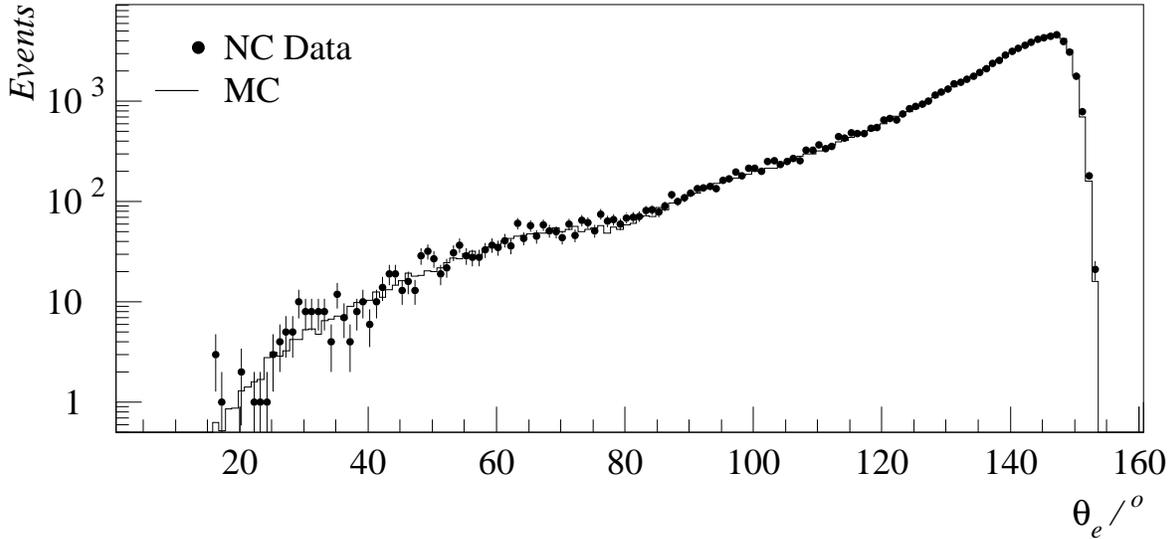,width=16cm,
bbllx=75pt,bblly=300pt,bburx=475pt,bbury=475pt}
}
\end{picture}
\end{center} 
\caption[]{
\sl {Distribution of polar angle of the scattered positron.
The data (points) are compared to the simulation (histogram) which is
normalized to the integrated $ep$ luminosity. 
}}
\label{fig:e-theta}
\end{figure}
(with the calorimeter cluster and event vertex) 
is better than $1$ ($3$) ${\rm{mrad}}$.
The proportion of scattered positrons in which $\theta_e$ 
is determined from the cluster and event vertex
is about $40\%$ in the central region. 
This proportion increases at smaller $\theta_e$ 
and is $100\%$ for $\theta_e < 35{^\circ}$. 
The vertex is determined from the tracking detectors with a precision of 
approximately $3 \ {\rm{mm}}$ in $z$ and $1 \ {\rm{mm}}$
in $x$ and $y$. Because the mean of the distribution of 
event vertices depends on the characteristics of the 
stored positron and proton beams, the vertex distribution
is determined from the data for every beam storage, 
and in event simulation 
the vertex distribution is adjusted to follow these changes.

The $\theta_e$ distributions for the data and for 
the simulation are shown in fig.~\ref{fig:e-theta}. 
The simulation describes the data well throughout 
the complete angular range.

\subsubsection{Positron Energy Measurement}
\label{ecalib}
For the present cross-section analysis
the calibration constants and their uncertainties have been 
improved compared to the previous H1 measurements by
making use of the increased NC event sample and
exploiting the overconstrained kinematic reconstruction.

\begin{figure}[htb] 
\begin{center} 
\epsfig{file=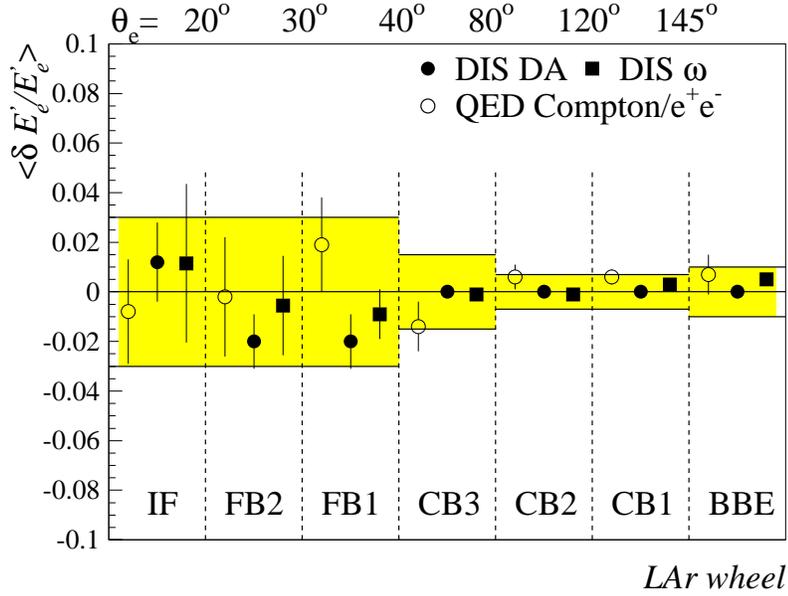,width=12cm}
\end{center} 
\caption[]{\label{fig:cal}
\sl {Comparison of the electromagnetic energy scale as determined by
different calibration methods. Shown is
 $\langle{\delta E_e^{\prime}/E_e^{\prime}} \rangle$, the mean
fractional energy shift of the different methods from the absolute energy
scale.
The shaded error band shows
the systematic uncertainty on the energy scale quoted on this measurement, 
which varies from $0.7$ to $3\%$, depending on the position in the
detector. }
}
\end{figure}

Before the {\it in situ} calibration discussed below, the measured
positron energy is corrected for energy loss in the material in front
of the calorimeter (between $0.7$ and $2.5X_{\circ}$). Further energy
loss can occur in the crack regions between the calorimeter modules in the
$z$ and $\varphi$ directions. To limit the size of the corrections
which occur because of the crack regions, the impact position of
the positron track on the calorimeter is required to lie outside a
fiducial area of $\pm 2^{\circ}$ around a $\varphi$-crack and $\pm 5 \ 
{\rm{cm}}$ around the $z$-crack located between the CB2 and CB3 wheels
of the LAr calorimeter (see fig.~\ref{fig:cal} for the angular
coordinates of the seven electromagnetic wheels of the LAr
calorimeter)~\cite{LARC}.

For the {\it in situ} calibration of the barrel region 
($\theta_e > 40^{\circ}$), only the NC events
with $y_{\Sigma} < 0.3 $ ($y_{\Sigma} < 0.5$)
in the region of 
$80^{\circ} \ \lapprox \ \theta_e \ \lapprox \ 153^{\circ}$ 
($40^{\circ} \ \lapprox \ \theta_e \ \lapprox \ 80^{\circ}$) are used.
For these $y$ values the energy of the scattered positron is predicted 
precisely by the double-angle (DA) method~\cite{DA}
in which the kinematic variables are determined solely
from $\theta_e$ and $\gamma_h$. 
The calibration is achieved by constraining 
the mean of the $E^{\prime}_e/ E_{\rm{DA}}$ distribution to $1$
via small local adjustments of the
calibration constants.
These constants are determined 
in 
finely segmented $z$ and $\varphi$ regions 
defined by the impact position of the positron track on 
the LAr calorimeter. An analogous procedure was performed
for the simulation.
The calibration constants
vary typically by $\pm 1\%$ around their average values, 
except in the regions close to the $z$-cracks,
where the corrections may reach up to $8\%$~\cite{beate}. 
Outside these regions the 
calibrated energy response is described by the
simulation within $0.5\%$.
The absolute calibration is obtained by applying 
in addition 
corrections of about $1\%$, derived from the simulation,
which take into account effects from 
initial state QED radiation and small biases originating from the 
imperfect $\gamma_h$ reconstruction.

Due to the limited number of events with positrons in the forward region
($\theta_e < 40^{\circ}$) two event samples, elastic QED
Compton and exclusive two photon $e^+e^-$ pair production, are used in
addition to the DIS events. 
The requirement of transverse
momentum balance allows the energy of the more forward
electromagnetic energy deposit to be determined from the well
calibrated backward cluster. 
For the DIS events the $\omega$ kinematic reconstruction
method~\cite{omega} is used to determine the calibration constants
instead of the DA method since it is by design less
sensitive to the effects of initial state QED radiation and is
therefore more reliable when there are low statistics.
A single calibration constant is
determined for the entire forward region.

\begin{figure}[ht] 
\begin{center} 
\begin{picture}(100,70)(0,0)
\setlength{\unitlength}{1 mm}
\put(-30,-8){\epsfig{file=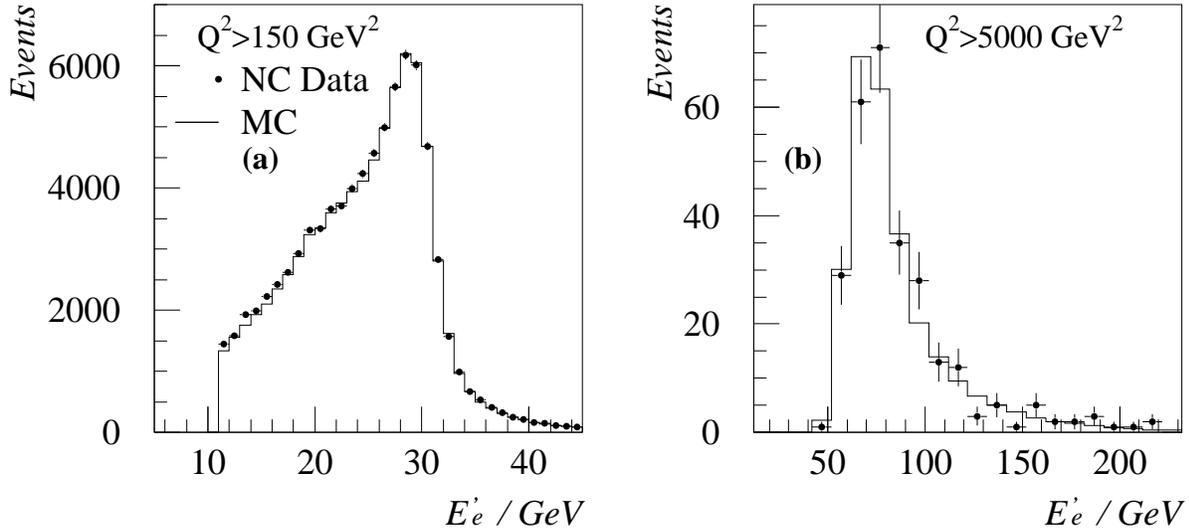,width=16cm,
bbllx=75pt,bblly=300pt,bburx=475pt,bbury=475pt}}
\put(3,42){\bf (a)}
\put(75,42){\bf (b)}
\end{picture}
\end{center} 
\caption[]{\sl Energy 
spectrum of the scattered positron at (a) $Q^2 > 150$~\Gevv , and
(b) $Q^2 > 5000$~\Gevv . 
The data (points) are compared to the simulation (histogram) which is
normalized to the integrated $ep$ luminosity.}
\label{fig:e-energy}
\end{figure}

After the application of these calibration procedures, the
positron energy scale is checked
for each calorimeter wheel
using the elastic QED-Compton and 
$e^+e^-$ event sample and, separately,
the $\omega$ method for the DIS sample.
The results from all the
different methods are found to be in good agreement,
as shown in fig.~\ref{fig:cal}. An error of $\pm
0.7$ ($1.0, 1.5, 3.0$)$\%$ on the absolute electromagnetic energy
scale of the CB1--CB2 (BBE, CB3, FB1--IF) wheels of the detector is
therefore assigned. The uncertainties in the electromagnetic energy
scale increase towards the forward region due to the decreasing number
of events. The resulting energy spectra are presented for $Q^2>150$
$(5000)$~\Gevv\ in fig.~\ref{fig:e-energy}a(b), and are well
described by the simulation within the normalization uncertainty of
$\pm 1.5\%$.

\subsubsection{Hadronic Energy Measurement}
\label{hcalib}

The optimal measurement of the hadronic final state energy
is obtained after applying specific techniques to the
reconstruction of the calorimeter and
tracking information, as described in the following.

\begin{figure}[b] 
\begin{center} 
\setlength{\unitlength}{1 mm}
\begin{picture}(140,140)(0,0)
\put(-10,-10){\epsfig{file=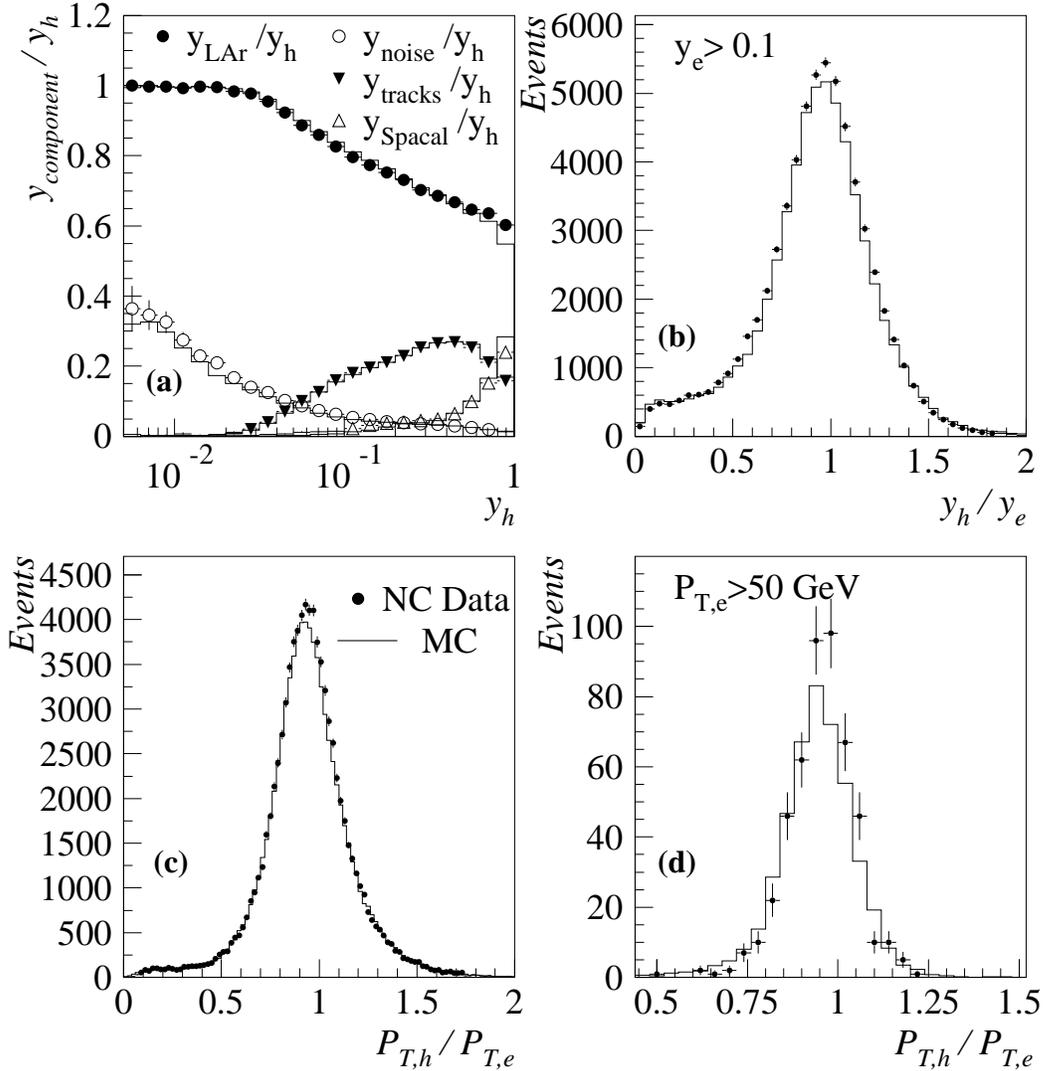,width=16cm}}
\put(17,84){\bf (a)}
\put(85,90){\bf (b)}
\put(18,20){\bf (c)}
\put(85,20){\bf (d)}
\end{picture}
\end{center} 
\caption[]{
\sl {(a) Distribution of the fraction of $y_h$ contributed by the tracks
($y_{tracks}$), the LAr ($y_{LAr}$) 
and the SPACAL calorimeters ($y_{Spacal}$), and the fractional 
contribution of the subtracted noise ($y_{noise}$). 
(b) Distribution of $y_h/y_e$ for $y_e>0.1$.
(c) Distribution of $P_{T,h}/P_{T,e}$ for the complete NC sample,
(d) for the sub-sample at $P_{T,e} > 50 $ \Gev . 
The data (points) are compared to the simulation (histogram) which is
normalized to the integrated $ep$ luminosity.}}
\label{fig:pth50}
\end{figure}

Since the H1 LAr calorimeter is non-compensating, weighting
algorithms are applied to the hadronic clusters in order to 
improve the energy resolution~\cite{ref.testbeam,weighting}.
A further improvement in
energy resolution of about $10$ to $20\%$, for events having a
$P_{T,h}$ between $10$ to $25$ \Gev , is obtained by using a
combination of the energies of low transverse momentum particles ($P_T
< 2 \ \Gev $) measured in the central tracking detector with the energies
deposited by other particles of the hadronic final state measured in
the calorimeter. To avoid ``double counting'', the energy measured in
the electromagnetic (hadronic) LAr calorimeter in a cylinder of $15 \,
(25) {\rm{cm}}$ in radius around the axis given by the direction of
a low transverse momentum track is not included, except if the total
energy in the cylinders is greater than the energy of the track, in
which case only the calorimetric measurement is used.
The fraction of $y_h$ measured
by each of the subdetectors (LAr, tracks, SPACAL)
is shown in fig.~\ref{fig:pth50}a 
to be well described by the simulation in
the range $0.005 \le y_h \le 0.9$. 
The contribution of the SPACAL calorimeter 
is below $10\%$ except at high $y$.

At low $y \ \lapprox \ 0.05$, where hadrons are produced in the forward 
direction and little energy is deposited in the
calorimeter, the 
measurement of the kinematic variables is distorted by the presence of
``noise'' in the calorimeter cells due either to the electronics of
the calorimeter readout, or to the secondary scattering of final state
particles into the calorimeter. Both sources of noise are included in
the simulation.
The noise is reduced by suppressing isolated low
energy deposits~\cite{diff-diss}, which results in a significant
improvement of the reconstruction of the kinematic variables at low $y$.
The fraction of $y_h$ identified as noise is shown 
in fig.~\ref{fig:pth50}a to be described by the simulation.
The effect of a variation of $\pm
25\%$ of the subtracted noise contribution is included in the
systematic error.

The {\it in situ} calibration of the hadronic energy scale~\cite{gayler}
is made by comparing the transverse momentum of the precisely
calibrated positron (section~\ref{ecalib}) to that of the
hadronic system in NC events. Calibration constants are determined
for each of the 7 electromagnetic and 8 hadronic wheels. The
calibration factor for each wheel is evaluated using the ratio
$P_{T,h}/P_{T,e}$ of each event weighted with the fraction of
$P_{T,h}$ carried by the wheel. The calibration constants are
adjusted iteratively until the average ratio $\langle P_{T,h}/P_{T,e}
\rangle$ for the data equals that of the simulation in all regions
of the detector.

Detailed studies of the dependence of $P_{T,h}/P_{T,e}$ and $y_h/y_e$ on
$P_{T}$ and $\gamma_h$ justify a systematic uncertainty on the relative
hadronic energy scale of the LAr calorimeter of $2\%$.
A further
confirmation of this scale uncertainty is obtained using the
topology of the NC events which can be divided in two samples.
In the sample of  events with only one well reconstructed jet, 
the jet direction directly determines the wheel containing the 
maximum amount of transverse energy, allowing the
corresponding wheel calibration constant
to be checked precisely. 
In the sample of events with a multijet topology
the $P_{T,h}/P_{T,e}$ distribution has also been observed to be better
described by the simulation after applying the hadronic
calibration.

The dependence of the calibration on the usage of two different
hadronic final state models in the simulation, one which assumes QCD
matrix elements and parton showers (MEPS) as implemented in
LEPTO and the other which assumes the colour dipole model in its
ARIADNE implementation, has been studied and found to be negligible. 

The quality of the resulting hadronic final state reconstruction is
illustrated in fig.~\ref{fig:pth50}b,c,d.
In fig.~\ref{fig:pth50}b the $y_h/y_e$ distribution for $y_e
> 0.1$ is shown. In this distribution the hadronic energy enters with a
different angular weight than in the $P_{T,h}/P_{T,e}$ distribution.
The agreement observed between data and simulation shows that the
hadronic calibration is valid for the energy itself, and not only for
the transverse energy. Fig.~\ref{fig:pth50}c(d) shows the
$P_{T,h}/P_{T,e}$ distribution in the complete NC event sample (with
$P_{T,e}> 50 \ \Gev $). In both distributions, the data are
described by the simulation within the quoted $2\%$ uncertainty. 

\section{Cross-Section Measurement Procedure}

\subsection{Cross-Sections and Structure Functions}
\label{theory}

In this section the cross-section definitions are introduced together
with the procedure adopted 
for the treatment of radiative corrections.
The measured cross-sections are:
\begin{itemize}
\item 
the NC (CC) double differential cross-section
$d^2\sigma_{NC(CC)}/d xd Q^2$;
\item
the NC (CC) single differential cross-sections 
$d \sigma_{NC(CC)}/ d Q^2$ and $d \sigma_{NC(CC)}/d x$.
\end{itemize}
These cross-sections are presented in this paper after corrections 
for the effects
of QED radiation have been made. They are derived from the ``initial" 
cross-sections which are determined using
the measurement procedure described in section~\ref{f2meas}. Thus
the double differential NC (CC) cross-sections are defined as
\begin{equation}
\label{init}
\frac{{ d}^2\sigma_{NC(CC)}}{{d}x\;{ d}\QQ} 
= 
\left(\frac{{ d}^2\sigma_{NC(CC)}}{{d}x\;{ d}\QQ}\right)_{initial}
\left[ 1+\delta^{qed}_{NC(CC)}(x,\QQ )\right]^{-1}
\, .
\end{equation}
The $\delta^{qed}_{NC}$ term includes 
the effects of photon emission from the lepton line, the effects
of the photonic lepton vertex corrections combined with the self energies 
of the external fermion lines,
and the effects of the fermion loops of the 
exchanged photon self energy.
The $\delta^{qed}_{CC}$ term 
includes the leptonic part of the
${\cal O}(\alpha)$ photonic correction to CC processes~\cite{spies2,bard}.
These radiative corrections\footnote{
The radiative corrections due to the exchange of two or more 
photons between the
lepton and the quark lines
are small and are included in the systematic uncertainty of the
radiative corrections.}
are calculated
using DJANGO and verified with the
HECTOR~\cite{hector} program.
The weak radiative corrections $\delta^{weak}_{NC(CC)}$,
which are defined in~\cite{spies} and which are small
(of the order of $1\%$), have not been 
applied to the measured cross-sections.

When extracting the structure functions of the proton
from cross-section measurements, the weak radiative corrections
are, however, applied. The Born cross-section is then defined as
\begin{equation}
\label{eq.toborn}
\left(\frac{{ d}^2\sigma_{NC(CC)}}{{d}x\;{ d}\QQ}\right)_{Born} 
= 
\frac{{ d}^2\sigma_{NC(CC)}}{{d}x\;{ d}\QQ} \hspace*{0.3cm}
\left[ 1+\delta^{weak}_{NC(CC)}(x,\QQ )\right]^{-1}
\, .
\end{equation}

The Born double differential NC cross-section 
for $e^{+}p\rightarrow e^{+}X$
can be written as 
\begin{eqnarray}
\label{Snc1}
\left( \frac{{ d}^2\sigma_{NC}}{{d}x\;{ d}\QQ}\right)_{Born}
& = & \frac{2\pi \alpha^2}{x} 
\left(\frac{1}{\QQ}\right)^2 \hspace*{0.2cm} 
\phi_{NC}(x,Q^2) \, ,
\end{eqnarray}
where
\begin{eqnarray}
\label{Snc}
\phi_{NC}(x,Q^2) & = & 
Y_+ \Ftwo (x,\QQ )-Y_{-}x\Fz (x,\QQ )-y^2 \FL (x,\QQ ) .
\end{eqnarray}
Here $\alpha$ is the fine structure constant taken to 
be $\alpha \equiv \alpha(Q^2=0)$.
The ``structure function term''
$\phi_{NC}(x,Q^2)$ 
is a linear combination of 
the \Ftwo\ structure function, 
the longitudinal structure 
function \FL , 
and the $x$\Fz\ structure function
which in the Standard Model is significant only when $Q^2$ is sufficiently
large to render $Z^\circ$ exchange non-negligible.
The helicity dependences of the electroweak interactions 
are contained in the functions $Y_{\pm} = 1 \pm (1-y)^2$. 

In leading order QCD, the structure function term is simply
related to the sum of the light quark densities,
weighted with the squared quark charges,
when neglecting $Z^\circ$ exchange: 
\begin{equation}
\left(\phi_{NC}\right)_{LO}= 
\left[1+(1-y)^2\right] \; x \; 
\left[ 
\frac{4}{9}\left(u+c+\bar{u}+\bar{c}\right)+
\frac{1}{9}\left(d+s+\bar{d}+\bar{s}\right)
\right] \, .
\label{SncLO}
\end{equation}
At high $x$ the structure function term $\phi_{NC}$
depends
predominantly on the valence distribution 
of the $u$ quark.

For unpolarized beams, 
the structure functions \Ftwo\ and $x$\Fz\ can be
decomposed, taking into account $Z^\circ$ exchange,
as~\cite{klein}
\begin{eqnarray}
\label{f2p}
\Ftwo \equiv & \Fem & - \ v \ \frac{\kappa_w \QQ}{(\QQ + M_Z^2)} 
\hspace*{0.2cm} \Fint
+ (v^2+a^2) \left(\frac{\kappa_w Q^2}{\QQ + M_Z^2}\right)^2 \Fwk \\
\label{f3p}
x\Fz \equiv & & - \ a \ \frac{\kappa_w \QQ}{(\QQ + M_Z^2)} x\Fzint
 + \hspace*{0.3cm} (2 v a) \hspace*{0.3cm}
 \left(\frac{\kappa_w Q^2}{\QQ + M_Z^2}\right)^2 x\Fzwk \, ,
\end{eqnarray}
where $M_Z$ is the mass of the \Zero ,
$\kappa_w= 1/(4 \sin^2{\theta_w} \cos^2{\theta_w}$)
is a function of the Weinberg angle ($\theta_w$),
and $v$ and $a$ are the vector and axial vector 
couplings of the electron to the
$Z^{\circ}$. They are related to the weak isospin of the electron, 
$I_3=-\frac{1}{2}$,
namely $v= I_3 + 2\sin^2{\theta_w}$ and $a = I_3$~\cite{gmu}.
The electromagnetic structure function \Fem\ originates from photon exchange
only, and the functions
\Fwk\ ($x\Fzwk $) and \Fint\ ($x\Fzint $) are the contributions to \Ftwo\
($x\Fz $) due to 
$Z^\circ$ exchange
and $\gamma Z^\circ$ interference respectively. 
Note that for unpolarized beams, 
\Ftwo\ is the same for electron and for 
positron scattering, while the $x$\Fz\ term in eq.~\ref{Snc} changes sign.

The NC ``reduced cross-section''
is defined from the measured ${{d}^2 \sigma_{NC}}/{{d}x{d}Q^2}$
in order to reduce the strong $Q^2$ dependence originating
from the propagator:
\begin{equation} 
\label{Rnc}
\tilde{\sigma}_{NC}(x,Q^2) \equiv \frac{1}{Y_+} \ 
\frac{ Q^4 \ x }{2 \pi \alpha^2}
 \  \frac{{d}^2 \sigma_{NC}}{{d}x{d}Q^2} \, .
\end{equation}
In the major part of the $(x,Q^2)$ domain
\Fem\ is the dominant
component of the structure function term $\phi_{NC}(x,Q^2)$,
and $\tilde{\sigma}_{NC}$ is conveniently expressed as 
\begin{equation}
\label{Rnc2}
\tilde{\sigma}_{NC}
\ = \ \Fem \ {(1 
+ \Delta_{F_2} + \Delta_{F_3}+ \Delta_{F_L})}
\ { (1+\delta^{weak}_{NC})} \
= \ \Fem \ (1 
+ \Delta_{all}) \, ,
\end{equation}
where the $\Delta_{F_2}$ and $\Delta_{F_3}$ terms originate from the 
\Fint , \Fwk\ and
\Fzint, \Fzwk\ functions defined in eq.~\ref{f2p} and~\ref{f3p},
and the $\Delta_{F_L}$ term from the longitudinal 
structure function \FL . 
Values of each of these terms obtained from the NLO QCD Fit described 
in section~\ref{QCDA} are given in table~\ref{tabf2}.

In the kinematic range investigated the effects of $Z^\circ$ exchange
($\Delta_{F_2} + \Delta_{F_3}$) on $\tilde{\sigma}_{NC}$ are expected
to be $ \ \le \ 5\%$ for $Q^2 < 5000$~\Gevv\ (table~\ref{tabf2}).
It is thus possible to extract \Fem\ from the
measured cross-section with little uncertainty in this $Q^2$ range. At
higher $Q^2$ values the contribution of the $x$\Fz\ term 
results in a significant reduction of 
the $e^{+}p$ cross-section. The determination of $F_2$
then relies strongly on the calculation of $\Delta_{F_2}$ and
$\Delta_{F_3}$. In QCD calculations the $\Delta_{F_L}$ term is small
and decreases at constant $y$ with increasing $Q^2$. It reaches
$6\%$ for $y \ge 0.65 $ and $Q^2 \le 1500 \ {\rm{GeV}}^2$ but is
negligible for $y \ \lapprox \ 0.4$.

The Born double differential CC cross-section 
for $e^+ p \rightarrow {\bar{\nu}} X$ can be written as
\begin{equation}
\left(\frac{{ d}^2\sigma_{CC}}{{d}x\;{ d}\QQ}\right)_{Born} =
 \frac{G_F^2}{2\pi x} 
\left(\frac{M_W^2}{M_W^2+Q^2} \right)^2 
\phi_{CC}(x,Q^2) \, , 
\label{Scc}
\end{equation}
where $G_F$ is the Fermi coupling constant and
the structure function term $\phi_{CC}(x,Q^2)$ can be decomposed 
into structure functions in a similar way as 
$\phi_{NC}(x,Q^2)$~\cite{blublu}.

For CC interactions a reduced cross-section is also introduced:
\begin{equation}
\label{Rcc}
\tilde{\sigma}_{CC}(x,Q^2) \equiv 
\frac{2 \pi x}{ G_F^2}
\left(\frac {M_W^2+Q^2} {M_W^2} \right)^2
  \frac{{d}^2 \sigma_{CC}}{{d}x{d}Q^2} \, .
\end{equation}
It is directly related to the CC structure function term by
\begin{equation}
\label{Pcc}
 \phi_{CC}(x,Q^2) 
\left[ 1+\delta^{weak}_{CC}(x,\QQ ) \right] 
= \tilde{\sigma}_{CC}(x,Q^2) 
\, .
\end{equation}

In leading order QCD, neglecting the effect of quark mixing
and the contribution of heavier quarks, the 
CC structure function term for $e^+p$ scattering is related 
to the quark densities:
\begin{equation}
\label{SccLO}
\left( \phi_{CC}\right)_{LO}= x \;
\left[ (\bar{u}+\bar{c})+(1-y)^2(d+s) \right] \, .
\end{equation}
At high $x$ the structure function term $\phi_{CC}$
depends predominantly on the valence distribution of the $d$ 
quark.

\subsection{QCD Analysis Procedure}
\label{QCDA}
 
Comparison of the Standard Model with the measurements of the NC and
CC $ep$ cross-sections depends both on the model's explicit
predictions for the interaction of a positron with a quark and on the
partonic content of the proton. The parameters of the electroweak
theory which describes positron-quark scattering in the Standard
Model have been measured precisely, and are therefore fixed 
to their world average values~\cite{gmu} in this
comparison. The parton distribution functions (PDFs), which describe
the partonic structure of the proton, are not predicted by QCD and so
must be obtained from the data.
In order to obtain the PDFs together with their uncertainties,
two NLO QCD fits are performed:
\begin{itemize}
\item
the first fit (Low $Q^2$ Fit) is made with published low $Q^2$ DIS data;
the proton (\Fem ) and deuteron ($F_2^d$)
data from the BCDMS~\cite{bcdms} and NMC~\cite{nmc} experiments are used, 
together with
the 1994 \Fem~measurements of H1~\cite{H194} 
at $Q^2 < 150$~\Gevv ;
\item
the second fit (NLO QCD Fit) includes the high $Q^2$ NC and CC double
differential cross-sections presented in this paper in addition to the
datasets used in the Low $Q^2$ Fit.
\end{itemize}

Since the emphasis of this study is on the new data
entering the fit which are at high $Q^2$, far above the squared 
masses of the charm ($c$) and bottom ($b$) 
quark, an approach is used in which all quarks are taken to be massless
within the DGLAP equations
and a cut of $Q^2 > 10$
${\rm GeV}^2$ is applied to the datasets. At high $x$ and low $W^2$ 
($W^2 \equiv {Q^2[1-x]}/{x}$)
non-perturbative effects may have a large influence. Therefore only the data
having $W^2 \geq 20$~\Gevv\ and $x<0.7$ are
used in the fits. The fixed target data are corrected for 
target mass effects using the Georgi-Politzer approach~\cite{tmc}, and
for deuteron binding effects using the parameterization 
obtained with the method of~\cite{deuteron} 
applied to SLAC measurements~\cite{gomez}.
The effect of the deuteron corrections on the fit result 
are negligible for NC and up to $7\%$ (at $x=0.4$) for CC.

For these fits, the DGLAP evolution equations~\cite{bb.dglap} are solved in the NLO $\overline{MS}$ factorization scheme using the
QCDNUM~\cite{qcdnum} program. The results obtained have been cross-checked
using an independent program~\cite{zomi}.
The strong coupling constant $\alpha_s$ is evolved according to QCD with 
the constraint $\alpha_s(M_Z^2)=0.118$. 
A starting scale of $Q_0^2=4$~\Gevv\ is
taken at which four PDFs are parameterized. These are the
$u$ and $d$ valence quarks ($xu_v$ and $xd_v$), the gluon ($xg$), and
the sea quark densities ($x{S} \equiv 
2 x [\bar{u}+\bar{d}+\bar{s}+\bar{c}]$).
An asymmetry between the $\bar{d}$ and $\bar{u}$ PDFs is enforced by 
using the $\bar{d}-\bar{u}$ parameterization from~\cite{mrst}
taking into account the different starting scale.
The strange ($s$) quark density is constrained to be
$\bar{s}=\bar{u}/2$ at $Q_0^2$~\cite{sbyu}.
The $xc$ contribution is normalized to $2\%$ of the sea quark density at
$Q_0^2$ since this gives a good description of the H1
measurements~\cite{fcharm} of the charm induced
structure function $F_2^{c}$. 
The $xb$ density is evolved according to the DGLAP equations
assuming that $b(x,\QQ)=0$ for \QQ$< 25$~${\rm GeV^2}$. 

The functional forms of the parton densities are parameterized as
\begin{eqnarray}
xu_v(x,Q_0^2) &=& A_{u_v}x^{B_{u_v}}(1-x)^{C_{u_v}}(1+D_{u_v}x^{E_{u_v}})\\
xd_v(x,Q_0^2) &=& A_{d_v}x^{B_{d_v}}(1-x)^{C_{d_v}}(1+D_{d_v}x^{E_{d_v}})\\
x{S}(x,Q_0^2) &=& A_Sx^{B_S}(1-x)^{C_S}\\
xg (x,Q_0^2) &=& A_gx^{B_g}(1-x)^{C_g} \, .
\end{eqnarray}
The parameters 
$A_{u_v}$ and $A_{d_v}$ are determined
by enforcing the valence counting rules which require
\mbox{$\int_0^1u_v dx=2$} and \mbox{$\int_0^1d_v dx=1$}.
The momentum sum rule
allows the determination of one further normalization parameter,
taken to be $A_g$. 

The fits are performed using the 
MINUIT~\cite{minuit} program which minimizes the $\chi^2$ defined
from the data value (${f^{data}_{i,j}}$) 
and the theoretical expectation (${f^{theo}_{i,j}}$)
of the measured point~$i$ in the dataset~$j$, 
normalized by the quadratic sum ($\oplus$) of its statistical 
($\delta f_{i,j}^{sta}$) and 
uncorrelated systematic  ($\delta f_{i,j}^{unc}$) errors:
\begin{equation}
\label{eq.chi2}
\chi^2 = \sum_{j=1}^{N_{dataset}} \left[
\sum_{i=1}^{N^{data}_j} \left(\frac{{f^{data}_{i,j}}\times 
(1+\delta {\cal{L}}_j/{\cal{L}}_j) - {f^{theo}_{i,j}} }
{\delta f_{i,j}^{sta} \oplus \delta f_{i,j}^{unc}}\right)^2
+ \left(\frac{\delta {\cal{L}}_j}
{\delta {\cal{L}}^0_j}\right)^2 \right] \, .
\end{equation}
The number of datasets and the number of data points in a dataset $j$
are defined here as $N_{dataset}$ and $N^{data}_j$.
The terms ${\delta{\cal{L}}^0_j}/{\cal{L}}_j $
are the  luminosity uncertainties of each dataset $j$
($1.5\%$ for the high $Q^2$ data, 
$1.5\%$ for the H1 1994 data, $3\%$ for BCDMS, $2.5\%$ for NMC).
The terms $(1+\delta {\cal{L}}_j{/\cal{L}}_j)$ 
are the normalizations of the datasets
which are allowed to vary 
according to the
quoted luminosity uncertainties

The results of the Low $Q^2$ Fit are presented in table~\ref{tab:tabfit}
in which the $\chi^2$ is given for each dataset, together
with their optimal relative normalization, according to the criteria
discussed above. 
The total $\chi^2$ per degree of freedom 
(${\rm{ndf}}$)
is $548/(529-13)=1.06$ when considering the uncorrelated
error of the data (obtained from the quadratic sum of the
statistical and systematic errors which are uncorrelated from one 
bin to another) as given in eq.~\ref{eq.chi2}. 
If the $\chi^2 / {\rm{ndf}}$ is
recalculated using the total error of the data (obtained by adding the
bin to bin correlated systematic error in quadrature to the
uncorrelated error) its value decreases to $0.78$.
The results of the NLO QCD Fit are presented in section~\ref{CSM2}.
\begin{table}[ht]
\begin{center}
\begin{tabular}{|c|c|c|c|c|c|c|} \hline
Experiment & {\small{H1 94}} & {\small{BCDMS-p}} & {\small{BCDMS-D}} &
 {\small{NMC-p}}& {\small{NMC-D}} & Total \\ \hline
data points  & $ 77 $& $ 139$& $ 133$& $ 90$& $ 90$& $529$\\ 
\hline
$\chi^2$ (unc. err.) & $ 67 $& $ 102$& $ 111$& $143 $& $125$& $548$\\ 
\hline
$\chi^2$ (total err.)& $ 39 $& $ 89$& $ 98$& $ 93$& $ 77$& $396$\\ 
\hline
normalization & $1.01$& $0.97$& $0.98$& $0.99$&$0.99$&\multicolumn{1}{c}{} \\ 
\cline{1-6}
\end{tabular}
\end{center}
\caption{\sl Results of the Low $Q^2$ Fit. For each experiment 
the following quantities are given: the number of data points, 
the contribution to the $\chi^2$ using the
uncorrelated errors of the data (unc. err.) as obtained from the
statistical errors and uncorrelated systematic errors added in quadrature,
the contribution to the $\chi^2$ using 
the total errors and the optimal normalization according to the fit. }
\label{tab:tabfit}
\end{table}

The uncertainty on the Standard Model expectation which is
used to interpret the data in section 4 is estimated 
from the experimental errors of the data points and
by varying the theoretical assumptions of the QCD fit.

The ``experimental error
of the fit'' is obtained by adding in quadrature the error from the QCD
fit (performed with uncorrelated errors) to the contributions due to 
each bin to bin correlated systematic errors on the measurement. 
These correlated systematic errors are 
taken into account by repeating
the QCD fit after varying the data points coherently under the influence
of each error source separately.

The ``theoretical error of the fit'' is obtained by repeating
the QCD fit after varying each of the fit assumptions in turn:
the value of 
$\alpha_s(M_z)$ is varied by $\pm 0.003$; the $s/\bar{s}$ contribution 
is changed by $\pm 25\%$; the $c/\bar{c}$ contribution at the starting scale
is multiplied by a factor of $2$;
an uncertainty of $\pm 50\%$ of the deuteron binding corrections 
is considered; 
the treatment of the $\bar{d}/\bar{u}$ asymmetry is changed 
to that given in~\cite{mrsr2} taking into account the different starting
scale; 
the $Q^2$ cut applied to the data is raised to $15$~\Gevv .
All the resulting differences, with respect to the nominal fit, 
are added in quadrature to form
an estimate of the { theoretical} error of the fit.

The ``total error of the fit''
is obtained by summing in quadrature these 
experimental and theoretical errors and is taken as the
uncertainty on the Standard Model expectation.
This procedure is
also used in the determination of the QCD uncertainty for the fit of
the CC cross-section to extract the $W$ boson mass as described in
section~\ref{EW}.

\subsection{Experimental Procedure for the Cross-Section Measurement} 
\label{f2meas}
The NC and CC cross-sections are evaluated in bins of the
$(x,Q^2)$ plane from the number of events which
pass the selection criteria (section~\ref{sec:select}), normalized
to the integrated $ep$ luminosity,
and corrected for
acceptance and bin to bin migrations  with the simulation. 
The simulation is found to reproduce well the
resolution of the measured kinematic variables,
as well as the efficiencies of the selection cuts 
within the errors described in 
section~\ref{systemaa}.
Whenever there is a difference,
the selection efficiency in the
simulation is adjusted to that of the data.
These bin averaged cross-sections are then
converted to cross-sections at chosen
bin centres using  corrections obtained from the NLO QCD Fit.

The NC data are binned 
in $Q^2$ with 10 bins per order of magnitude, 
except at $Q^2 \ge 3000$~\Gevv , for which a 
binning twice as large 
is adopted to account for the rapidly decreasing number of events.
The data are binned in $x$ with 5 bins per order of magnitude, 
except at $x>0.13 $ and $Q^2 \le 400$~\Gevv , for which a coarser 
binning is chosen to accommodate 
the degradation of the $x$ resolution at
very low $y$ ($< 0.02$).
The CC data are binned with 3 bins per order of magnitude in both 
$Q^2$ and $x$. The coarser CC binning is due to the smaller 
statistics of the CC sample and the inferior
resolution of the kinematic reconstruction 
of the $h$
method compared with the $e\Sigma$ method used for NC events.
The bins which are used in this measurement have to satisfy two quality
criteria which have been studied with the simulation:
their stability and purity\footnote{The stability (purity) 
is defined as the number of simulated events which originate from a bin 
and which are reconstructed in it, divided by the number of generated 
(reconstructed) events in that bin.}
are required to be larger than $30\%$.

\begin{figure}[hbt]
\begin{center} 
 \begin{picture}(160,150)(0,0)
 \put(-2,0){\epsfig{file=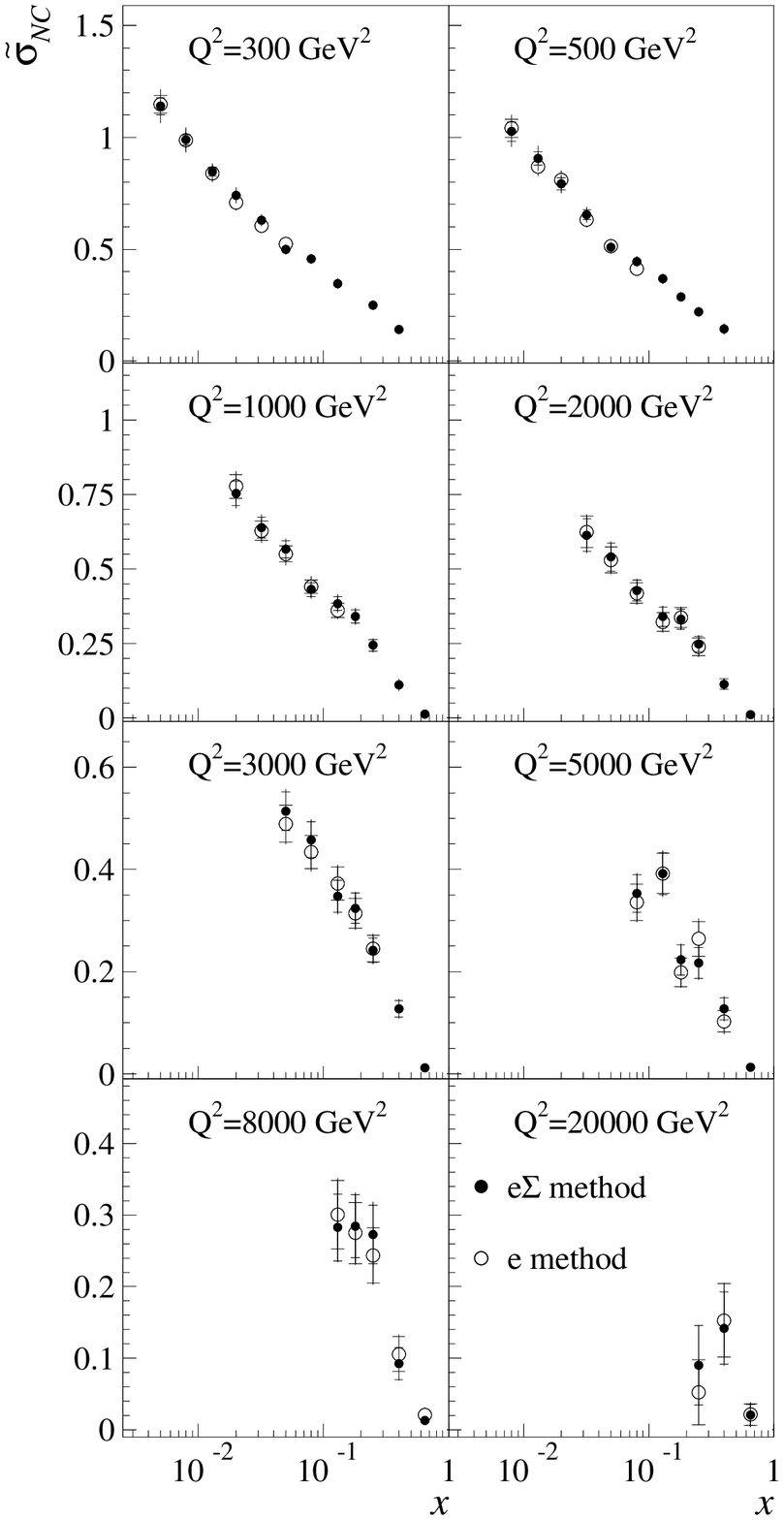,height=15.cm,
 bbllx=100pt,bblly=60pt,bburx=450pt,bbury=740pt}
 \epsfig{file=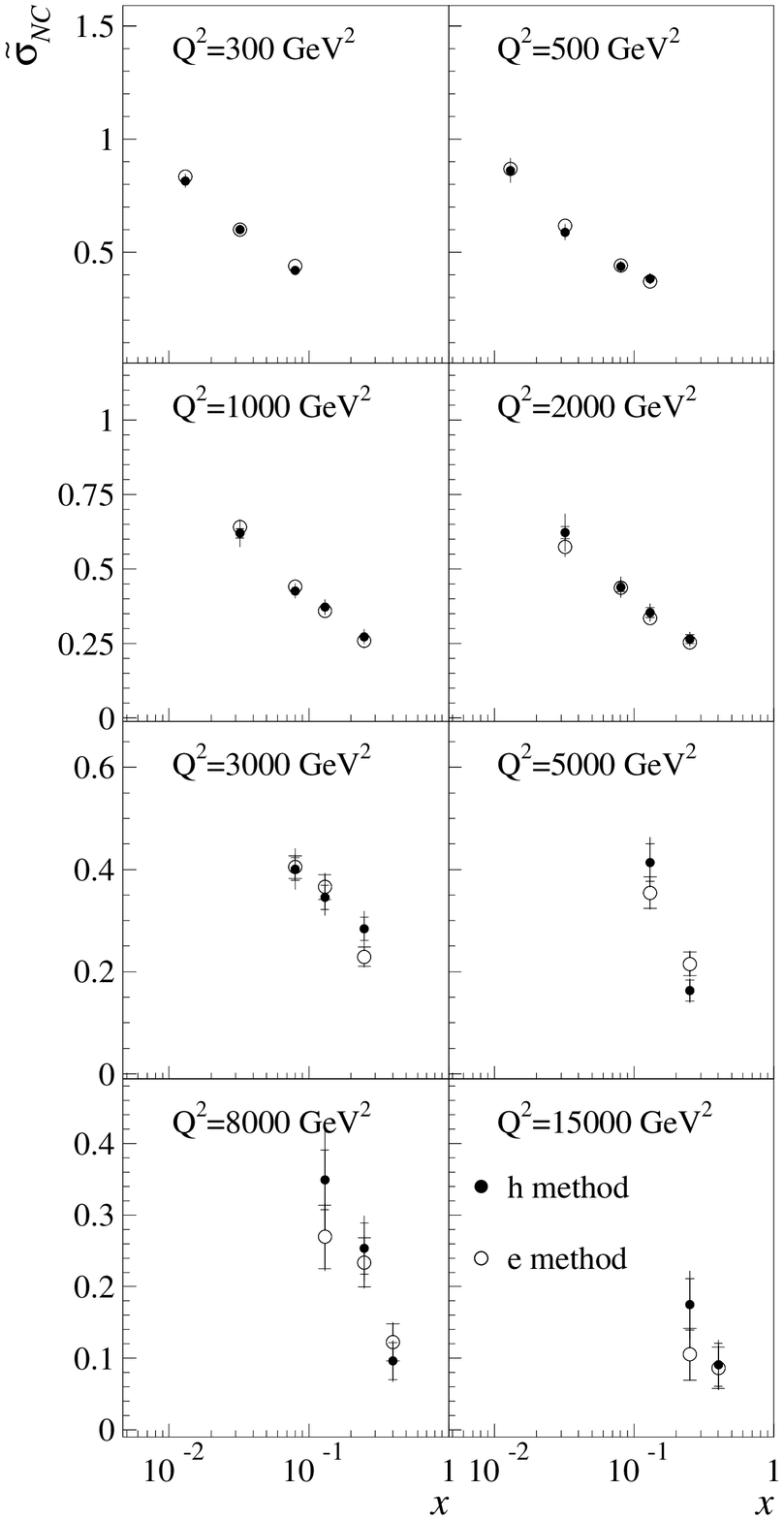,height=15.cm,
 bbllx=90pt,bblly=60pt,bburx=440pt,bbury=740pt}}
 \put(15,9){\bf (a)}
 \put(96,9){\bf (b)}
 \end{picture}
\end{center} 
\caption[]{\label{fig:ehad}
\sl Comparison of the NC reduced cross-section $\tilde{\sigma}_{NC}$ 
measured (a) in the NC binning at eight different $Q^2$ values, with the
$e$ (open points) and the $e\Sigma$ method (solid points), 
and (b) in the CC binning
with the $e$ (open points) and the $h$ method (solid points). 
}
\end{figure}

The reliability of the cross-section measurements is checked by comparing
the results obtained from different kinematic reconstruction methods.
Fig.~\ref{fig:ehad}a shows that there is good agreement between the
measurements of $\tilde{\sigma}_{NC}$ with the $e$ and the $e\Sigma$
methods in the region where the $e$ method is precise ($y \gapprox \ 
0.1$).
Good agreement is also found between the $e\Sigma$, the $\Sigma$ and
the DA methods (not shown) over the whole $y$ range.
Fig.~\ref{fig:ehad}b shows the NC reduced cross-section in the CC
binning, measured using the $h$ and $e$ methods. The good agreement
between these two independent kinematic reconstruction methods
demonstrates the reliability of the $h$ method in the CC analysis.

\subsection{Systematic Errors on the Cross-Section Measurement}
\label{systemaa}

The uncertainties in the measurement lead
to systematic errors on the cross-sections which can 
be bin to bin correlated, or uncorrelated. 
All the correlated systematic errors were checked to
be symmetric to a good approximation and are assumed 
so in the following\footnote
{For instance the effect of a $+0.5\%$ shift in the positron 
energy gives a systematic shift on the cross-section which is opposite
to the effect of a $-0.5\%$ shift.}. 
The correlated systematic errors
and the main uncorrelated systematic errors of the 
NC and CC cross-section measurements are
given in tables~\ref{longtabnc} and \ref{longtabcc} 
and their origin is discussed in the following.
\begin{itemize}
\item{The uncertainty of the positron energy 
is $1\%$ if the $z$ position of its impact on
the calorimeter is in the backward part
($z < -145 \ {\rm{cm}}$), 
$0.7\%$ in the CB1 and CB2 wheels ($-145 < z < 20 \ {\rm{cm}}$), 
$ 1.5\%$ for $20 < z < 100 \ {\rm{cm}}$ 
and $3\%$ in the forward part ($z > 100 \ {\rm{cm}}$).
These uncertainties are obtained 
by the quadratic sum of an uncorrelated
uncertainty and a bin to bin correlated uncertainty.
This correlated uncertainty comes mainly from the potential
bias of the calibration method and is estimated to be
$0.5\%$ in the whole LAr calorimeter.
The resulting correlated (uncorrelated) systematic error
on the NC cross-section 
is $\lapprox \ 3 \ (5)\%$ except for the measurement 
at the two highest $x$ values.}

\item
The correlated (uncorrelated) uncertainty  on the positron polar angle is 
$1 (2) \ {\rm {mrad}}$.
The uncorrelated uncertainty is the average of the different 
uncertainties when using the tracking system or the cluster for the
polar angle determination.
The resulting correlated (uncorrelated) 
systematic error is small, typically $\lapprox \ 1 \ (2)\%$.

\item
The uncertainty on the hadronic energy in the LAr calorimeter
is $2\%$. It is
obtained from the quadratic sum of an uncorrelated
uncertainty of $1.7\%$ and a
correlated uncertainty of $1\%$ originating from the 
calibration method,
and from the uncertainty of the reference scale ($P_{T,e}$). 
The resulting correlated systematic error increases at low $y$, 
and is typically $\lapprox \ 4\%$ except at high $Q^2$ for the CC
measurements.

\item
The uncertainty on the energy of 
the hadronic final state measured in the SPACAL (tracking system)
is $7$ ($3$)\%.
Their influence on the cross-section is small compared to the
uncorrelated uncertainty of the LAr calorimeter energy, and so the three
contributions (LAr, SPACAL, tracks) have been 
added quadratically, giving rise to the uncorrelated hadronic error.

\item
The correlated uncertainty on the energy identified as noise 
in the LAr calorimeter is  $25\%$. The resulting systematic error 
is largest at low $y$, reaching
$10$ to $15\%$ at $x=0.65$ and $Q^2 \le 2000$~\Gevv\ in 
the NC measurements.

\item 
The variation of the $V_{ap}/V_p$ cut by $\pm 0.02$ leads to 
a correlated systematic error which 
reaches a maximum of $12\%$ at low $x$ and $Q^2$ in the CC analysis.

\item
The uncertainty on the subtracted 
photoproduction background is $30\%$. The 
resulting correlated systematic error 
is always smaller than $5\%$ in the NC and CC analysis.

\end{itemize}
The following uncertainties are found to give rise to
uncorrelated systematic errors on the cross-sections:
\begin{itemize}
\item
a $2\%$ error ($4\%$ at $y > 0.5$ and $Q^2 < 500$~\Gevv )
from the positron identification efficiency in the NC analysis;

\item a 1\% error from the efficiency of the track-cluster 
link requirement in the NC analysis;

\item
a $0.5$ ($3$ to $8$)$\%$ error from the trigger efficiency 
in the NC (CC) analysis;

\item
a $1$ ($3$)$\%$ error from the QED radiative corrections
in the NC (CC) analysis;

\item
a $3\%$ error from the efficiency of the non-$ep$ background finders, 
in the CC analysis;

\item
a $2\%$ error ($5\%$ for $y < 0.1$) 
from the vertex finding efficiency in the CC analysis.
\end{itemize}
Overall the typical total systematic error for the NC (CC) double 
differential cross-section is about $4$ ($8$)\%. 
In addition a $1.5\%$ normalization error, due to the luminosity 
uncertainty averaged over the years, has to be considered, 
but is not added in the systematic error of the measurements 
given henceforth in the tables, or shown in the figures.

\section{Results and Interpretation}
\label{CSM}

\subsection{Measurement of the NC Cross-Section
\boldmath{$d^2\sigma_{NC}/dxdQ^2$} }
\label{CSM2}

\begin{figure}[htbp] 
\begin{center} 
\epsfig{file=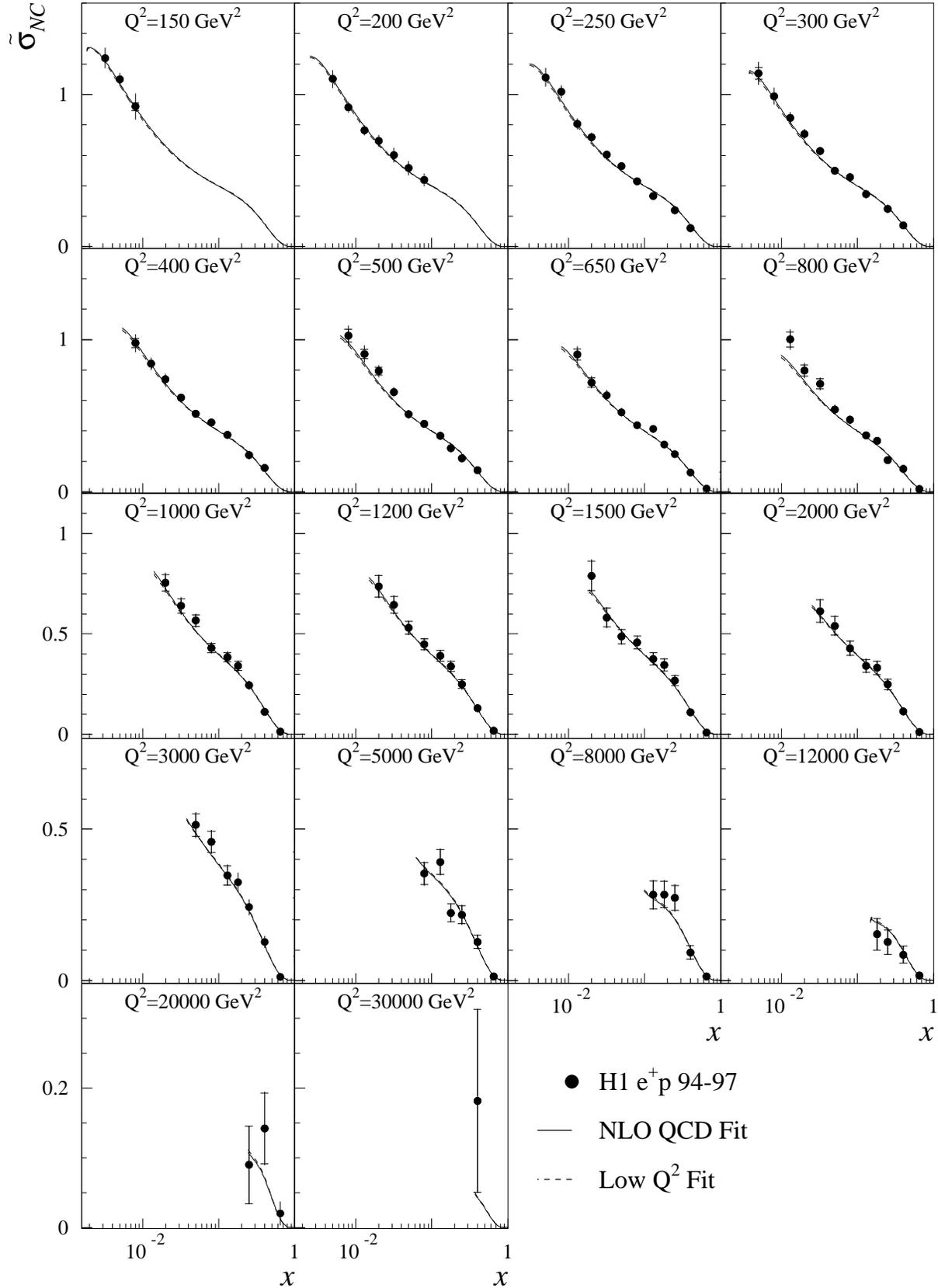
,width=16cm,bbllx=30pt,bblly=70pt,bburx=540pt,bbury=750pt} 
\caption[]{\label{fig:dxdq2.nc}
\sl {NC reduced cross-section $\tilde{\sigma}_{NC}$
measured as a function of $x$ for different values of 
$Q^2$ (points) compared with the NLO QCD Fit (solid curves).
Also shown is the Low $Q^2$ Fit (dashed curves).
The inner (outer) error bars represent the statistical (total) errors.
}}
\end{center} 
\end{figure}

The NC reduced cross-section (eq.~\ref{Rnc}) 
is shown in fig.~\ref{fig:dxdq2.nc} as a function of $x$ 
for fixed $Q^2$ values
and listed in table~\ref{tabf2}.
The measurement covers the range in $y$ between $0.007$ and $0.88$.
At $Q^2 \ \lapprox \ 500$~\Gevv\ the total error 
is dominated by the systematic 
uncertainties in the energy scale and identification efficiency of the
scattered positron and by the uncertainty in the energy scale 
of the hadronic final state. In this region the systematic
error is typically $4\%$. 
At higher $Q^2$ the statistical error becomes increasingly dominant. 

The kinematic domain of the NC cross-section measurements 
is significantly extended compared to previous HERA measurements
both in $Q^2$ (from $5000$ to $30\,000$~\Gevv ) and towards higher $x$,
with measurements at $x=0.65$ for $Q^2$
between $650$ and $20\,000$~\Gevv .
The reduced cross-section 
rises steeply with decreasing $x$, corresponding to
the increase of the sea quark and gluon densities at low $x$. 
As expected a sharp decrease of the cross-section is also observed 
in the valence quark region at high $x$.

The NLO QCD Fit, described in section~\ref{QCDA}, is compared with the
data in fig.~\ref{fig:dxdq2.nc}. It provides a good description of
the new measurements throughout the kinematic plane. The fit results
in a value of
\mbox{$\chi^2/\rm{ndf} = 1.02 $} for a total number of
data points (${\rm{ndp}}$) of $684$, 
when considering the uncorrelated error.
If the total errors are used to determine the
$\chi^2$ a value of $\chi^2/\rm{ndf}=0.77$ is obtained. 
The $\chi^2$ values for each dataset 
of the NLO QCD Fit are given in table~\ref{tab:tabfit2}.
The $\chi^2/\rm{ndp}$ of the new high $Q^2$ (NC+CC) datasets is
$(114+19)/(130+25)=0.86$.
The normalizations $(1+\delta {\cal{L}}_j/{\cal{L}}_j) $ 
obtained by the fit for the different datasets 
are also given in the table. All datasets agree to within $2\%$ with their
nominal normalization, with the exception of {\small{BCDMS-p}} 
which, however, has a luminosity uncertainty of $3\%$.

The parameters of the NLO QCD Fit are given in table~\ref{tab:lqpar}. 
Since only DIS data was used in the fit the gluon density at $x>0.2$ 
is not well constrained.  In this kinematic region the valence quark 
densities are strongly influenced by the BCDMS data, which still have a higher
precision than the new measurement.

\begin{table}[ht]
\begin{center}
\begin{tabular}{|c|c|c|c|c|c|c|c|c|} \hline
Experiment & {\small{H1 NC}} & {\small{H1 CC}} & 
{\small{H1 94}} & {\small{BCDMS-p}} & {\small{BCDMS-D}} &
{\small{NMC-p}}& {\small{NMC-D}} & Total \\ \hline 
data points  &$130$&$25$&$77$&$139$&$133$&$ 90$&$ 90$&$684$\\ 
\hline
$\chi^2$ (unc. err.) &$114$&$19$&$65$&$104$&$112$&$143$&$126$&$683$\\ 
\hline
$\chi^2$ (total err.)&$ 99$&$18$&$38$&$ 98$&$ 91$&$ 93$&$ 77$&$514$\\ 
\hline
normalization &\multicolumn{2}{c|}{$0.98 $}
  &$1.02$&$0.96$&$0.98$&$0.98$&$0.98 $&
\multicolumn{1}{c}{} \\ 
\cline{1-8}
\end{tabular}
\end{center}
\caption{\sl Results of the NLO QCD Fit.
For each experiment the following quantities are given: 
number of data points, 
contribution to the $\chi^2$ using the
uncorrelated errors (unc. err.) as obtained from the
statistical errors and uncorrelated systematic errors added in quadrature,
contribution to the $\chi^2$ using 
the total errors and the normalization required by the fit. }
\label{tab:tabfit2}
\end{table}

\begin{table}[ht]
\begin{center}
\begin{tabular}{|c|c|c|c|c|c|} \hline
   {$ _{PDF}$} & $A_{PDF}$ &$ B_{PDF}$\hspace*{0.4cm} 
&$ C_{PDF}$\hspace*{0.4cm} &$ D_{PDF}$ \hspace*{0.4cm}
& $E_{PDF}$\hspace*{0.4cm} \\ 
\hline 
$u_v$    &  $3.49$  & $\phantom{-}0.673$  &  $3.67$  & $1.24$ & 
 $\phantom{-}0.921$  \\ 
\hline
$d_v$   &  $1.04$  &  $\phantom{-}0.763$  &  $4.09$  & $1.43$  & $-0.067$ \\ 
\hline
$S$           &  $0.69$ & $-0.185$ &  $6.04$  & \multicolumn{2}{c}{} \\ 
\cline{1-4}
$g$         &  $2.64$  & $-0.095$ &  $7.18$  &  \multicolumn{2}{c}{} \\ 
\cline{1-4}
\end{tabular}
\end{center}
\caption{\sl Parameters of the NLO QCD Fit. The
parameters $A_g$, $A_{u_v}$, and $A_{d_v}$ are obtained from
the sum rules.}
\label{tab:lqpar}
\end{table}

The Low $Q^2$ Fit, which is described in section~\ref{QCDA}, is also compared
with the data in fig.~\ref{fig:dxdq2.nc}. The prediction of the 
Low $Q^2$ Fit agrees
well with the high $Q^2$ data. 
Compared to the Low $Q^2$ Fit the NLO QCD Fit, which includes also the
high $Q^2$ data,
results in a cross-section expectation 
that is higher by a maximum of $2\%$ at low
$x$ and is lower by a maximum of $3\%$ at high $x$. These differences
are, however, smaller than the uncertainty of the fit.
At high $Q^2$ this uncertainty is reduced when including 
the high $Q^2$ data in the QCD fit, for example from $7$ to $6\%$ at
$Q^2 \approx 10\,000$~\Gevv\ and $x = 0.4$.

\begin{figure}[hptb] 
\begin{center} 
\epsfig{file=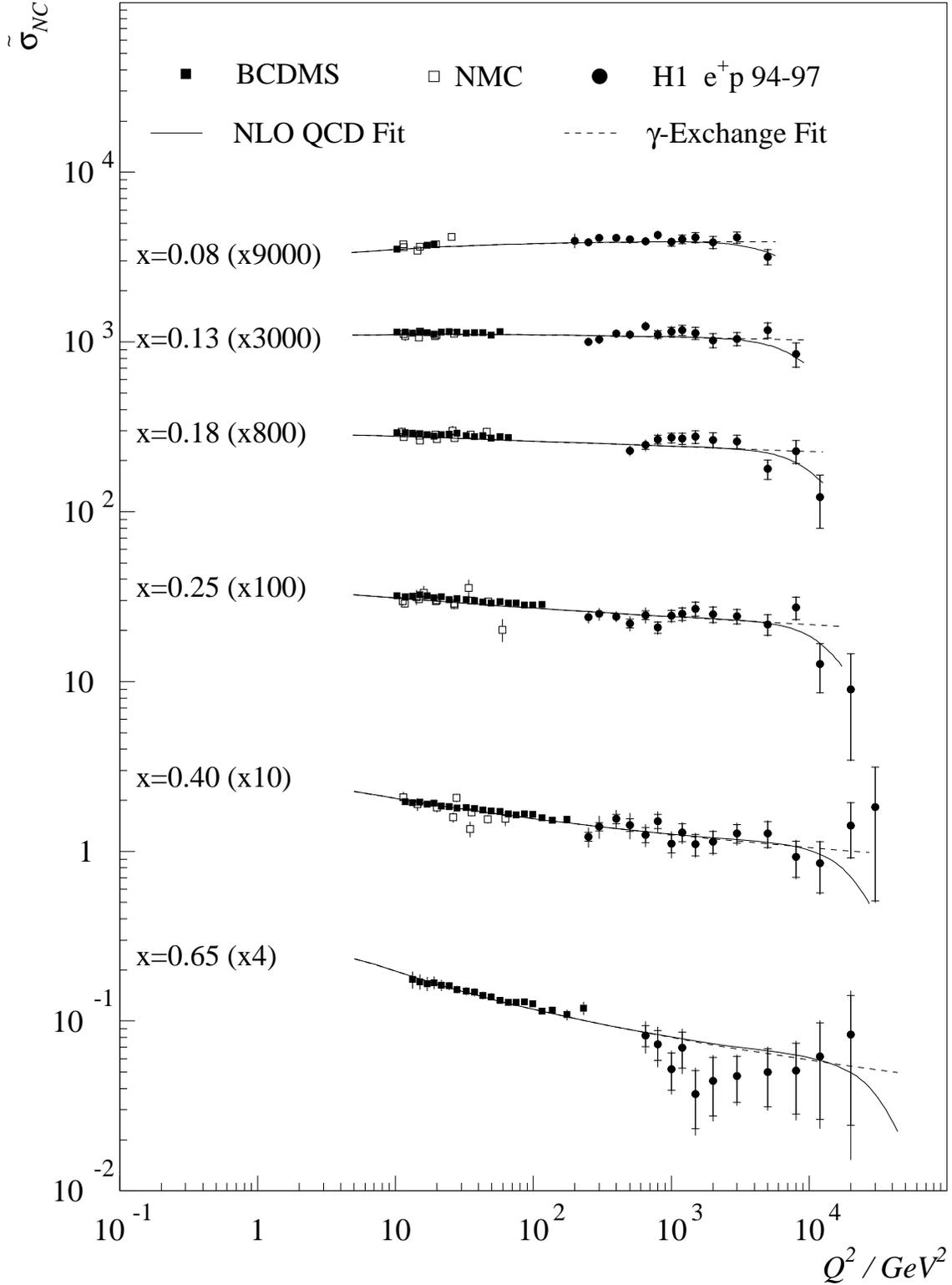,
width=15.5cm,bbllx=40pt,bblly=65pt,bburx=540pt,bbury=725pt} 
\caption[]{\label{fig:dxdq2.x}
\sl {NC reduced cross-section $\tilde{\sigma}_{NC}$
measured at high $x$ (solid points)
compared with Standard Model expectations as given by the NLO QCD
Fit (solid curves) and with the $\gamma$-Exchange Fit (dashed curves). 
The inner (outer) error bars represent the statistical 
(total) errors.
Also shown are the NMC data(open squares),
and the BCDMS data (solid squares).
}}
\end{center} 
\end{figure}

In fig.~\ref{fig:dxdq2.x} 
the reduced cross-section is shown as a function of $Q^2$ at fixed
values of $x$ for $ 0.08 \le x \le 0.65$. It can be seen that the H1
data are consistent with the fixed target data, in particular at
$x=0.25, 0.40$ in which the measurements are made in contiguous
kinematic regions. These measurements
test the QCD evolution at high $Q^2$, and render possible
the study of the structure function scaling violations at high $x$,
in a region where non-perturbative effects are negligible.

At $x= 0.40$ an enhancement of the cross-section above the Standard
Model expectation, as given by the NLO QCD Fit,
is visible for the highest $Q^2$ values ($Q^2 > 15\,000$~\Gevv ).
This corresponds to the accumulation of events
around an inclusive invariant mass of the lepton quark system of about
$200\ \Gev $, which was already reported with the 1994--1996
data~\cite{h1hiq296}. The significance of this excess decreases when
1997 data are included. A detailed analysis of these events is
presented in~\cite{yves}.

At $x=0.65$ and for $Q^2 < 10\,000$~\Gevv\ the NLO QCD
Fit lies above the H1 data.
This difference can be due either to a too high expectation at $x=0.65$
since the main constraint on the fit comes from the BCDMS data which
are known to favour a lower $\alpha_s$ value than the world average,
or to the H1 data which share a correlated error of about $12\%$ at
this $x$ value (table~\ref{longtabnc}). Furthermore, this
difference is rendered 
less significant by the $7\%$ uncertainty on the cross-section
expectation.

The destructive $\gamma Z^\circ$ interference expected 
in the Standard Model
for $e^+p$ collisions reduces at HERA the cross-section at 
$Q^2 \ \gapprox \ M_Z^2$. This reduction is
observed with the highest $Q^2$ measurements for $ 0.08 \le x \le
0.25$ as shown in fig.~\ref{fig:dxdq2.x}. To determine the extent to
which $Z^0$ exchange is seen in the NC data, the NLO QCD Fit is
repeated but allowing only for pure photon exchange ($\gamma$-Exchange
Fit), i.e. \Fz\ $= 0$, $\Ftwo=\Fem $ and \FL\ $= F_L$, $F_L$ being the 
electromagnetic part of the longitudinal structure function. 
The $\gamma$-Exchange
Fit, also shown in fig.~\ref{fig:dxdq2.x}, is observed to have a
larger $\chi^2$ than that of the standard NLO QCD Fit
(table~\ref{tab:tabfit2})
by $14$ units, $11$ of which
are from the NC data at $Q^2 \geq 5000$~\Gevv . The description 
and $\chi^2$ contributions of 
all other data are unchanged thereby showing that 
the effects of the $\gamma Z^\circ$ interference are visible in DIS 
$ep$ scattering at high values of $Q^2$.

\subsection{Extraction of the 
Proton Structure Function \boldmath{$F_2(x,Q^2)$} 
at High \boldmath{$Q^2$}}

Assuming the validity of the electroweak sector of the Standard Model,
and of the DGLAP equations at high $Q^2$,
\begin{figure}[htbp] 
\begin{center} 
\epsfig{file=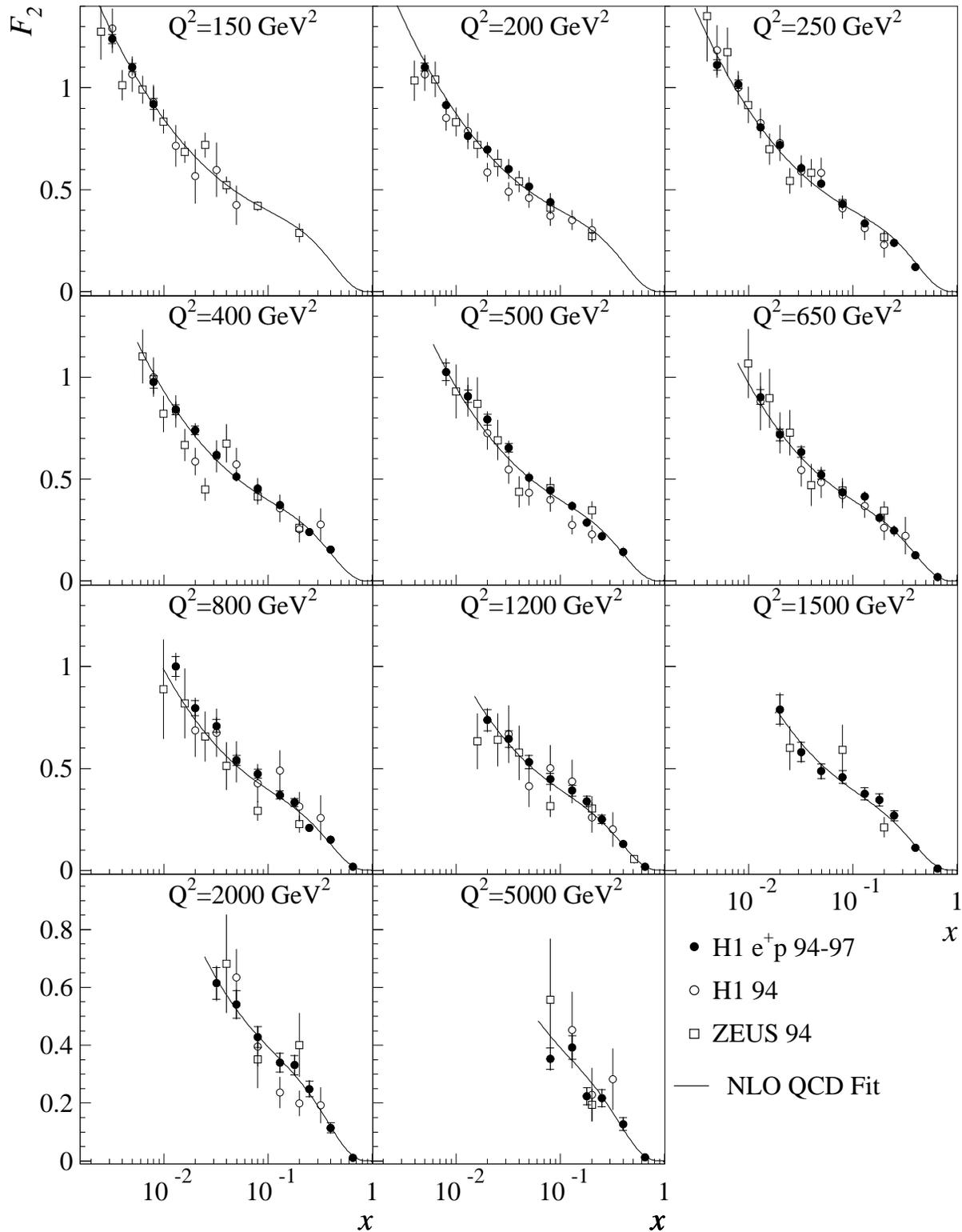,
width=16cm,bbllx=30pt,bblly=50pt,bburx=540pt,bbury=710pt} 
\caption[]{\label{fig:f294}
\sl {Measurement of the electromagnetic structure function \Fem $(x,Q^2)$
with the data taken between 
1994 and 1997 (solid points). 
The inner (outer) error bars represent the statistical (total) errors.
Also shown are the results obtained by H1 and ZEUS
with the 1994 data (open symbols) together with their total errors.
The NLO QCD Fit is represented by the solid curves.
}}
\end{center} 
\end{figure} 
the electromagnetic proton structure function 
\Fem\ is extracted from the 
double differential NC cross-section (eqs.~\ref{Snc},\ref{f2p})
using
the NLO QCD Fit calculations for 
$\Delta_{F_L}$, $\Delta_{F_2}$, $\Delta_{F_3}$ and $\delta^{weak}_{NC}$.

A comparison of the $F_2$ data at high $Q^2$ with the corresponding
H1~\cite{H194} and ZEUS~\cite{ZEUS94} 
results based on the 1994 data is shown in fig.~\ref{fig:f294}. Only the 
data at $Q^2$ values which were 
measured in 1994 are shown here.
The complete set of
$F_2$ values is listed in table~\ref{tabf2}.
The extension in kinematic coverage at low $y$ (high $x$) is 
visible. A reduction of the systematic error of
the new measurement by more than a factor of two with respect to the 1994
results is achieved. 
The new measured points are in agreement with the 1994
data.
Due to its superior precision
the new measurement
supersedes the H1 1994 data at 
$Q^2 \ge 250 $~\Gevv , at $Q^2 =200$~\Gevv\ for $x < 10^{-1}$,  and
 at $Q^2 =150$~\Gevv\ for $x < 10^{-2}$.

\begin{figure}[bht] 
\begin{center}
\epsfig{file=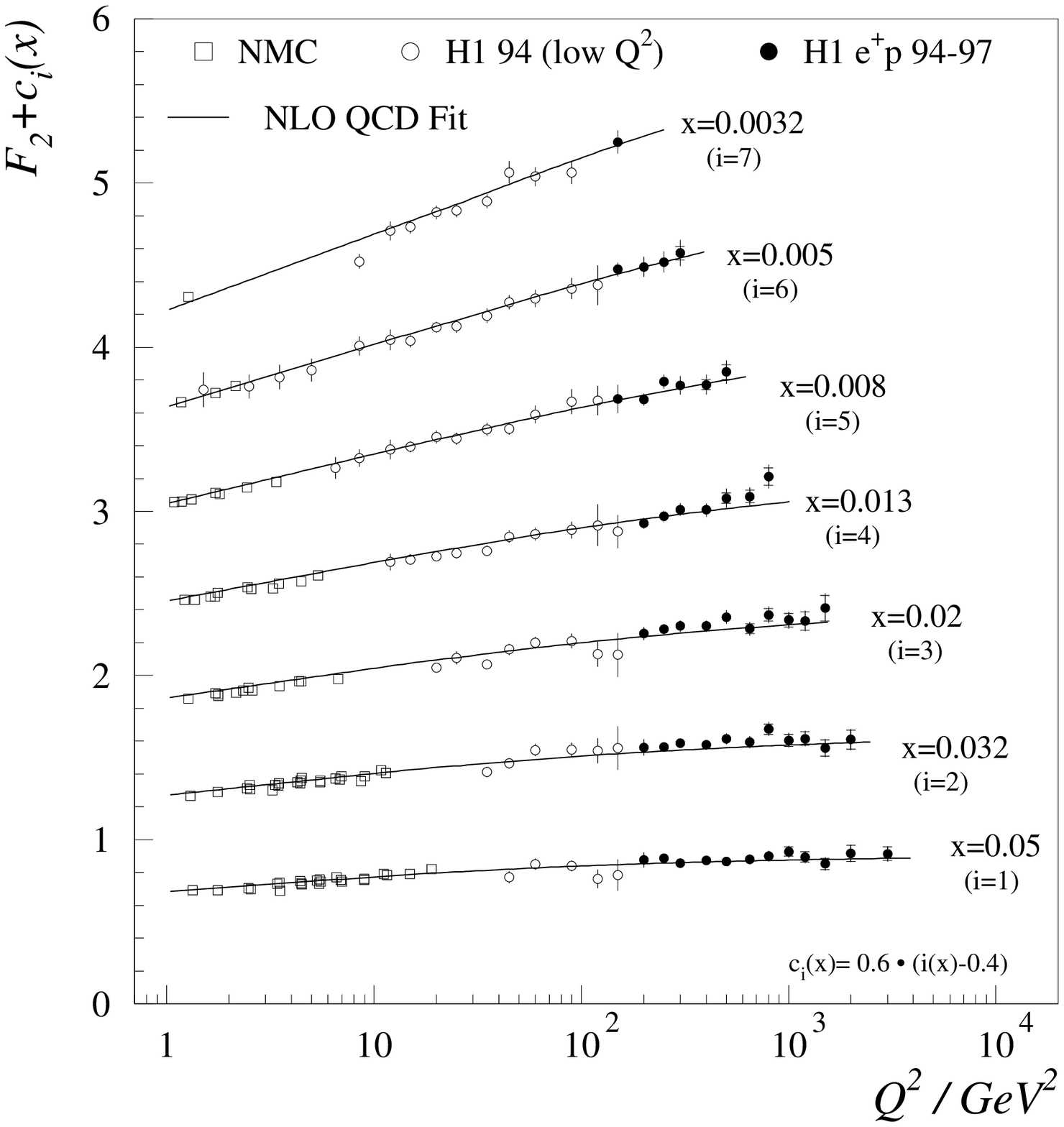,width=16.5cm,
bbllx=25pt,bblly=5pt,bburx=545pt,bbury=515pt} 
\caption[]{\label{fig:f2.nc}
\sl {Measurement of the electromagnetic structure function \Fem $(x,Q^2)$
with the data taken between 
1994 and 1997 (solid points)
as a function of $Q^2$ 
for $x$ values between $0.0032$ and $0.05$.
The inner (outer) error bars represent the statistical (total) errors.
Also shown are the results
obtained by H1 at $Q^2 < 150$~\Gevv\ with 
the 1994 data (open points) and the NMC data (open squares),
together with their total errors. 
The NLO QCD Fit is represented by the solid curves.
}}
\end{center} 
\end{figure}

The $F_2$ measurements at low $x$ ($x \le 0.05$) are
shown in fig.~\ref{fig:f2.nc}
as a function of $Q^2$.
The high $Q^2$ data are 
compared with 
the published H1 1994 data~\cite{H194} at $Q^2 < 150 $~\Gevv\ and 
with the NMC~\cite{nmc} proton data.
The measured data points are well described by the $Q^2$ evolution
of $F_2$ predicted by the NLO DGLAP equations
from $Q^2 \approx 1$~\Gevv\ up to the highest measured $Q^2$.
A positive slope as a function of $Q^2$ is
visible for the low $x$ data points and this slope 
decreases with increasing $x$ as expected from QCD.

\subsection{Measurement of the CC 
Cross-Section \boldmath{$d^2\sigma_{CC}/dxdQ^2$} }

\begin{figure}[t] 
\begin{center}
\epsfig{file=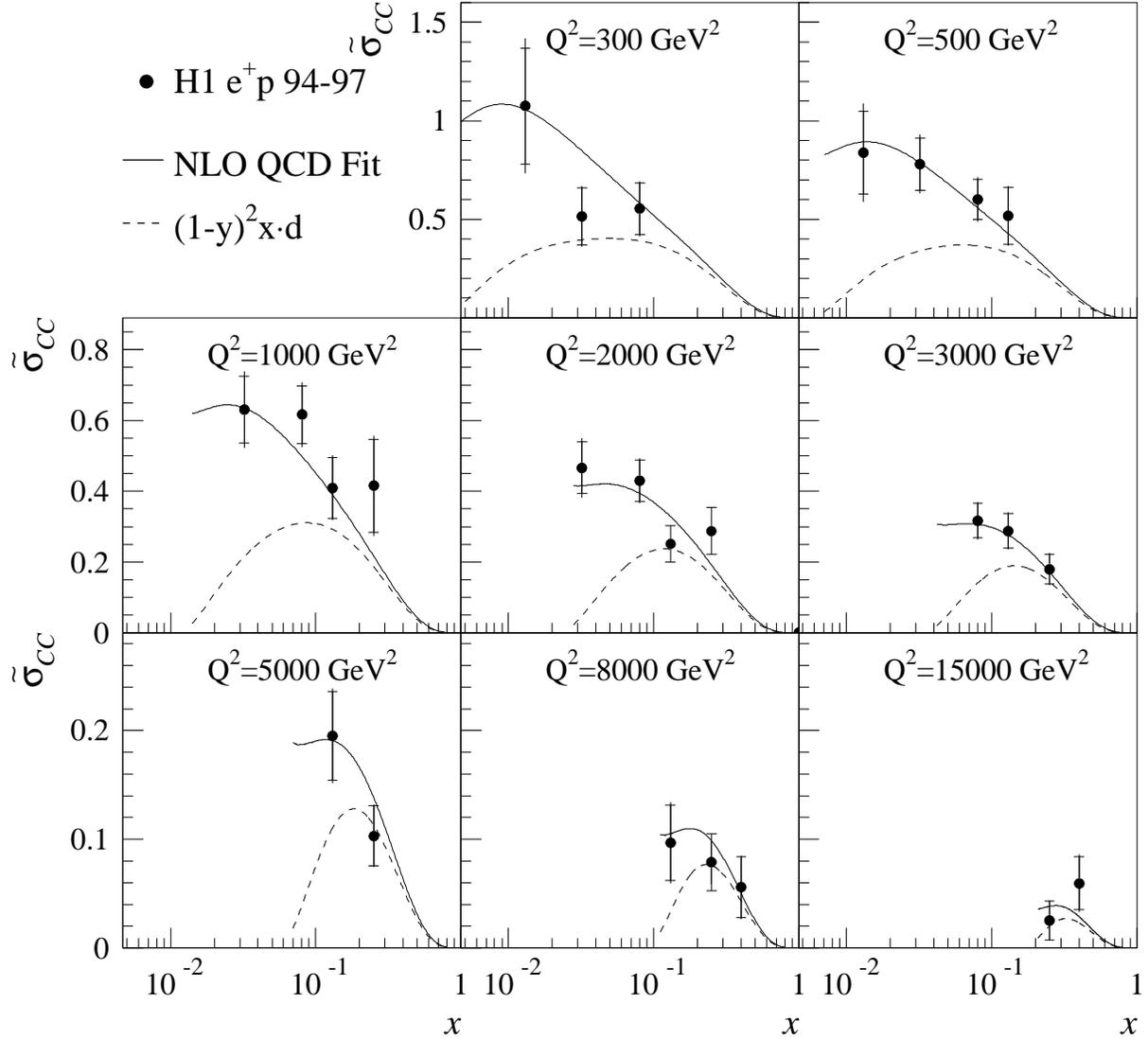,width=16.cm,
bbllx=55pt,bblly=170pt,bburx=525pt,bbury=600pt}
\caption[]{\label{fig:dxdq2.cc}
\sl {CC reduced cross-section $\tilde{\sigma}_{CC}$
obtained from the measured double differential cross-section 
$d^2\sigma_{CC}/dxdQ^2$  shown as a function of $x$ 
for different $Q^2$ values (points) and compared with 
the NLO QCD Fit (solid curves). Also shown is the 
$d$ quark contribution to the NLO QCD Fit (dashed curves).
The inner (outer) error bars represent the statistical (total) errors.
}}
\end{center} 
\end{figure}

The double differential cross-section $d^2\sigma_{CC}/dxdQ^2$, 
measured for $ 300 \le Q^2 \le 15\,000$~\Gevv\ and 
for $0.013 \le x \le 0.4$, is listed in table~\ref{tabf2cc}. It is
displayed in fig.~\ref{fig:dxdq2.cc} in the form of the reduced cross
section (eq.~\ref{Rcc}).
The uncertainties of the measurements are dominated by 
the statistical errors.
The largest systematic errors come from the uncertainty in the 
energy scale of the hadronic final state
and from the uncertainty in the trigger efficiency.
The Standard Model
cross-section, calculated using the NLO QCD Fit parton distributions,
is found to agree well with the data.

At high $x$ the
$e^+p$ CC cross-section is expected to be dominated by $d$ quark
scattering as is shown in fig.~\ref{fig:dxdq2.cc}, which includes the
expected $d$ quark contribution to the CC cross-section obtained from
the NLO QCD Fit. The observed rise of $\tilde{\sigma}_{CC}$ as $x$
decreases can be explained by the expected contribution of $\bar{u}$ and
$\bar{c}$ quarks from the sea. The contribution of $d$ and $s$ quarks
is small at low $x$  due to the suppression at high $y$ by
the $(1-y)^2$ term (eq.~\ref{SccLO}).

\subsection{Helicity Structure of the NC and CC Interactions} 

The double differential NC and CC cross-section measurements test 
the predictions of electroweak theory for the scattering of two
fermions at large momentum transfer, and 
allow the contributions of individual quark flavours to be analysed.

\begin{figure}[b] 
\begin{center}
\begin{picture}(160,145)(0,0)
\put(-3,3){\epsfig{file=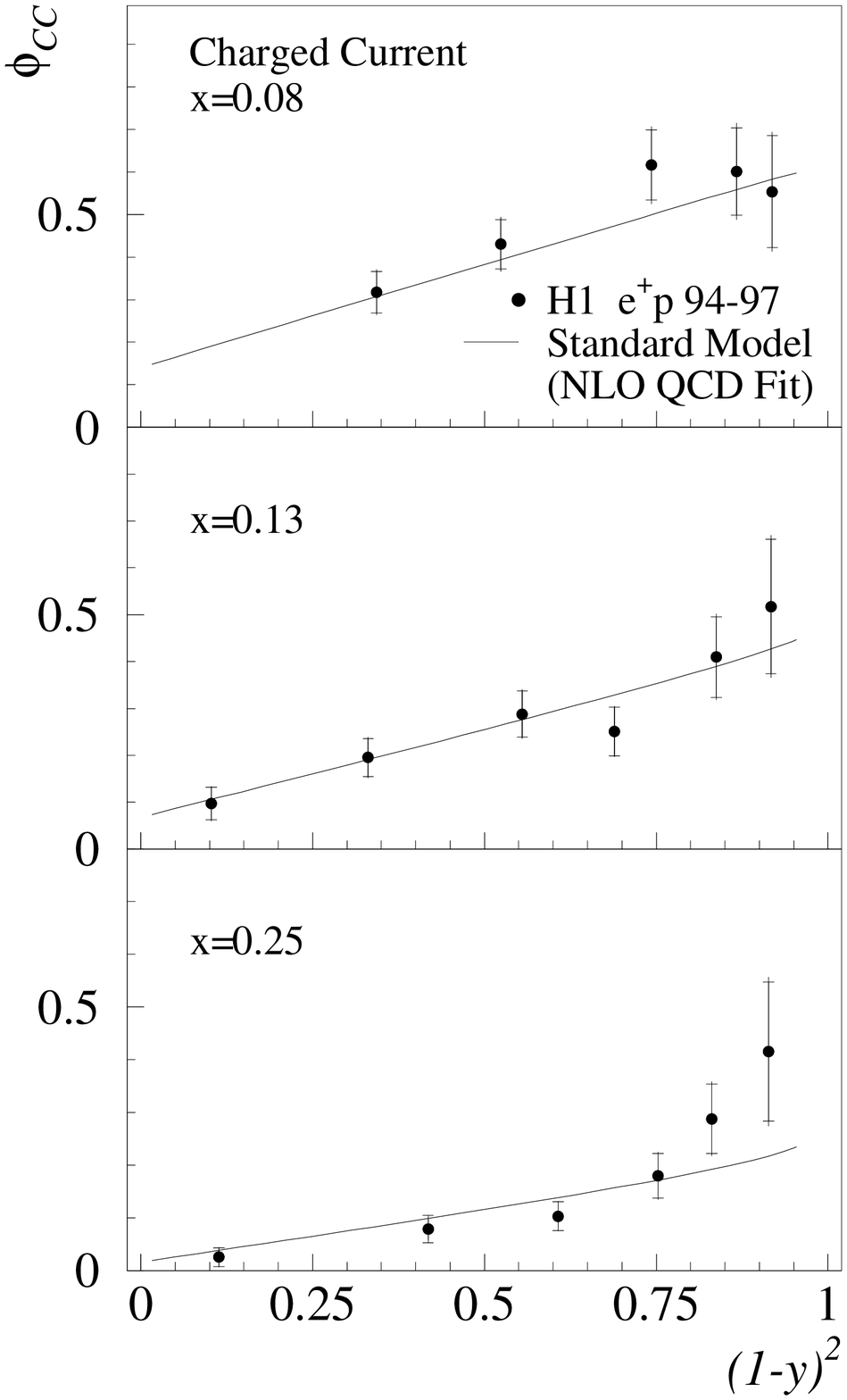,width=7.9cm,
 bbllx=60pt,bblly=50pt,bburx=480pt,bbury=640pt}
 \epsfig{file=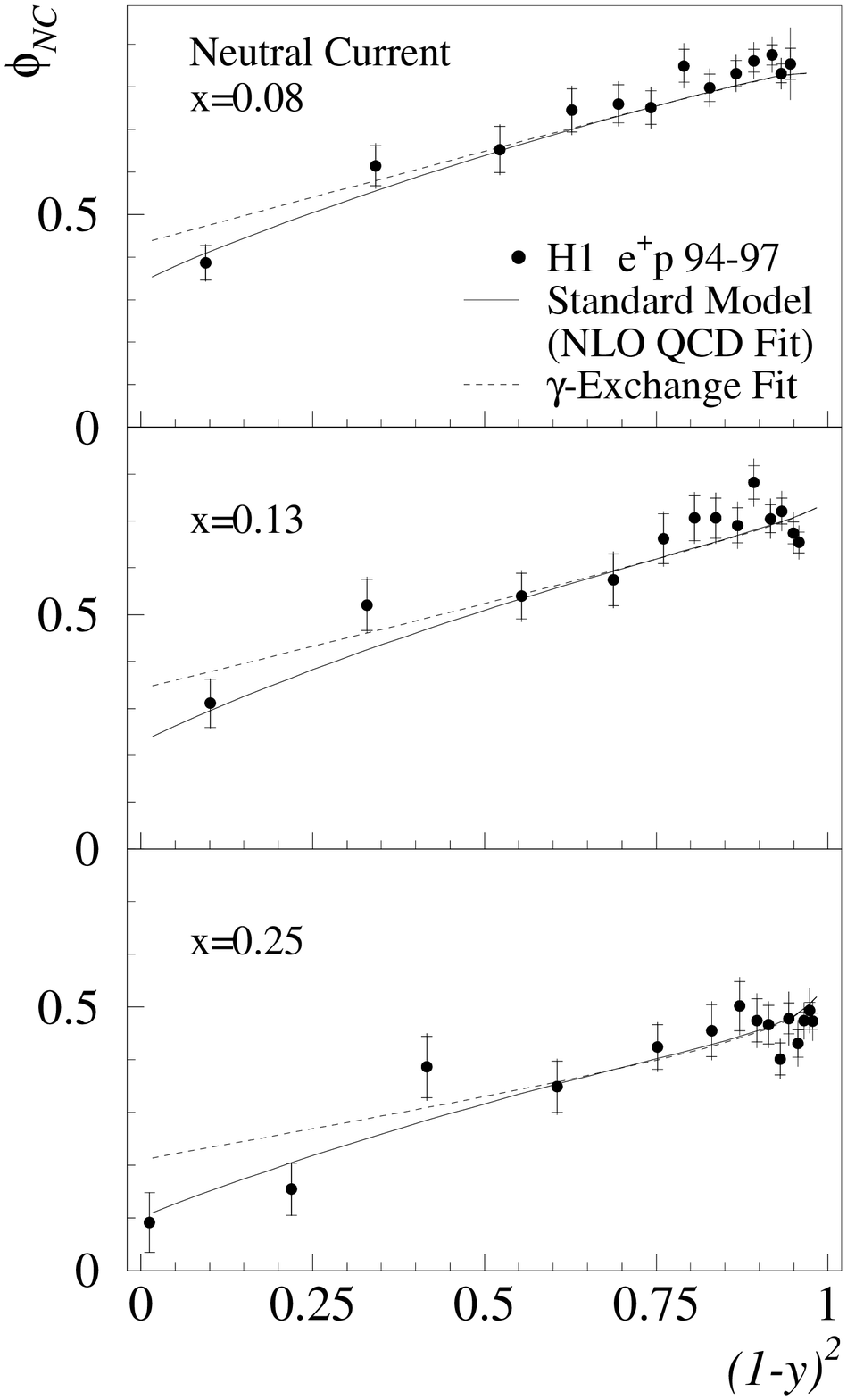,width=7.9cm,
 bbllx=60pt,bblly=50pt,bburx=480pt,bbury=640pt}}
 \put(17.,115.5){\bf (a)}
 \put(97.,115.5){\bf (d)}
 \put(17.,76.5){\bf (b)}
 \put(97.,76.5){\bf (e)}
 \put(17.,37.5){\bf (c)}
 \put(97.,37.5){\bf (f)}
 \end{picture}
\caption[]{\label{fig:me}
\sl {Structure function terms $\phi_{CC}$ of the CC (a,b,c) 
and $\phi_{NC}$ of the NC (d,e,f) measured double differential 
cross-sections $d^2\sigma_{NC(CC)}/dxdQ^2$
as a function of $(1-y)^2$ at $x=(0.08,0.13,0.25)$ (points).
The inner (outer) error bars represent the statistical (total) errors.
The expectations for $\phi_{CC}$ and $\phi_{NC}$ from the NLO QCD
Fit are shown (solid curves) in (a,b,c) and (d,e,f) respectively.
The $\phi_{NC}$ obtained from the $\gamma$-Exchange Fit is also 
shown (dashed curves) in (d,e,f).
}}
\end{center} 
\end{figure}

In the region of approximate Bjorken scaling, $x\approx 0.1$, 
the helicity dependence of the positron-quark interactions
can be separated from the quark density distributions.
In fig.~\ref{fig:me} the measured structure function
terms $\phi_{CC}$ and $\phi_{NC}$ are shown as a function of $(1-y)^2$.
The inelasticity $y$ is related to the positron scattering angle 
$\theta^*_e$ in the positron-quark centre of mass system through
$\cos^2{\frac{\theta^*_e}{2}} = 1-y$.

The measurements of $\phi_{CC}$ are shown in figs.~\ref{fig:me}a-c.
They are consistent with a linear dependence on $(1-y)^2$.
In leading order (eq.~\ref{SccLO})
these dependences are expected to result from two components 
reflecting the helicity structure of the CC interactions:
an isotropic distribution from positron-antiquark 
($\bar{u}, \bar{c}$) scattering,
and a distribution linearly rising with $(1-y)^2$ from 
positron-quark ($d, s$) scattering.

The curves in figs.~\ref{fig:me}a-c
represent the expectation for $\phi_{CC}$ from 
the NLO QCD Fit and show the two
helicity components to be of different magnitude.
At $x=0.08$ the contribution of the antiquarks,
which dominates as
$(1-y)^2\rightarrow 0$,
is sizeable but decreases as $x$ increases.
The component rising with $(1-y)^2$ reflects the quark contribution
which is larger than that of the antiquarks.

The measurements of $\phi_{NC}$ are shown in figs.~\ref{fig:me}d-f.
Two helicity components are also expected to contribute,
but with similar magnitude  (eq.~\ref{SncLO}) 
since NC processes are insensitive 
to the difference between quarks and antiquarks.

The interference between the photon and $Z^\circ$ contributions
in the NC measurement discussed 
in section~\ref{CSM2} is also visible in figs.~\ref{fig:me}d-f.
In the region of large $(1-y)^2$ the data follow the 
$\gamma$-Exchange Fit
reflecting the two helicity components expected from photon 
exchange between the positron and the (anti)quarks.
However, at small values of $(1-y)^2=\left(1-Q^2/s x\right)^2$
the data lie significantly below this fit hypothesis, 
in agreement with the Standard Model expectation.

\subsection{Quark Densities from NC and CC Results at High 
\boldmath{$x$}}

The behaviour of the $d/u$ ratio in the valence quark region 
at high $x$ is still controversial~\cite{bodek}. 
At HERA the $u$ and $d$ quark distributions can be
extracted from the measured high $Q^2$ NC and CC cross-sections
with minimal assumptions. Such an
extraction is presented here for $x$ values of $0.25$ and $0.4$.

The density of the sea quarks is expected to be small
at high $x$, as can be inferred for CC interactions 
from fig.~\ref{fig:me}c.
At high $x$ the structure function term $\phi_{NC}$ is primarily
sensitive to the $u$ quark density since the contribution from the $d$
quark is suppressed due to the quark charge squared (eq.\ref{SncLO}).
Conversely, $\phi_{CC}$ is sensitive mainly to the $d$ quark density, since
$u$ quark scattering does not contribute in $e^+p$ CC interactions. 

\begin{figure}[htb] 
\begin{center}
\begin{picture}(140,132)(0,0)
\put(-10,0){\epsfig{file=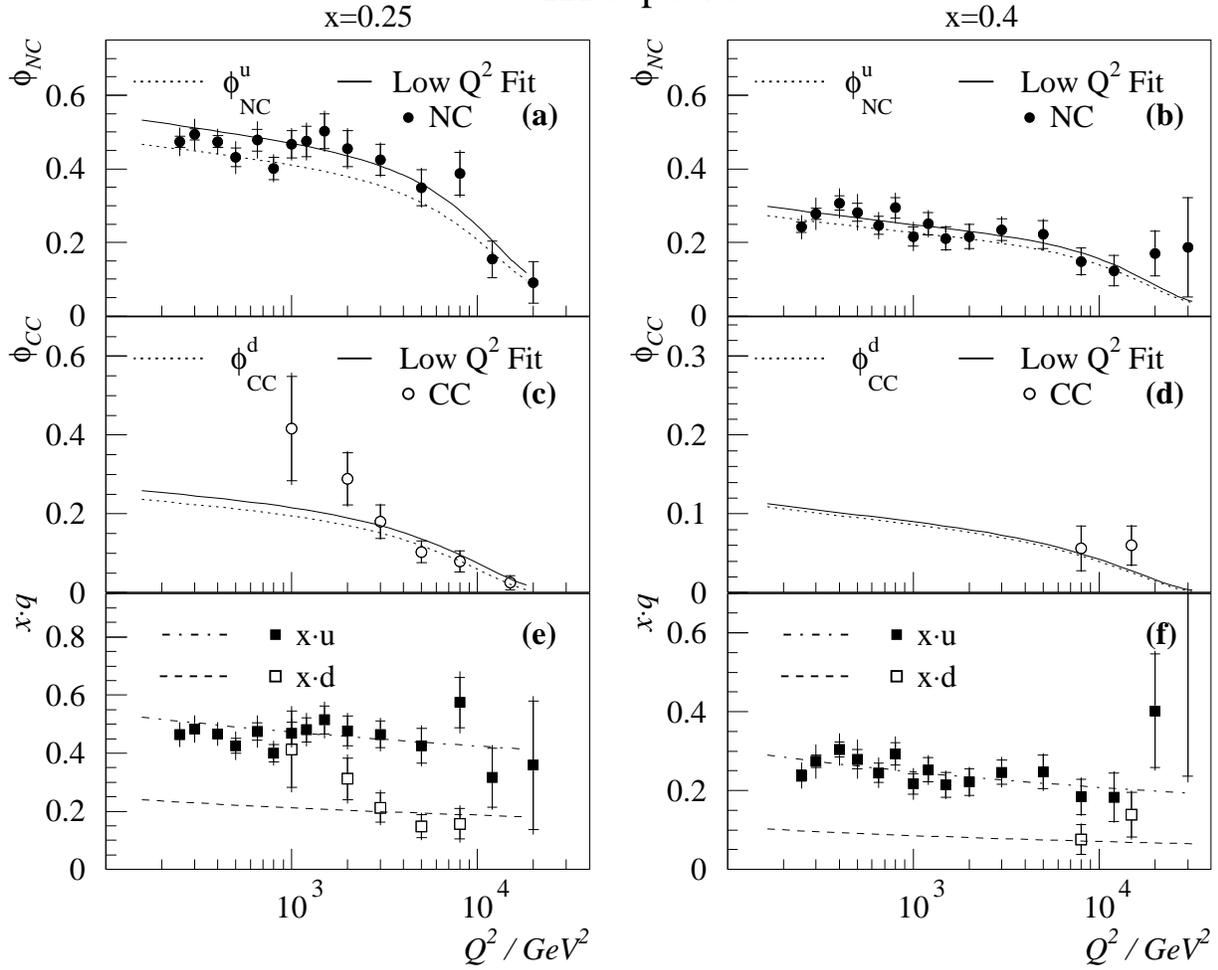,width=15.8cm,
bbllx=65pt,bblly=160pt,bburx=545pt,bbury=550pt}}
 \put(58,112){\bf (a)}
 \put(141,112){\bf (b)}
 \put(58,75){\bf (c)}
 \put(141,75){\bf (d)}
 \put(58,43){\bf (e)}
 \put(141,43){\bf (f)}
\end{picture}
\caption[]{
\label{fig:val}
\sl 
{Structure function terms $\phi_{NC}$ (solid points) and $\phi_{CC}$
 (open points) of the measured NC and CC double differential 
 cross-sections $d^2\sigma_{NC,CC}/dxdQ^2$
 as a function of $Q^2$ at $x=0.25$ (a,c) and $x=0.40$ (b,d).
 The expectation for $\phi_{CC}$ and $\phi_{NC}$ from the Low $Q^2$
 Fit are shown (solid curves) in (a,b,c,d). 
 The dominant $u$ quark contribution $\phi^u_{NC}$ (a,b) and 
 $d$ quark contribution $\phi^d_{CC}$ (c,d) are also shown 
 as dotted curves.
 The extracted $u$ quark density from the NC cross-section (solid squares) 
 and the $d$ quark 
 density from the CC  cross-section (open squares) are 
 shown at $x=0.25$ in (e) and $x=0.4$ 
 in (f), and are compared to their QCD expectation which are obtained from the 
 Low $Q^2$ Fit for the $u$ quark (dash-dotted curves) and $d$ quark
 (dashed curves).
 The inner (outer) error bars represent the statistical (total) errors.
}}
\end{center} 
\end{figure} 

The NC (CC) structure function term at $x=0.25$ and $x=0.40$
and its prediction from the Low $Q^2$ Fit, 
which only  uses data with $Q^2<150$~${\rm GeV}^2$,
are shown in fig.~\ref{fig:val}a(c) and b(d).
Also shown is the dominant contribution  $\phi^u_{NC}$ ($\phi^d_{CC}$)
to the structure function term, which is obtained from the fit 
and  which originates from the $xu$ ($xd$) density.

The data can also be displayed as $xu$ and $xd$ densities.
The extraction of the $xu$ ($xd$) 
density in the $\overline{MS}$ scheme 
is made
by multiplying the measured structure function term
$\phi_{NC}$ ($\phi_{CC}$) by the ratio of the $xu$ ($xd$) density
to $\phi_{NC}$ ($\phi_{CC}$)
obtained from the Low $Q^2$ Fit:
\begin{equation} 
x u = \phi_{NC} { \left[ \frac{x u }{\phi_{NC}} \right] }_{\rm {
Low \ Q^2 \ Fit}}
\hspace*{2cm} x d = \phi_{CC} 
\left[\frac{ x d}{\phi_{CC}}\right]_{\rm {Low \ Q^2 \ Fit}} \, .
\end{equation}

The results are shown in
fig.~\ref{fig:val}e,f as a function of $Q^2$ together with the NLO
QCD expectation of these densities.
For this extraction the uncertainties due to
the other quark densities 
were estimated 
by varying these densities by $\pm 50\%$.
They
are generally
below $2\%$ for the NC case and $7\%$ for the CC case,
and are added in quadrature to the total errors of the data.
This measurement of the $d$ quark density, using only $e^+p$
scattering data, agrees well with results from other DIS experiments
where different targets have to be combined.


\subsection{Measurement of the 
\boldmath{$x$} Dependence of the NC and CC Cross-Sections}

\begin{figure}[htb]
\setlength{\unitlength}{1 mm}
\begin{center}
\begin{picture}(160,105)(0,0)
\put(-2,40){\epsfig{file=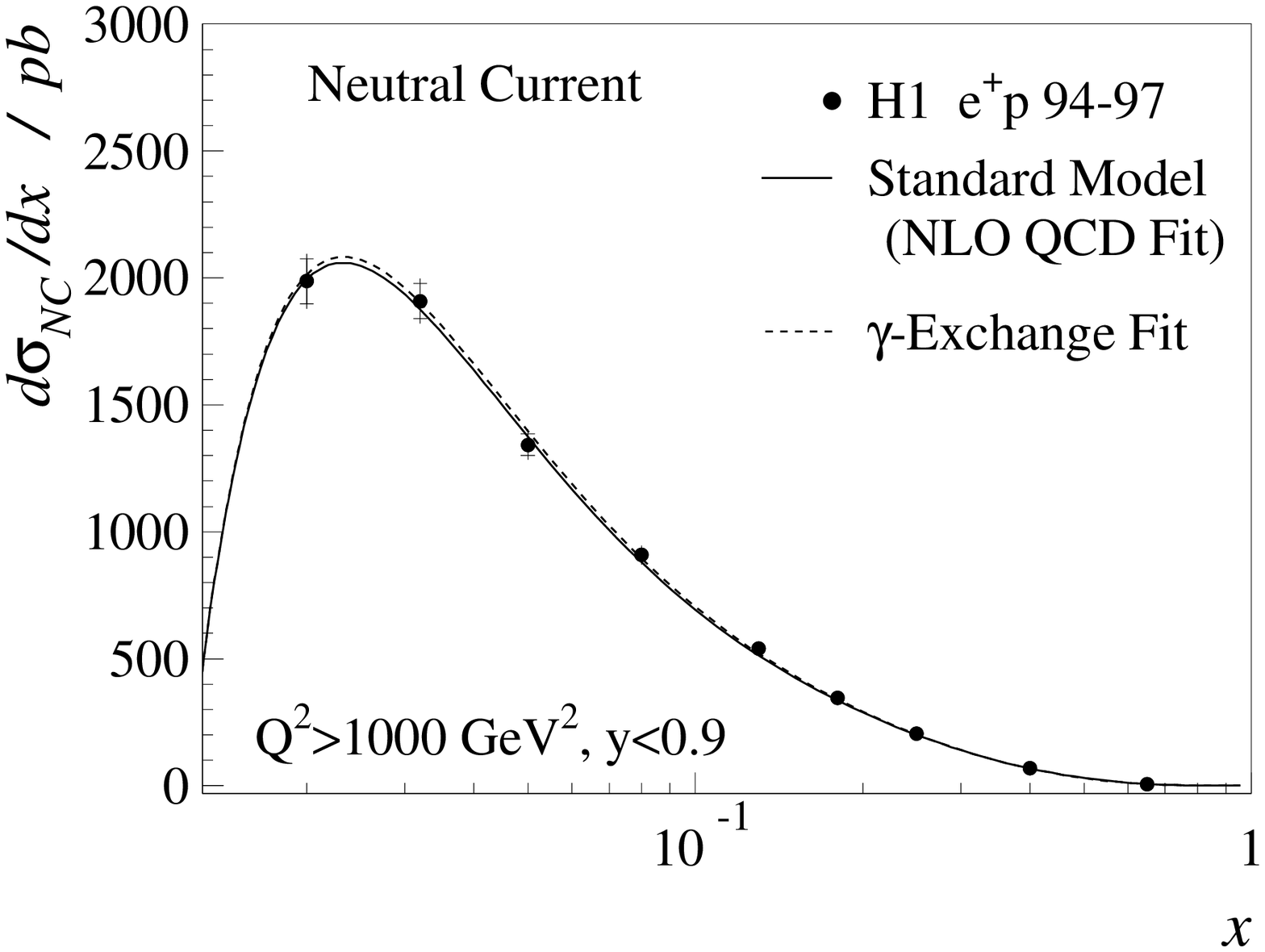,width=8.7cm}}
\put(-2,-10){\epsfig{file=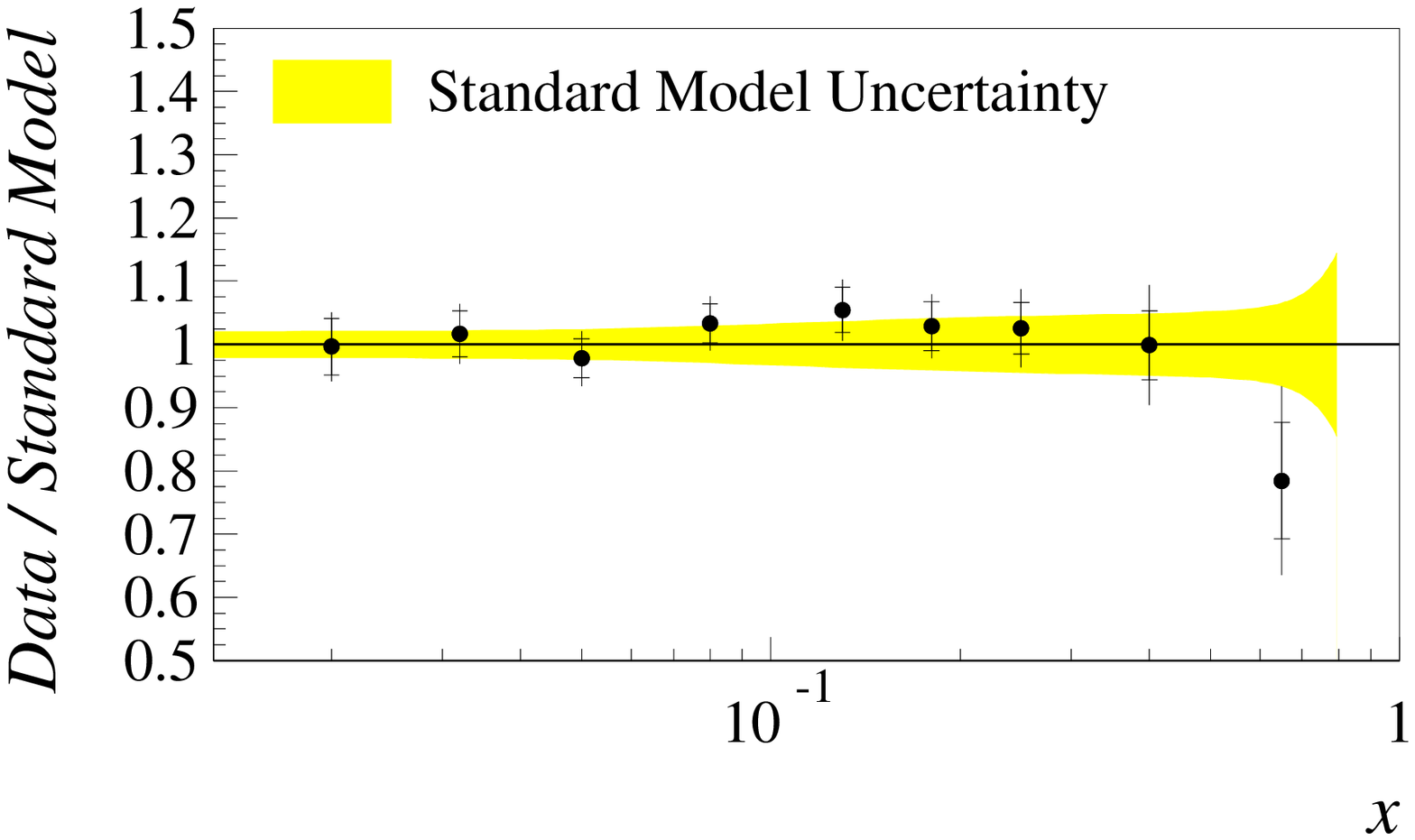,width=8.7cm}}
\put(18,92){\bf (a)}
\put(18,5){\bf (c)}
\put(80,40){\epsfig{file=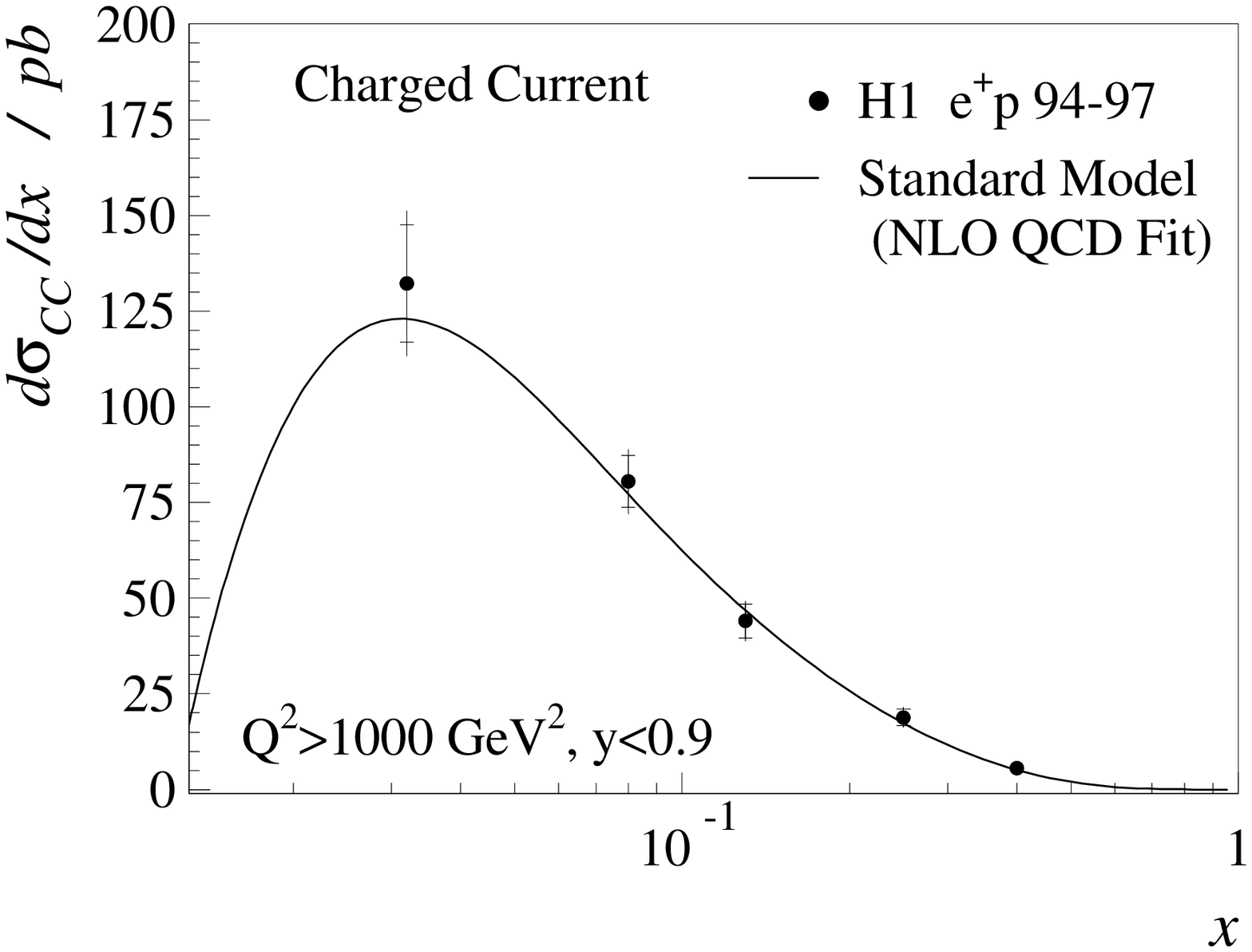,width=8.7cm}}
\put(80,-10){\epsfig{file=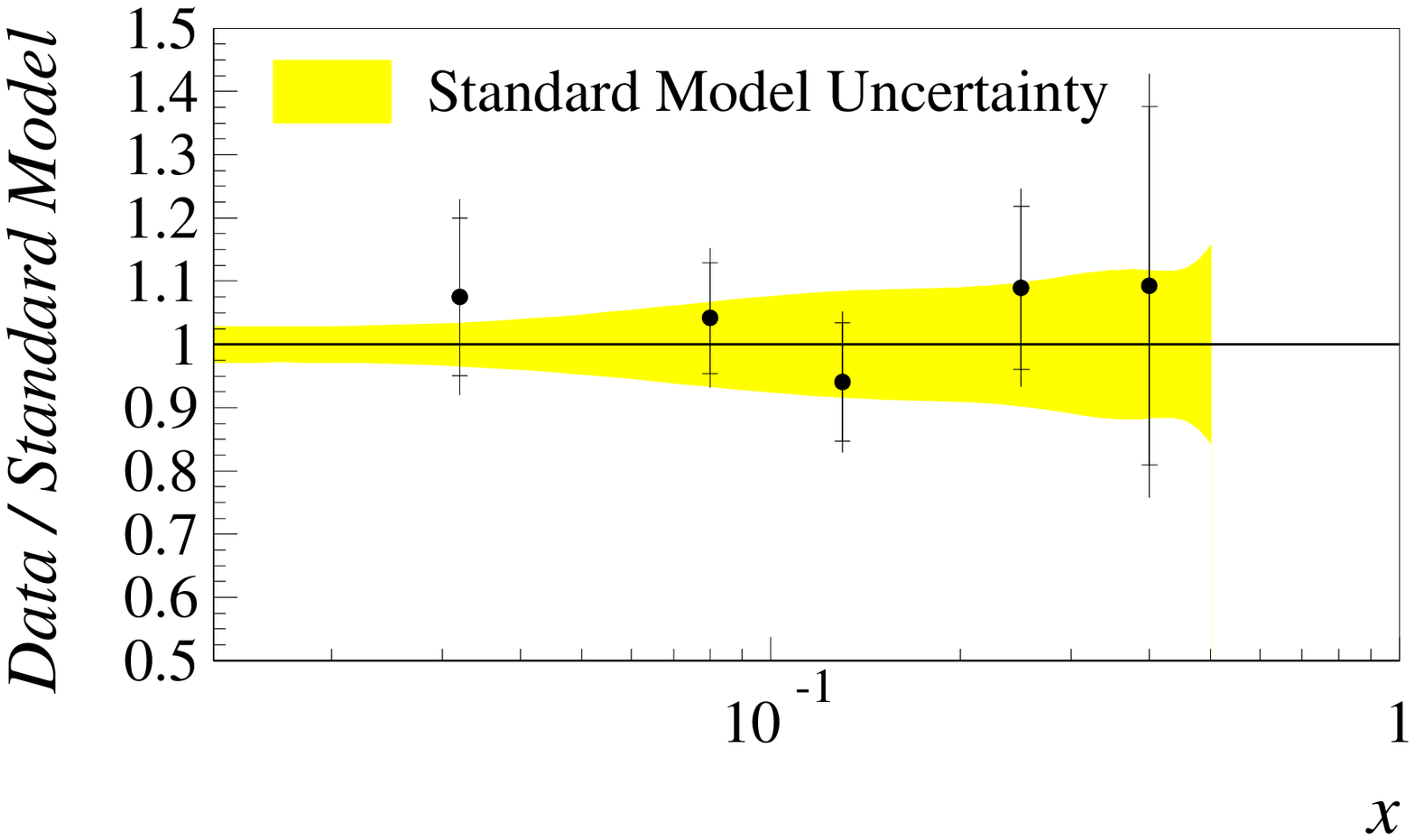,width=8.7cm}}
\put(100,92){\bf (b)}
\put(100,5){\bf (d)}
\end{picture}
\end{center}
\caption[]{\label{fig:dx.nccc}\sl 
{(a) Measurement of the NC cross-section $d\sigma_{NC}/dx$ (points) 
compared with the $\gamma$-Exchange Fit (dashed curve) and
with the Standard Model expectation as obtained from the 
NLO QCD Fit (solid curve).
(b) Measurement of the CC cross-section $d\sigma_{CC}/dx$ (points)
compared with the Standard Model expectation (solid curve). 
(c,d) NC, CC cross-sections (points) divided by the 
Standard Model expectation.
The shaded band in (c,d) represents the Standard Model uncertainty.
The cross-sections are given for $Q^2 > 1000$~\Gevv\ and $y< 0.9$.
The inner (outer) error bars represent the statistical (total) errors.}}
\label{fig:dx}
\end{figure}

The NC (CC) single differential cross-sections 
$d\sigma_{NC(CC)}/dx$ are shown for $Q^2 > 1000$~\Gevv\
in fig.~\ref{fig:dx.nccc}a(b). The measurement extends in $x$ from
$0.013$ to $0.65$ ($0.025$ to $0.4$). 
The cross-sections rise towards low $x$. 
The decrease of the cross-section
at $x < 3 \cdot 10^{-2}$ is due to the kinematic requirements $y < 0.9$ and
$Q^2 > 1000$~\Gevv . 
In this $Q^2$ range the 
NC and CC cross-sections are still 
dominated by positron scattering on low $x$ sea partons.

The ratio of the measured cross-sections $d\sigma_{NC(CC)}/d x$ to
the Standard Model expectation is shown in fig.~\ref{fig:dx.nccc}c(d).
Also shown is the uncertainty on the Standard Model expectation which was
determined using the procedure described in section~\ref{QCDA}. 
This uncertainty for the NC cross-section is about $2.5\%$ at $x=0.02$ 
and increases to about $7.0\%$ at $x=0.65$. 
It is larger for the CC cross-section, rising from about $3.5\%$
at $x=0.03$ to about $12.0\%$ at $x=0.4$. 
At high $x$ the CC cross-section depends mainly on the $d$ quark density
which is less constrained than the $u$ quark density. 
The main contributions to the uncertainty of the 
$d$ quark density in this 
region originate from the experimental errors of the BCDMS deuteron data 
and from the theoretical assumptions for the deuteron binding correction.
All data agree well with the Standard Model expectation. 
The significance of the difference between the measurement of 
$d\sigma_{NC}/d x$ and the expectation at $x=0.65$ is small when taking 
into account the systematic error and the uncertainty of the expectation.

The $\gamma$-Exchange Fit, also 
displayed in fig.~\ref{fig:dx}a, shows 
almost no difference from the
Standard Model expectation.
This shows that NC scattering is still dominated
by photon exchange at $Q^2 \approx 1000$~\Gevv . 

\begin{figure}[htb] 
\begin{center}
\epsfig{file=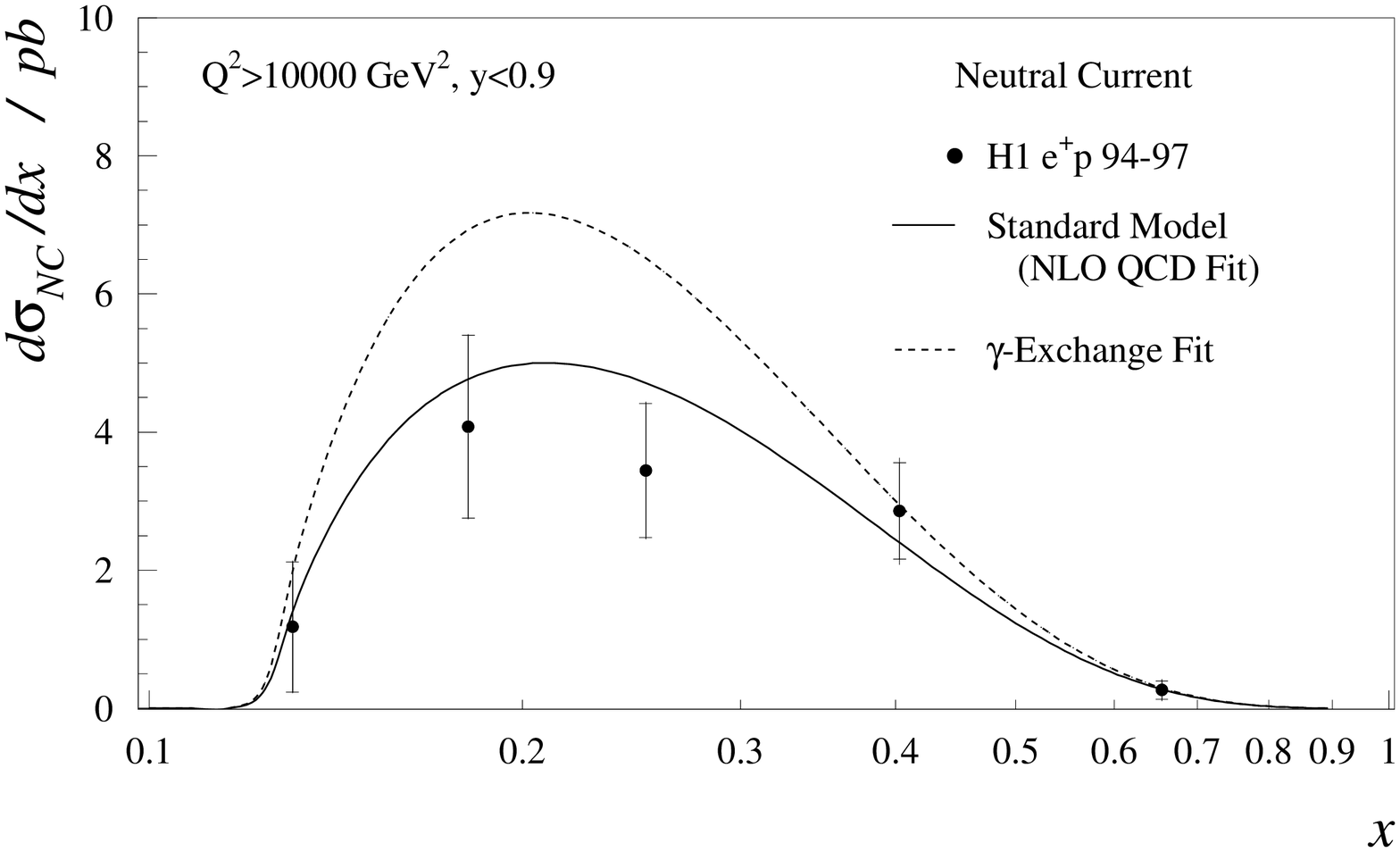,
width=14cm,bbllx=30pt,bblly=30pt,bburx=650pt,bbury=420pt} 
\caption[]{\label{fig:dxhq}
\sl {Measurement of the NC cross-section $d\sigma_{NC}/dx$ (points) 
at $Q^2 > 10\,000$~\Gevv\ compared with the Standard
Model expectation (solid curve) and with the expectation 
when the coupling to the $Z^{\circ}$ boson is not taken into account
(dashed curve).
The inner (outer) error bars represent the statistical (total) errors.
}}
\end{center} 
\end{figure}

The results of the two fits are compared with $d\sigma_{NC}/dx$ 
for $Q^2\geq 10\,000$~\Gevv\ in fig.~\ref{fig:dxhq}.
In contrast with what is observed for $Q^2 > 1000$~\Gevv , 
the two expectations are significantly different. 
The Standard Model expectation gives a good description of the measurements.
The $\gamma$-Exchange Fit fails to do so.

\subsection{Measurement of the 
\boldmath{$Q^2$} Dependence of the NC and CC Cross-Sections}

\begin{figure}[tp]
\begin{center}
\epsfig{file=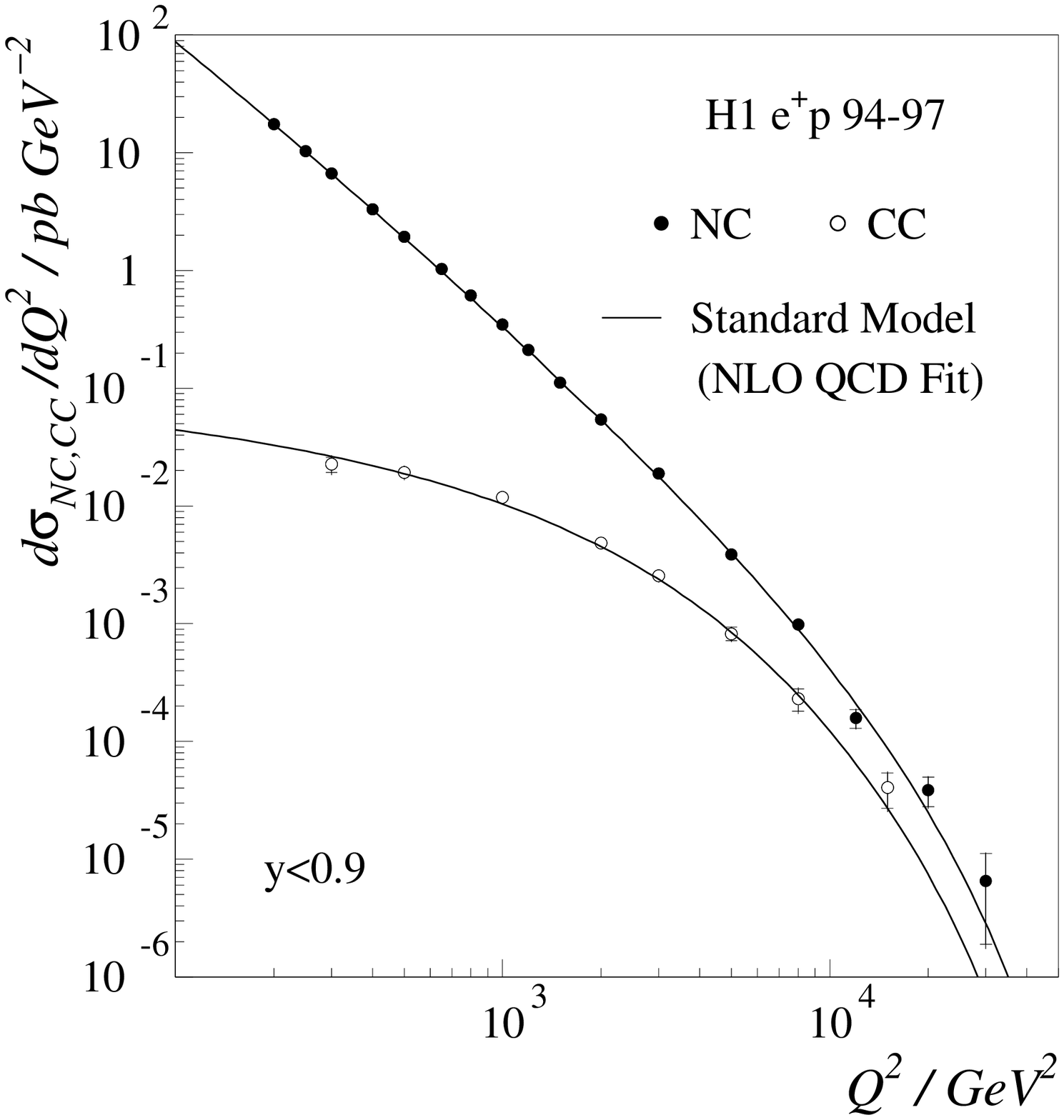,
width=9.6cm,bbllx=30pt,bblly=30pt,bburx=490pt,bbury=530pt}
\caption[]{\label{fig:dq2.nccc}
\sl 
{Measurement of the NC (solid points) and CC (open points) cross-sections 
$d\sigma_{NC}/ d Q^2$ 
and $d\sigma_{CC}/ d Q^2$  compared with the Standard Model
expectation (solid curves).
The cross-sections are given for $y< 0.9$.
The inner (outer) error bars represent the statistical (total) errors.}}
\begin{picture}(160,80)(0,0)
\put(-3,-5){\epsfig{file=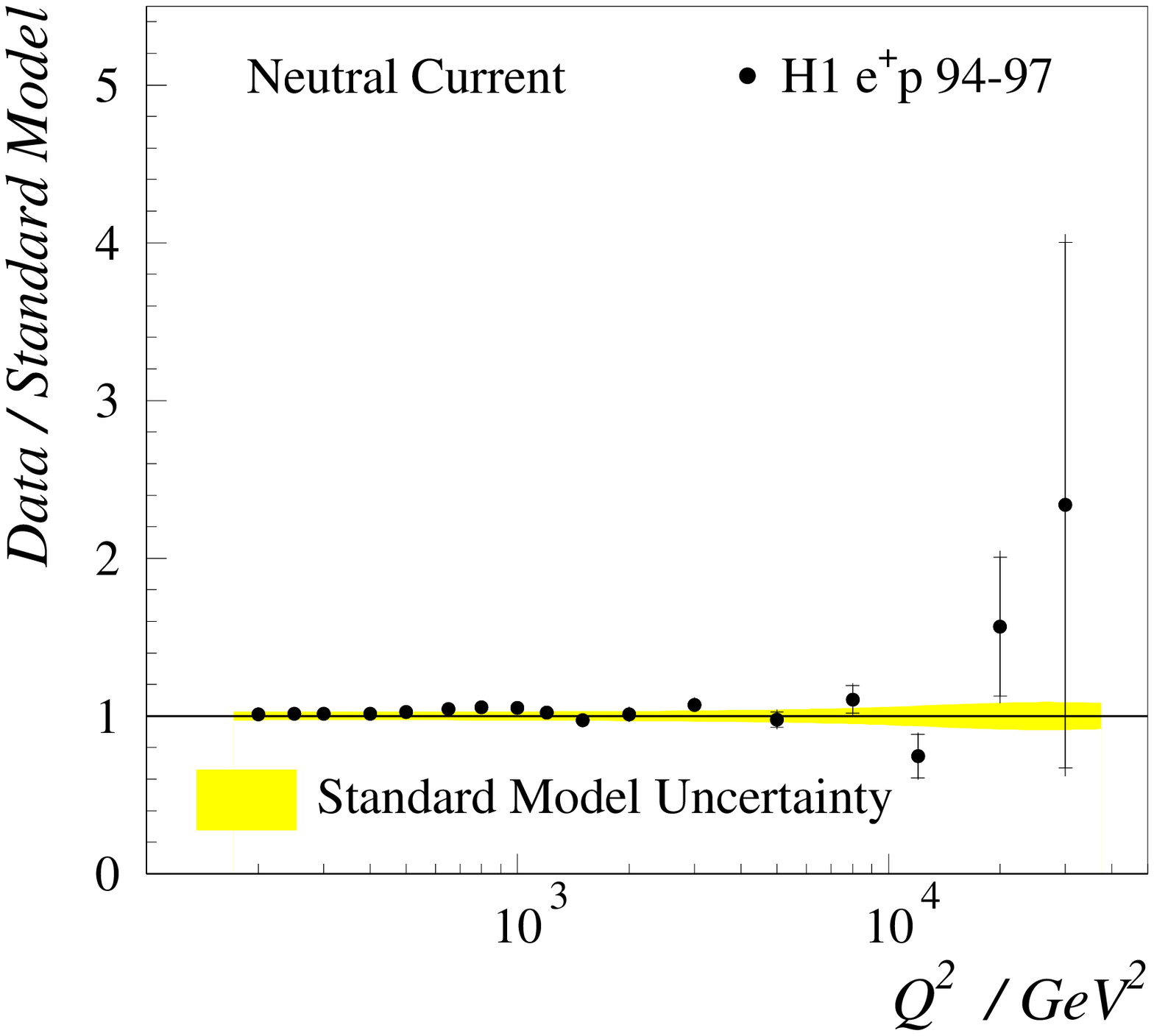,
width=9.cm}}
\put(13,22){\epsfig{file=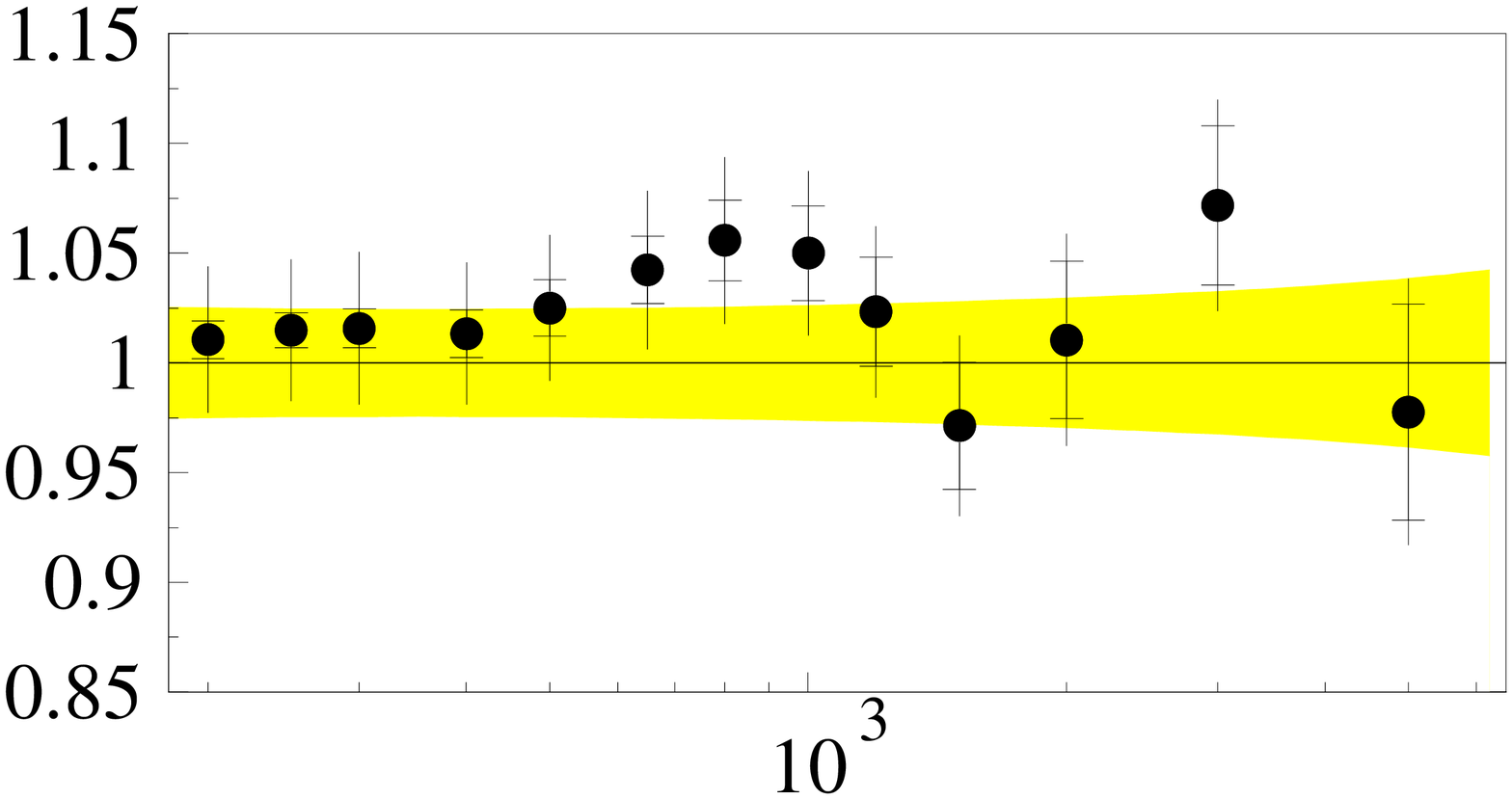,
width=5.4cm}}
\put(17,57){\bf (a)}
\put(77,-5){\epsfig{file=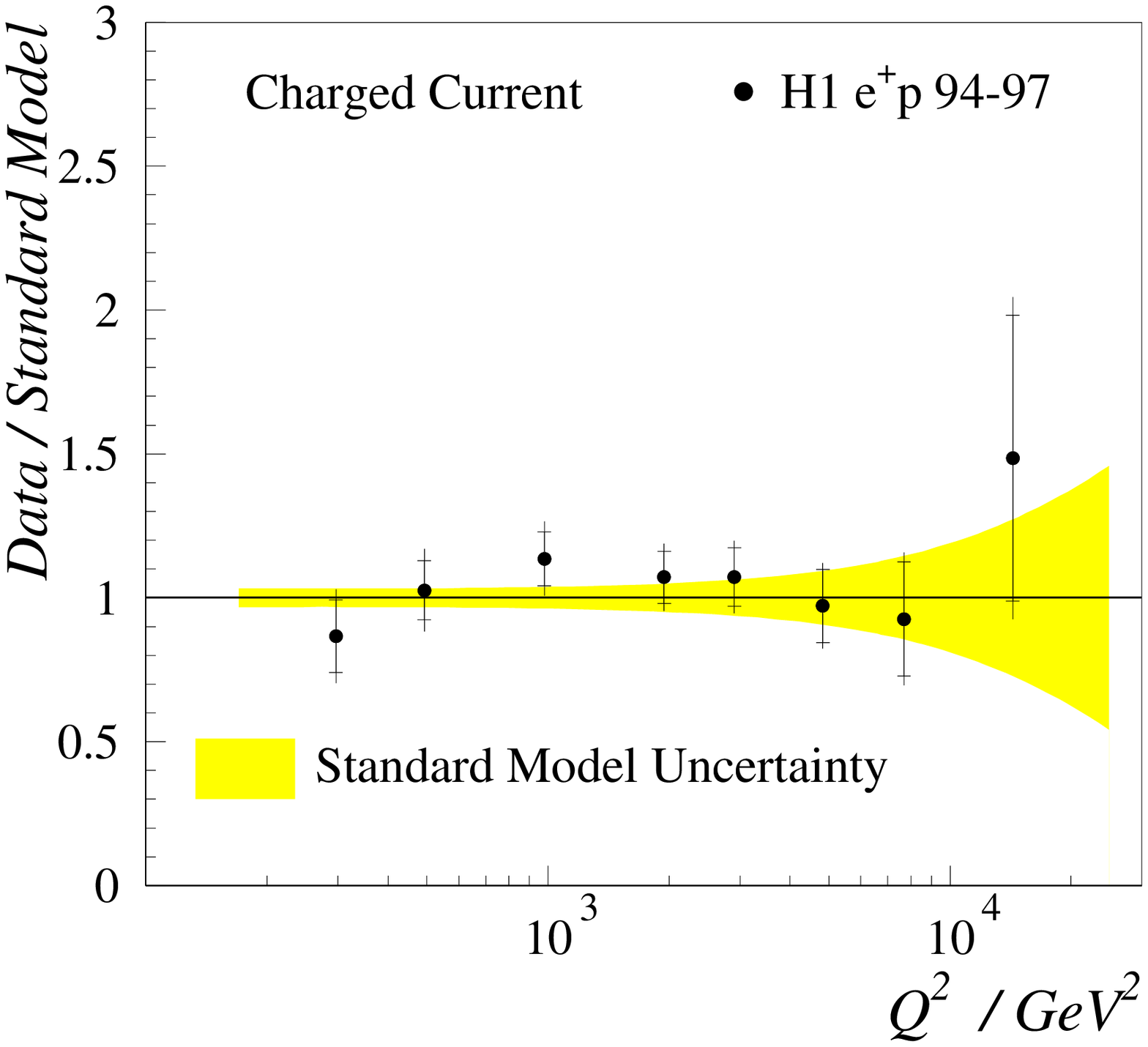,
width=9.cm}}
\put(97,57){\bf (b)}
\end{picture}
\end{center}
\caption[]{\label{fig:dq2.ncccr}\sl 
{NC (a) and  CC (b) cross-sections (points) divided by 
the Standard Model expectation.
The shaded bands represent the Standard Model uncertainty.
The inner (outer) error bars represent the statistical (total) errors.}}
\end{figure}

The NC  and CC single differential cross-sections $d\sigma_{NC}/dQ^2$ 
and $d\sigma_{CC}/dQ^2$ 
are shown in fig.~\ref{fig:dq2.nccc} and are listed in
tables~\ref{tabsnc} and \ref{tabscc} respectively.
Also shown is the Standard Model expectation given by the NLO QCD Fit. 
The cross-sections have been corrected
for a part of the cross-section that is unmeasured due to
kinematic requirements.
With these corrections (see tables~\ref{tabsnc} and \ref{tabscc}) 
the NC and CC measurements are
presented for the same kinematic range of $y<0.9$.
The statistical uncertainty is the dominating error at $Q^2$
above $1000 $~\Gevv\ for the NC cross-section, 
and at all $Q^2$ for the CC cross-section.
The systematic errors on these cross-sections 
are about $3$ ($7$)$\%$ in the NC (CC) case. 

The measurement of the NC cross-section spans more than 
two orders of magnitude~in~$Q^2$. The  cross-section falls
with $Q^2$ by about $6$ orders of magnitude.
Due to the propagator mass term and to the different coupling the CC
cross-section is smaller than the NC
cross-section, 
and it falls less steeply, 
by about $3$ orders of magnitude, between $Q^2 = 300$ and 
$15\,000$~\Gevv . The shape and magnitude of the NC and CC 
cross-sections are well described by
the Standard Model expectation. 

The ratio of the measured NC (CC) cross-section to the Standard Model 
expectation
is shown in fig.~\ref{fig:dq2.ncccr}a(b). 
The NC data at $Q^2 \le 5000$~\Gevv , 
shown also in an inserted figure 
in fig.~\ref{fig:dq2.ncccr}a, are well described 
by the NLO QCD Fit.
The enhancement of the cross-section, visible in the two highest
$Q^2$ measurements, corresponds
to the excess discussed
in section~\ref{CSM2}.
The Standard Model uncertainty shown in fig.~\ref{fig:dq2.ncccr}a and 
b is determined from
the total errors of the fit discussed in section~\ref{QCDA}.

\subsection{Measurement of the \boldmath{$W$} 
Boson Mass from the CC Cross-Section}
\label{EW}

CC interactions are understood in  the Standard Model 
in terms of the exchange  in the
$t$-channel of a $W$ boson.
Therefore the dependence on $Q^2$ of the CC cross-section and a
determination from it of the $W$ mass arising in the $t$-channel propagator
(eq.~\ref{Scc}) makes possible an important test of the space-like predictions
of the Standard Model~\cite{bardin}.
By comparing the propagator mass with the mass of the $W$ boson
measured in experiments in which the $W$ decays are observed (time-like),
it is then possible to test the universality of the   Standard Model.

\begin{figure}[hbt]
\begin{center}
\epsfig{file=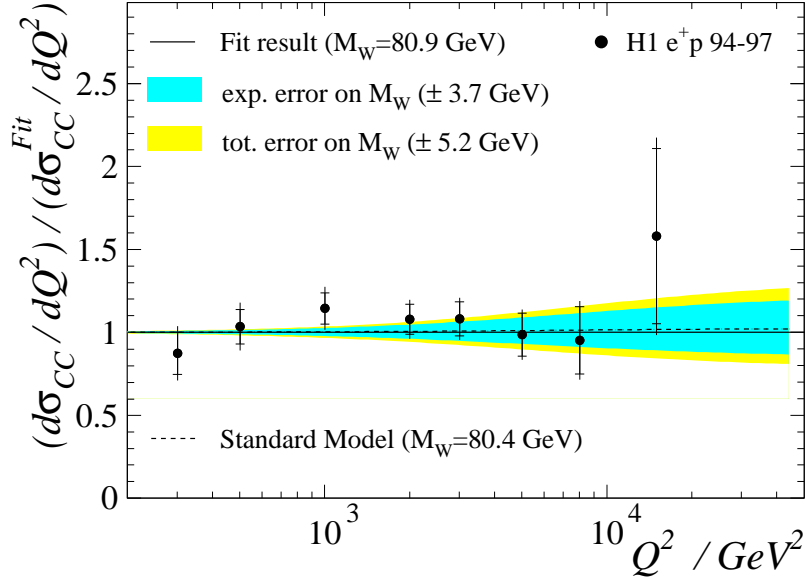,
width=10.cm,bbllx=55pt,bblly=250pt,bburx=520pt,bbury=610pt} 
\caption[]{\label{fig:mw}
\sl {Measured CC cross-section $d\sigma_{CC}/dQ^2$ (points)
divided by the Propagator Mass Fit (see text).
The dashed curve shows in comparison the Standard Model expectation
divided by the Propagator Mass Fit.
The shaded bands indicate the uncertainties on the Propagator Mass Fit 
due to 
the experimental ($\pm 3.7 \ \Gev $) and 
the total ($\pm 5.2 \ \Gev $) error on $M_W$.}}
\end{center} 
\end{figure}

A fit of the CC cross-section which is sensitive only to the value of
$M_{W}$ from the propagator term is performed by taking the
Standard Model expectation of the CC cross-section 
(eq.~\ref{Scc}) and allowing only $M_{W}$ to vary. 
The Fermi constant $G_{F}$ is set to its experimentally determined
value $G_{\mu}$~\cite{gmu}. The expectation is calculated using the
HECTOR program with $\phi_{CC}$ evaluated using the PDFs
from the Low $Q^2$ Fit\footnote{The weak radiative corrections
have been taken into account for the theoretical predictions and
have a negligible
effect on the results.}.
The resulting Propagator Mass Fit, made using the double
differential CC cross-section data, 
has a $\chi^2/{\rm ndf}$ of $19.9/(25-1)=0.83$ and gives the
value:
\begin{equation}
M_{W}=80.9 \pm 3.3 ({\rm stat.}) \pm 1.7 
 ({\rm syst.}) \pm 3.7 ({\rm theo.}) {\rm GeV} \, .
\end{equation}
The Standard Model uncertainty (theo.) 
is evaluated by varying the assumptions for the 
input Low $Q^2$ Fit as described in section~\ref{QCDA}. 
The largest contribution
to this uncertainty comes from the parameterization of the 
$\bar{d}/\bar{u}$ asymmetry which leads
to an error\footnote{
As described in section~\ref{QCDA}, the effect of an uncertainty of 
$\pm 50\%$ of the deuteron binding corrections is included
in the theoretical error. However, 
if the deuteron binding correction is not applied, 
the resulting $W$ mass
is shifted by $-0.7$ \Gev . If the correction proposed in~\cite{melnit} 
is applied the $W$ mass is shifted by $-1.7$ \Gev .}
 on the $W$ mass of $1.4$ \Gev.
The value of $M_{W}$ extracted in the space-like regime is thus found
to be in agreement with time-like determinations~\cite{mwworld}.
This result is illustrated in fig.~\ref{fig:mw}, 
where are shown the ratio of
the measured CC cross-section  $d\sigma_{CC}/dQ^2$,
and the ratio of the
Standard Model expectation from the Low $Q^2$ Fit,
to the result of the Propagator Mass Fit.

\section{Summary}

Neutral
and charged current processes with $Q^2$ between $150$ and 
$30\,000$~\Gevv\ and $x$ between $0.0032$ and $0.65$ 
have been investigated in $e^+ p$ collisions 
with the H1 detector at HERA using the data taken from 1994 to 1997.
The increased integrated luminosity, combined with
progress in the understanding of the detector response, 
has permitted significantly more precise measurements at high $Q^2$.
The double differential cross-section $d^2\sigma/dx dQ^2$ has been 
measured for NC and CC interactions in new kinematic domains.

The cross-section $d^2\sigma_{NC}/dx dQ^2$ has a typical precision
of $4\%$ for the bulk of the measurements. 
They are well described by a NLO
QCD fit performed on the low $Q^2$ H1 and 
fixed target data (BCDMS and NMC).
The inclusion of the high $Q^2$ data in the fit reduces the
Standard Model expectation 
at high $x$ and high $Q^2$ by about $3\%$ and its
uncertainty to for example  $6\%$ at $Q^2 \approx 10\,000$~\Gevv\
and $x = 0.4$.
The test of perturbative QCD in DIS is extended 
with this measurement to higher
$Q^2$, showing that the validity of the DGLAP equations extends over 
$4$ orders of magnitude in $Q^2$. 

The decrease of the cross-section, which is expected 
in $e^+p$ collisions at $Q^2 \ \gapprox \ 8000 $~\Gevv\ 
due to the $\gamma Z^\circ$ interference,
is observed at high $Q^2$ for
$ 0.08 \le x \le 0.25$.
In contrast at $x=0.4$ and $Q^2 > 15\,000$~\Gevv\ an enhancement 
of the cross-section relative to 
the expectation is visible.
This effect was reported in a previous publication using 
the 1994--1996 data alone.
It has become less significant 
with the addition of the 1997 data.
At the highest $x$ value of $0.65$, the cross-section is
slightly below the expectation which is mainly 
constrained by the BCDMS data.

The cross-section $d^2 \sigma_{CC}/dxdQ^2$ has been measured for
$Q^2$ between $300$ and $15\,000$~\Gevv\ and for $x$ between $0.013$ 
and $0.4$. The uncertainties of the measurements are dominated 
by the statistical errors.
The Standard Model expectation agrees well with the data.

An extraction of the $u$ and $d$ quark densities
at high $x$ ($x=0.25,0.4$) has been made 
from the NC and CC cross-sections, giving
complementary information compared to the
previous extractions of the valence quark densities from the 
deep-inelastic scattering  of leptons 
off hydrogen and deuterium targets.

The NC and CC single differential cross-sections $d\sigma_{NC}/dx$
and $d\sigma_{CC}/dx$ have
been presented for $Q^2 > 1000$~\Gevv\ and $y<0.9$.
The Standard Model expectation has been found to be in good agreement
with both measurements. This remains true
at $Q^2 >10\,000$~\Gevv\, where the effects of the 
$Z^{\circ}$ become manifest.
If the $Z^{\circ}$ exchange is removed from the Standard Model
calculation, the prediction fails to describe the
measurements.

The NC and CC single differential cross-sections $d\sigma_{NC}/dQ^2$ 
and $d\sigma_{CC}/dQ^2$ 
have been shown to be described by the Standard Model expectation. 
A fit to the $Q^2$ dependence of the CC double differential cross-section
gives a mass $M_{W} = 80.9 \pm  3.7$ (exp.) $\pm 3.7$ (theo.)  \Gev.
This value agrees
well with the mass of the $W$ boson measured in time-like processes, 
thereby confirming the electroweak sector of the Standard
Model in space-like lepton nucleon scattering.

\section*{Acknowledgements}

We are grateful to the HERA machine group whose outstanding
efforts have made and continue to make this experiment possible. 
We thank
the engineers and technicians for their work in constructing and now
maintaining the H1 detector, our funding agencies for 
financial support, the
DESY technical staff for continual assistance, 
and the DESY directorate for the
hospitality which they extend to the non-DESY 
members of the collaboration.
We thank H. Spiesberger for useful discussions on the 
radiative corrections aspects of this measurement.

\newpage

\include{mytable}
\include{longtab}

\end{document}

%% file: h1auts.tex
 C.~Adloff$^{33}$,                
 V.~Andreev$^{24}$,               
 B.~Andrieu$^{27}$,               
 V.~Arkadov$^{34}$,               
 A.~Astvatsatourov$^{34}$,        
 I.~Ayyaz$^{28}$,                 
 A.~Babaev$^{23}$,                
 J.~B\"ahr$^{34}$,                
 P.~Baranov$^{24}$,               
 E.~Barrelet$^{28}$
% ...

\vspace*{1cm}

\noindent
% ...
$ ^{23}$ Institute for Theoretical and Experimental Physics,
          Moscow, Russia \\
 $ ^{24}$ Lebedev Physical Institute, Moscow, Russia$^{f,k}$ \\
% ...
$ ^{27}$ LPNHE, \'{E}cole Polytechnique, IN2P3-CNRS, Palaiseau, France \\
 $ ^{28}$ LPNHE, Universit\'{e}s Paris VI and VII, IN2P3-CNRS,
          Paris, France \\
% ...
$ ^{33}$ Fachbereich Physik, Bergische Universit\"at Gesamthochschule
          Wuppertal, Wuppertal, Germany$^a$ \\
 $ ^{34}$ DESY, Zeuthen, Germany$^a$ \\
% ...
$ ^a$ Supported by the Bundesministerium f\"ur Bildung, Wissenschaft,
        Forschung und Technologie, FRG,
        under contract numbers 7AC17P, 7AC47P, 7DO55P, 7HH17I, 7HH27P,
        7HD17P, 7HD27P, 7KI17I, 6MP17I and 7WT87P \\
% ...
$ ^f$ Supported by the Deutsche Forschungsgemeinschaft \\
% ...
$ ^k$ Supported by Russian Foundation for Basic Research 
       grant no. 96-02-00019 \\
% ...

%% file: mytable.tex
\begin{table}[ht]
  \begin{center}
    \small
    \begin{tabular}{|*{3}{c|}|*{4}{c|}|c|*{4}{r|}|r|}
      \hline
      $Q^2$  &$x$  & $y$ & $\tilde{\sigma}_{NC}$ & 
      $\delta_{sta}$ & $\delta_{sys}$ &
      $\delta_{tot}$ &
      $F_2$ & $\Delta_{all}$ & $\Delta_{F_2}$ &$\Delta_{F_3}$ &$\Delta_{F_L}$ 
      & $\delta^{qed}_{NC}$ 
      \\[-1mm]
      \footnotesize $(\rm GeV^2)$ & & & &
      \footnotesize $(\%)$ &
      \footnotesize $(\%)$ &\footnotesize $(\%)$ &
      &
      \multicolumn{1}{c|}{\footnotesize $(\%)$}&
      \multicolumn{1}{c|}{\footnotesize $(\%)$}&
      \multicolumn{1}{c|}{\footnotesize $(\%)$}&
      \multicolumn{1}{c||}{\footnotesize $(\%)$}&
      \multicolumn{1}{c|}{\footnotesize $(\%)$} \\
 \hline
$   150$&$  0.003$&$  0.518$&$  1.240$&$  1.8$&$  5.2$&$  5.5$&$  1.291$&$ -4.0$&$  0.1$&$ -0.1$&$ -4.0$&$ 7.0 $\\
$   150$&$  0.005$&$  0.331$&$  1.100$&$  1.8$&$  3.3$&$  3.8$&$  1.115$&$ -1.3$&$  0.1$&$ -0.1$&$ -1.3$&$ 6.8 $\\
$   150$&$  0.008$&$  0.207$&$  0.920$&$  2.9$&$  8.9$&$  9.3$&$  0.924$&$ -0.4$&$  0.1$&$ -0.1$&$ -0.4$&$ 6.7 $\\
 \hline                                                                                                    
$   200$&$  0.005$&$  0.442$&$  1.102$&$  1.8$&$  5.0$&$  5.3$&$  1.130$&$ -2.5$&$  0.2$&$ -0.1$&$ -2.5$&$ 7.3 $\\
$   200$&$  0.008$&$  0.276$&$  0.915$&$  1.9$&$  3.5$&$  4.0$&$  0.922$&$ -0.8$&$  0.2$&$ -0.1$&$ -0.8$&$ 7.1 $\\
$   200$&$  0.013$&$  0.170$&$  0.765$&$  2.2$&$  3.7$&$  4.3$&$  0.767$&$ -0.2$&$  0.2$&$ -0.1$&$ -0.2$&$ 7.0 $\\
$   200$&$  0.020$&$  0.110$&$  0.696$&$  2.6$&$  4.9$&$  5.5$&$  0.696$&$ -0.1$&$  0.2$&$ -0.1$&$ -0.1$&$ 7.0 $\\
$   200$&$  0.032$&$  0.069$&$  0.601$&$  3.2$&$  7.5$&$  8.1$&$  0.601$&$  0.0$&$  0.2$&$ -0.1$&$  0.0$&$ 7.0 $\\
$   200$&$  0.050$&$  0.044$&$  0.516$&$  3.7$&$  8.2$&$  9.0$&$  0.516$&$  0.0$&$  0.2$&$ -0.1$&$  0.0$&$ 7.1 $\\
$   200$&$  0.080$&$  0.028$&$  0.439$&$  4.2$&$  9.0$&$  9.9$&$  0.439$&$  0.0$&$  0.1$&$ -0.1$&$  0.0$&$ 7.0 $\\
 \hline                                                                                                    
$   250$&$  0.005$&$  0.552$&$  1.113$&$  2.3$&$  5.1$&$  5.6$&$  1.161$&$ -4.1$&$  0.2$&$ -0.2$&$ -4.1$&$ 7.6 $\\
$   250$&$  0.008$&$  0.345$&$  1.018$&$  2.0$&$  3.7$&$  4.2$&$  1.031$&$ -1.2$&$  0.2$&$ -0.1$&$ -1.2$&$ 7.5 $\\
$   250$&$  0.013$&$  0.212$&$  0.807$&$  2.1$&$  3.9$&$  4.4$&$  0.810$&$ -0.4$&$  0.2$&$ -0.1$&$ -0.4$&$ 7.3 $\\
$   250$&$  0.020$&$  0.138$&$  0.721$&$  2.1$&$  3.6$&$  4.1$&$  0.721$&$ -0.1$&$  0.2$&$ -0.1$&$ -0.1$&$ 7.3 $\\
$   250$&$  0.032$&$  0.086$&$  0.606$&$  2.2$&$  3.6$&$  4.3$&$  0.606$&$  0.0$&$  0.2$&$ -0.1$&$  0.0$&$ 7.3 $\\
$   250$&$  0.050$&$  0.055$&$  0.529$&$  2.4$&$  3.4$&$  4.2$&$  0.529$&$  0.0$&$  0.2$&$ -0.1$&$  0.0$&$ 7.3 $\\
$   250$&$  0.080$&$  0.035$&$  0.430$&$  2.7$&$  3.6$&$  4.5$&$  0.430$&$  0.0$&$  0.2$&$ -0.1$&$  0.0$&$ 7.3 $\\
$   250$&$  0.130$&$  0.021$&$  0.334$&$  3.4$&$  4.3$&$  5.5$&$  0.334$&$  0.0$&$  0.2$&$ -0.1$&$  0.0$&$ 7.1 $\\
$   250$&$  0.250$&$  0.011$&$  0.240$&$  3.3$&$  7.4$&$  8.1$&$  0.239$&$  0.1$&$  0.2$&$  0.0$&$  0.0$&$ 6.1 $\\
$   250$&$  0.400$&$  0.007$&$  0.122$&$  5.9$&$ 12.1$&$ 13.4$&$  0.122$&$  0.1$&$  0.2$&$  0.0$&$  0.0$&$ 4.0 $\\
 \hline                                                                                                    
$   300$&$  0.005$&$  0.663$&$  1.139$&$  3.4$&$  5.6$&$  6.5$&$  1.214$&$ -6.2$&$  0.3$&$ -0.2$&$ -6.2$&$ 7.8 $\\
$   300$&$  0.008$&$  0.414$&$  0.989$&$  2.4$&$  5.1$&$  5.7$&$  1.008$&$ -1.9$&$  0.3$&$ -0.2$&$ -1.9$&$ 7.7 $\\
$   300$&$  0.013$&$  0.255$&$  0.846$&$  2.4$&$  3.8$&$  4.5$&$  0.851$&$ -0.6$&$  0.3$&$ -0.2$&$ -0.6$&$ 7.6 $\\
$   300$&$  0.020$&$  0.166$&$  0.740$&$  2.4$&$  3.9$&$  4.6$&$  0.742$&$ -0.2$&$  0.3$&$ -0.2$&$ -0.2$&$ 7.5 $\\
$   300$&$  0.032$&$  0.104$&$  0.629$&$  2.4$&$  3.7$&$  4.4$&$  0.630$&$  0.0$&$  0.2$&$ -0.1$&$ -0.1$&$ 7.5 $\\
$   300$&$  0.050$&$  0.066$&$  0.499$&$  2.6$&$  3.6$&$  4.5$&$  0.499$&$  0.0$&$  0.2$&$ -0.1$&$  0.0$&$ 7.5 $\\
$   300$&$  0.080$&$  0.041$&$  0.456$&$  2.7$&$  3.9$&$  4.8$&$  0.456$&$  0.0$&$  0.2$&$ -0.1$&$  0.0$&$ 7.5 $\\
$   300$&$  0.130$&$  0.025$&$  0.346$&$  3.4$&$  5.8$&$  6.8$&$  0.346$&$  0.1$&$  0.2$&$ -0.1$&$  0.0$&$ 7.3 $\\
$   300$&$  0.250$&$  0.013$&$  0.250$&$  3.1$&$  8.1$&$  8.7$&$  0.250$&$  0.1$&$  0.2$&$ -0.1$&$  0.0$&$ 6.2 $\\
$   300$&$  0.400$&$  0.008$&$  0.140$&$  5.7$&$ 14.5$&$ 15.6$&$  0.140$&$  0.1$&$  0.2$&$  0.0$&$  0.0$&$ 4.2 $\\
 \hline                                                                                                    
$   400$&$  0.008$&$  0.552$&$  0.976$&$  3.1$&$  5.1$&$  6.0$&$  1.013$&$ -3.6$&$  0.4$&$ -0.3$&$ -3.5$&$ 8.2 $\\
$   400$&$  0.013$&$  0.340$&$  0.841$&$  2.8$&$  3.9$&$  4.8$&$  0.850$&$ -1.1$&$  0.4$&$ -0.3$&$ -1.0$&$ 8.0 $\\
$   400$&$  0.020$&$  0.221$&$  0.739$&$  2.8$&$  3.7$&$  4.7$&$  0.742$&$ -0.4$&$  0.4$&$ -0.3$&$ -0.3$&$ 7.9 $\\
$   400$&$  0.032$&$  0.138$&$  0.619$&$  2.8$&$  3.6$&$  4.6$&$  0.619$&$ -0.1$&$  0.4$&$ -0.2$&$ -0.1$&$ 7.9 $\\
$   400$&$  0.050$&$  0.088$&$  0.513$&$  3.0$&$  3.8$&$  4.8$&$  0.513$&$  0.0$&$  0.4$&$ -0.2$&$  0.0$&$ 7.9 $\\
$   400$&$  0.080$&$  0.055$&$  0.455$&$  3.1$&$  4.0$&$  5.1$&$  0.455$&$  0.0$&$  0.3$&$ -0.2$&$  0.0$&$ 7.8 $\\
$   400$&$  0.130$&$  0.034$&$  0.373$&$  3.8$&$  4.5$&$  5.9$&$  0.373$&$  0.1$&$  0.3$&$ -0.1$&$  0.0$&$ 7.6 $\\
$   400$&$  0.250$&$  0.018$&$  0.241$&$  3.5$&$  6.5$&$  7.4$&$  0.241$&$  0.1$&$  0.3$&$ -0.1$&$  0.0$&$ 6.5 $\\
$   400$&$  0.400$&$  0.011$&$  0.155$&$  6.2$&$ 11.6$&$ 13.2$&$  0.155$&$  0.1$&$  0.3$&$ -0.1$&$  0.0$&$ 4.4 $\\ 
 \hline
    \end{tabular}
\end{center}
\caption{\small \sl NC reduced cross-section
  $\tilde{\sigma}_{NC}(x,Q^2)$ 
obtained by dividing $d^2\sigma_{NC}/dx dQ^2$ by the kinematic factor  
$x Q^4/(Y_+ 2\pi \alpha^2)$,  
with statistical error ($\delta_{sta}$), 
systematic error ($\delta_{sys}$)
and  total error ($\delta_{tot}$).
The electromagnetic proton  structure function $F_2(x,Q^2)$ is then given, together with
$\Delta_{F_2}$, $\Delta_{F_3}$, $\Delta_{F_L}$ 
(which are the corrections  due to \Ftwo, 
  \Fz\  and \FL  used to calculate $F_2$)  and  $\Delta_{all}$ as defined
in eq.~\ref{Rnc2}, i.e. 
$1+\Delta_{all}=(1+\Delta_{F_2}+\Delta_{F_3}+\Delta_{F_L})
(1+\delta_{NC}^{weak})$. 
The correction $\delta^{qed}_{NC}$ due to QED radiation effects, 
as defined in eq.~\ref{init}, is also given. 
The normalization uncertainty, 
which is not included in the systematic error, is $1.5\%$.   
The table continues on the next 2 pages.}
\label{tabf2}
\end{table}

\begin{table}[ht]
  \begin{center}
    \small
    \begin{tabular}{|*{3}{c|}|*{4}{c|}|c|*{4}{r|}|r|}
      \hline
      $Q^2$  &$x$  & $y$ & $\tilde{\sigma}_{NC}$ & 
      $\delta_{sta}$ & $\delta_{sys}$ &
      $\delta_{tot}$ &
      $F_2$ & $\Delta_{all}$ & $\Delta_{F_2}$ &$\Delta_{F_3}$ &$\Delta_{F_L}$ 
      & $\delta^{qed}_{NC}$
      \\[-1mm]
      \footnotesize $(\rm GeV^2)$ & & & &
      \footnotesize $(\%)$ &
      \footnotesize $(\%)$ &\footnotesize $(\%)$ &
      &
      \multicolumn{1}{c|}{\footnotesize $(\%)$}&
      \multicolumn{1}{c|}{\footnotesize $(\%)$}&
      \multicolumn{1}{c|}{\footnotesize $(\%)$}&
      \multicolumn{1}{c||}{\footnotesize $(\%)$}&
      \multicolumn{1}{c|}{\footnotesize $(\%)$} \\  
 \hline
$   500$&$  0.008$&$  0.690$&$  1.026$&$  4.2$&$  5.1$&$  6.6$&$  1.091$&$ -6.0$&$  0.5$&$ -0.5$&$ -5.8$&$ 8.5 $\\  
$   500$&$  0.013$&$  0.425$&$  0.906$&$  3.3$&$  5.2$&$  6.2$&$  0.922$&$ -1.8$&$  0.5$&$ -0.5$&$ -1.6$&$ 8.4 $\\ 
$   500$&$  0.020$&$  0.276$&$  0.792$&$  3.3$&$  3.9$&$  5.2$&$  0.797$&$ -0.6$&$  0.5$&$ -0.4$&$ -0.5$&$ 8.3 $\\ 
$   500$&$  0.032$&$  0.173$&$  0.654$&$  3.3$&$  4.0$&$  5.2$&$  0.655$&$ -0.2$&$  0.5$&$ -0.4$&$ -0.2$&$ 8.2 $\\ 
$   500$&$  0.050$&$  0.110$&$  0.508$&$  3.5$&$  4.1$&$  5.4$&$  0.509$&$  0.0$&$  0.5$&$ -0.3$&$ -0.1$&$ 8.1 $\\
$   500$&$  0.080$&$  0.069$&$  0.445$&$  3.6$&$  3.7$&$  5.2$&$  0.445$&$  0.0$&$  0.5$&$ -0.3$&$  0.0$&$ 8.0 $\\
$   500$&$  0.130$&$  0.042$&$  0.368$&$  4.3$&$  4.3$&$  6.1$&$  0.367$&$  0.1$&$  0.4$&$ -0.2$&$  0.0$&$ 7.8 $\\
$   500$&$  0.180$&$  0.031$&$  0.287$&$  4.9$&$  5.4$&$  7.3$&$  0.286$&$  0.1$&$  0.4$&$ -0.2$&$  0.0$&$ 7.4 $\\
$   500$&$  0.250$&$  0.022$&$  0.220$&$  5.9$&$  8.5$&$ 10.4$&$  0.220$&$  0.1$&$  0.4$&$ -0.1$&$  0.0$&$ 6.7 $\\
$   500$&$  0.400$&$  0.014$&$  0.143$&$  8.6$&$ 15.3$&$ 17.5$&$  0.143$&$  0.2$&$  0.4$&$ -0.1$&$  0.0$&$ 4.5 $\\
 \hline                                                                                                    
$   650$&$  0.013$&$  0.552$&$  0.903$&$  4.0$&$  4.3$&$  5.9$&$  0.933$&$ -3.2$&$  0.7$&$ -0.8$&$ -3.0$&$ 8.8 $\\
$   650$&$  0.020$&$  0.359$&$  0.718$&$  4.1$&$  3.9$&$  5.7$&$  0.727$&$ -1.2$&$  0.7$&$ -0.8$&$ -0.9$&$ 8.7 $\\
$   650$&$  0.032$&$  0.224$&$  0.633$&$  4.0$&$  4.0$&$  5.7$&$  0.635$&$ -0.4$&$  0.7$&$ -0.7$&$ -0.3$&$ 8.5 $\\
$   650$&$  0.050$&$  0.144$&$  0.521$&$  4.1$&$  3.9$&$  5.7$&$  0.522$&$ -0.1$&$  0.7$&$ -0.6$&$ -0.1$&$ 8.5 $\\
$   650$&$  0.080$&$  0.090$&$  0.436$&$  4.0$&$  4.0$&$  5.7$&$  0.436$&$  0.0$&$  0.7$&$ -0.5$&$  0.0$&$ 8.3 $\\
$   650$&$  0.130$&$  0.055$&$  0.413$&$  4.6$&$  4.7$&$  6.6$&$  0.413$&$  0.1$&$  0.6$&$ -0.4$&$  0.0$&$ 8.1 $\\
$   650$&$  0.180$&$  0.040$&$  0.309$&$  5.3$&$  5.8$&$  7.9$&$  0.309$&$  0.1$&$  0.6$&$ -0.3$&$  0.0$&$ 7.7 $\\
$   650$&$  0.250$&$  0.029$&$  0.246$&$  6.2$&$  8.7$&$ 10.6$&$  0.246$&$  0.2$&$  0.6$&$ -0.2$&$  0.0$&$ 6.9 $\\
$   650$&$  0.400$&$  0.018$&$  0.125$&$  9.9$&$ 11.5$&$ 15.2$&$  0.125$&$  0.2$&$  0.5$&$ -0.2$&$  0.0$&$ 4.7 $\\
$   650$&$  0.650$&$  0.011$&$  0.021$&$ 14.3$&$ 15.7$&$ 21.3$&$  0.020$&$  0.3$&$  0.5$&$ -0.1$&$  0.0$&$-0.4 $\\
 \hline                                                                                                    
$   800$&$  0.013$&$  0.680$&$  1.000$&$  5.0$&$  4.7$&$  6.8$&$  1.055$&$ -5.2$&$  1.0$&$ -1.2$&$ -4.7$&$ 9.1 $\\
$   800$&$  0.020$&$  0.442$&$  0.796$&$  4.6$&$  4.3$&$  6.3$&$  0.812$&$ -1.9$&$  1.0$&$ -1.1$&$ -1.5$&$ 9.0 $\\
$   800$&$  0.032$&$  0.276$&$  0.709$&$  4.5$&$  4.0$&$  6.0$&$  0.714$&$ -0.7$&$  1.0$&$ -1.0$&$ -0.4$&$ 8.9 $\\
$   800$&$  0.050$&$  0.177$&$  0.540$&$  4.6$&$  3.9$&$  6.0$&$  0.542$&$ -0.3$&$  0.9$&$ -0.9$&$ -0.1$&$ 8.7 $\\
$   800$&$  0.080$&$  0.110$&$  0.474$&$  4.6$&$  4.2$&$  6.2$&$  0.474$&$ -0.1$&$  0.9$&$ -0.7$&$  0.0$&$ 8.6 $\\
$   800$&$  0.130$&$  0.068$&$  0.370$&$  5.4$&$  4.8$&$  7.2$&$  0.369$&$  0.1$&$  0.9$&$ -0.6$&$  0.0$&$ 8.3 $\\
$   800$&$  0.180$&$  0.049$&$  0.333$&$  6.0$&$  4.9$&$  7.8$&$  0.333$&$  0.2$&$  0.8$&$ -0.4$&$  0.0$&$ 7.9 $\\
$   800$&$  0.250$&$  0.035$&$  0.208$&$  7.5$&$  5.8$&$  9.4$&$  0.208$&$  0.2$&$  0.8$&$ -0.3$&$  0.0$&$ 7.1 $\\
$   800$&$  0.400$&$  0.022$&$  0.150$&$  9.6$&$ 10.5$&$ 14.2$&$  0.150$&$  0.3$&$  0.7$&$ -0.2$&$  0.0$&$ 4.9 $\\
$   800$&$  0.650$&$  0.014$&$  0.018$&$ 19.6$&$ 18.4$&$ 26.9$&$  0.018$&$  0.3$&$  0.6$&$ -0.2$&$  0.0$&$-0.3 $\\
 \hline                                                                                                    
$  1000$&$  0.020$&$  0.552$&$  0.754$&$  5.4$&$  3.8$&$  6.6$&$  0.779$&$ -3.2$&$  1.4$&$ -1.8$&$ -2.5$&$ 9.4 $\\
$  1000$&$  0.032$&$  0.345$&$  0.639$&$  5.6$&$  4.1$&$  6.9$&$  0.647$&$ -1.2$&$  1.3$&$ -1.6$&$ -0.7$&$ 9.2 $\\
$  1000$&$  0.050$&$  0.221$&$  0.566$&$  5.1$&$  3.8$&$  6.4$&$  0.569$&$ -0.5$&$  1.3$&$ -1.4$&$ -0.2$&$ 9.1 $\\
$  1000$&$  0.080$&$  0.138$&$  0.431$&$  5.3$&$  3.7$&$  6.5$&$  0.432$&$ -0.2$&$  1.2$&$ -1.1$&$ -0.1$&$ 8.9 $\\
$  1000$&$  0.130$&$  0.085$&$  0.385$&$  6.1$&$  4.8$&$  7.7$&$  0.384$&$  0.0$&$  1.2$&$ -0.9$&$  0.0$&$ 8.5 $\\
$  1000$&$  0.180$&$  0.061$&$  0.341$&$  6.7$&$  4.3$&$  7.9$&$  0.340$&$  0.2$&$  1.1$&$ -0.7$&$  0.0$&$ 8.1 $\\
$  1000$&$  0.250$&$  0.044$&$  0.244$&$  7.8$&$  5.4$&$  9.5$&$  0.243$&$  0.3$&$  1.0$&$ -0.5$&$  0.0$&$ 7.3 $\\
$  1000$&$  0.400$&$  0.028$&$  0.111$&$ 12.1$&$ 13.4$&$ 18.1$&$  0.111$&$  0.4$&$  1.0$&$ -0.3$&$  0.0$&$ 5.0 $\\
$  1000$&$  0.650$&$  0.017$&$  0.013$&$ 25.0$&$ 15.1$&$ 29.2$&$  0.013$&$  0.5$&$  0.9$&$ -0.2$&$  0.0$&$-0.2 $\\
 \hline                                                                                                    
$  1200$&$  0.020$&$  0.663$&$  0.737$&$  7.2$&$  3.7$&$  8.1$&$  0.774$&$ -4.8$&$  1.8$&$ -2.5$&$ -3.7$&$ 9.6 $\\
$  1200$&$  0.032$&$  0.414$&$  0.645$&$  6.4$&$  3.8$&$  7.4$&$  0.657$&$ -1.9$&$  1.7$&$ -2.3$&$ -1.0$&$ 9.5 $\\
$  1200$&$  0.050$&$  0.265$&$  0.531$&$  6.0$&$  3.5$&$  6.9$&$  0.536$&$ -0.9$&$  1.7$&$ -2.0$&$ -0.3$&$ 9.3 $\\
$  1200$&$  0.080$&$  0.166$&$  0.448$&$  5.9$&$  3.6$&$  6.9$&$  0.450$&$ -0.4$&$  1.6$&$ -1.6$&$ -0.1$&$ 9.1 $\\
$  1200$&$  0.130$&$  0.102$&$  0.391$&$  6.8$&$  3.7$&$  7.8$&$  0.391$&$  0.0$&$  1.5$&$ -1.2$&$  0.0$&$ 8.8 $\\
$  1200$&$  0.180$&$  0.074$&$  0.338$&$  7.5$&$  4.7$&$  8.9$&$  0.337$&$  0.1$&$  1.4$&$ -1.0$&$  0.0$&$ 8.3 $\\
$  1200$&$  0.250$&$  0.053$&$  0.250$&$  8.7$&$  6.7$&$ 10.9$&$  0.249$&$  0.3$&$  1.4$&$ -0.8$&$  0.0$&$ 7.5 $\\
\hline
    \end{tabular}
  \end{center}
\end{table}

\begin{table}[ht]
  \begin{center}
    \small
    \begin{tabular}{|*{3}{c|}|*{4}{c|}|c|*{4}{r|}|r|}
      \hline
      $Q^2$  &$x$  & $y$ & $\tilde{\sigma}_{NC}$ & 
      $\delta_{sta}$ & $\delta_{sys}$ &
      $\delta_{tot}$ &
      $F_2$ & $\Delta_{all}$ & $\Delta_{F_2}$ &$\Delta_{F_3}$ &$\Delta_{F_L}$ 
      & $\delta^{qed}_{NC}$ 
      \\[-1mm]
      \footnotesize $(\rm GeV^2)$ & & & &
      \footnotesize $(\%)$ &
      \footnotesize $(\%)$ &\footnotesize $(\%)$ &
      &
      \multicolumn{1}{c|}{\footnotesize $(\%)$}&
      \multicolumn{1}{c|}{\footnotesize $(\%)$}&
      \multicolumn{1}{c|}{\footnotesize $(\%)$}&
      \multicolumn{1}{c||}{\footnotesize $(\%)$}&
      \multicolumn{1}{c|}{\footnotesize $(\%)$} \\
 \hline
$  1200$&$  0.400$&$  0.033$&$  0.129$&$ 12.1$&$  8.5$&$ 14.8$&$  0.129$&$  0.5$&$  1.2$&$ -0.5$&$  0.0$&$ 5.2 $\\
$  1200$&$  0.650$&$  0.020$&$  0.017$&$ 24.2$&$ 17.5$&$ 29.9$&$  0.017$&$  0.6$&$  1.1$&$ -0.3$&$  0.0$&$-0.1 $\\
 \hline                                                                                                    
$  1500$&$  0.020$&$  0.828$&$  0.789$&$  9.2$&$  5.0$&$ 10.5$&$  0.855$&$ -7.7$&$  2.4$&$ -3.5$&$ -6.1$&$ 9.7 $\\
$  1500$&$  0.032$&$  0.518$&$  0.581$&$  8.1$&$  4.3$&$  9.2$&$  0.601$&$ -3.2$&$  2.4$&$ -3.5$&$ -1.7$&$ 9.9 $\\
$  1500$&$  0.050$&$  0.331$&$  0.486$&$  7.2$&$  3.8$&$  8.1$&$  0.494$&$ -1.6$&$  2.3$&$ -3.0$&$ -0.4$&$ 9.7 $\\
$  1500$&$  0.080$&$  0.207$&$  0.457$&$  6.8$&$  3.7$&$  7.8$&$  0.461$&$ -0.8$&$  2.2$&$ -2.5$&$ -0.1$&$ 9.4 $\\
$  1500$&$  0.130$&$  0.127$&$  0.376$&$  8.0$&$  3.9$&$  8.9$&$  0.377$&$ -0.2$&$  2.1$&$ -1.9$&$  0.0$&$ 9.0 $\\
$  1500$&$  0.180$&$  0.092$&$  0.345$&$  8.6$&$  4.2$&$  9.6$&$  0.345$&$  0.1$&$  2.0$&$ -1.5$&$  0.0$&$ 8.5 $\\
$  1500$&$  0.250$&$  0.066$&$  0.268$&$  9.4$&$  5.8$&$ 11.0$&$  0.267$&$  0.4$&$  1.9$&$ -1.1$&$  0.0$&$ 7.7 $\\
$  1500$&$  0.400$&$  0.041$&$  0.110$&$ 14.6$&$  7.8$&$ 16.6$&$  0.109$&$  0.7$&$  1.7$&$ -0.8$&$  0.0$&$ 5.3 $\\
$  1500$&$  0.650$&$  0.025$&$  0.009$&$ 37.8$&$ 19.6$&$ 42.6$&$  0.009$&$  0.8$&$  1.6$&$ -0.5$&$  0.0$&$ 0.0 $\\
 \hline                                                                                                    
$  2000$&$  0.032$&$  0.690$&$  0.614$&$  9.0$&$  4.1$&$  9.9$&$  0.653$&$ -6.1$&$  3.6$&$ -5.9$&$ -3.2$&$10.3 $\\
$  2000$&$  0.050$&$  0.442$&$  0.541$&$  8.7$&$  4.3$&$  9.7$&$  0.559$&$ -3.2$&$  3.5$&$ -5.3$&$ -0.9$&$10.2 $\\
$  2000$&$  0.080$&$  0.276$&$  0.428$&$  8.3$&$  3.9$&$  9.1$&$  0.436$&$ -1.7$&$  3.3$&$ -4.3$&$ -0.2$&$ 9.8 $\\
$  2000$&$  0.130$&$  0.170$&$  0.340$&$  9.6$&$  4.3$&$ 10.6$&$  0.343$&$ -0.7$&$  3.1$&$ -3.3$&$ -0.1$&$ 9.4 $\\
$  2000$&$  0.180$&$  0.123$&$  0.331$&$ 10.1$&$  4.8$&$ 11.1$&$  0.331$&$ -0.1$&$  3.0$&$ -2.6$&$  0.0$&$ 8.8 $\\
$  2000$&$  0.250$&$  0.088$&$  0.249$&$ 10.7$&$  5.9$&$ 12.2$&$  0.248$&$  0.4$&$  2.8$&$ -2.0$&$  0.0$&$ 8.0 $\\
$  2000$&$  0.400$&$  0.055$&$  0.114$&$ 15.1$&$  8.2$&$ 17.2$&$  0.113$&$  0.9$&$  2.6$&$ -1.3$&$  0.0$&$ 5.5 $\\
$  2000$&$  0.650$&$  0.034$&$  0.011$&$ 37.8$&$ 18.7$&$ 42.2$&$  0.011$&$  1.2$&$  2.3$&$ -0.8$&$  0.0$&$ 0.1 $\\
 \hline                                                                                                    
$  3000$&$  0.050$&$  0.663$&$  0.513$&$  7.3$&$  4.1$&$  8.4$&$  0.558$&$ -8.0$&$  6.0$&$-11.0$&$ -2.2$&$10.9 $\\
$  3000$&$  0.080$&$  0.414$&$  0.458$&$  7.7$&$  4.2$&$  8.7$&$  0.481$&$ -4.8$&$  5.8$&$ -9.3$&$ -0.5$&$10.6 $\\
$  3000$&$  0.130$&$  0.255$&$  0.347$&$  9.1$&$  4.8$&$ 10.2$&$  0.356$&$ -2.3$&$  5.4$&$ -7.0$&$ -0.1$&$ 9.9 $\\
$  3000$&$  0.180$&$  0.184$&$  0.324$&$  9.2$&$  4.1$&$ 10.0$&$  0.327$&$ -1.0$&$  5.1$&$ -5.5$&$  0.0$&$ 9.3 $\\
$  3000$&$  0.250$&$  0.133$&$  0.242$&$  9.9$&$  4.9$&$ 11.1$&$  0.242$&$  0.1$&$  4.8$&$ -4.2$&$  0.0$&$ 8.3 $\\
$  3000$&$  0.400$&$  0.083$&$  0.127$&$ 12.5$&$  9.0$&$ 15.4$&$  0.126$&$  1.3$&$  4.4$&$ -2.7$&$  0.0$&$ 5.8 $\\
$  3000$&$  0.650$&$  0.051$&$  0.012$&$ 30.1$&$ 14.9$&$ 33.6$&$  0.012$&$  2.0$&$  4.0$&$ -1.6$&$  0.0$&$ 0.2 $\\
 \hline                                                                                                    
$  5000$&$  0.080$&$  0.690$&$  0.353$&$ 10.4$&$  4.7$&$ 11.4$&$  0.412$&$-14.3$&$ 10.8$&$-22.3$&$ -1.8$&$11.6 $\\
$  5000$&$  0.130$&$  0.425$&$  0.392$&$ 10.4$&$  5.0$&$ 11.6$&$  0.429$&$ -8.7$&$ 10.1$&$-17.5$&$ -0.4$&$11.0 $\\
$  5000$&$  0.180$&$  0.307$&$  0.223$&$ 13.4$&$  4.5$&$ 14.1$&$  0.235$&$ -5.1$&$  9.6$&$-13.7$&$ -0.1$&$10.1 $\\
$  5000$&$  0.250$&$  0.221$&$  0.217$&$ 13.9$&$  6.6$&$ 15.4$&$  0.222$&$ -2.1$&$  9.0$&$-10.3$&$ -0.1$&$ 8.9 $\\
$  5000$&$  0.400$&$  0.138$&$  0.127$&$ 17.1$&$  8.8$&$ 19.3$&$  0.126$&$  1.1$&$  8.3$&$ -6.5$&$  0.0$&$ 6.1 $\\
$  5000$&$  0.650$&$  0.085$&$  0.012$&$ 37.8$&$ 14.9$&$ 40.6$&$  0.012$&$  3.0$&$  7.5$&$ -4.0$&$  0.0$&$ 0.3 $\\
 \hline                                                                                                    
$  8000$&$  0.130$&$  0.680$&$  0.283$&$ 16.5$&$  4.9$&$ 17.2$&$  0.367$&$-23.0$&$ 16.6$&$-37.1$&$ -1.2$&$12.4 $\\
$  8000$&$  0.180$&$  0.491$&$  0.284$&$ 15.5$&$  6.4$&$ 16.7$&$  0.338$&$-16.0$&$ 15.6$&$-30.1$&$ -0.4$&$11.5 $\\
$  8000$&$  0.250$&$  0.353$&$  0.273$&$ 15.1$&$  7.0$&$ 16.6$&$  0.300$&$ -9.0$&$ 14.7$&$-22.6$&$ -0.1$&$ 9.9 $\\
$  8000$&$  0.400$&$  0.221$&$  0.093$&$ 24.2$&$  9.9$&$ 26.2$&$  0.094$&$ -1.5$&$ 13.5$&$-14.2$&$  0.0$&$ 6.5 $\\
$  8000$&$  0.650$&$  0.136$&$  0.013$&$ 44.7$&$ 19.8$&$ 48.9$&$  0.012$&$  3.1$&$ 12.3$&$ -8.5$&$  0.0$&$ 0.3 $\\
 \hline                                                                                                    
$ 12000$&$  0.180$&$  0.736$&$  0.153$&$ 34.4$&$  4.3$&$ 34.6$&$  0.232$&$-34.3$&$ 22.2$&$-53.9$&$ -1.1$&$13.4 $\\
$ 12000$&$  0.250$&$  0.530$&$  0.127$&$ 32.1$&$  6.2$&$ 32.7$&$  0.165$&$-23.5$&$ 20.9$&$-42.6$&$ -0.4$&$11.6 $\\
$ 12000$&$  0.400$&$  0.331$&$  0.085$&$ 33.3$&$ 11.4$&$ 35.2$&$  0.093$&$ -8.8$&$ 19.2$&$-26.9$&$ -0.1$&$ 7.3 $\\
$ 12000$&$  0.650$&$  0.204$&$  0.015$&$ 57.7$&$ 24.2$&$ 62.6$&$  0.015$&$  0.8$&$ 17.4$&$-15.9$&$  0.0$&$ 0.4 $\\
 \hline                                                                                                    
$ 20000$&$  0.250$&$  0.884$&$  0.090$&$ 61.9$&$  5.5$&$ 62.2$&$  0.188$&$-52.0$&$ 29.2$&$-78.3$&$ -1.3$&$15.1 $\\
$ 20000$&$  0.400$&$  0.552$&$  0.142$&$ 35.7$&$  9.9$&$ 37.0$&$  0.206$&$-31.1$&$ 26.9$&$-56.3$&$ -0.3$&$10.1 $\\
$ 20000$&$  0.650$&$  0.340$&$  0.021$&$ 70.7$&$ 41.6$&$ 82.0$&$  0.023$&$-10.0$&$ 24.4$&$-33.4$&$  0.0$&$ 1.2 $\\
 \hline                                                                                                    
$ 30000$&$  0.400$&$  0.828$&$  0.182$&$ 71.9$&$  9.6$&$ 72.6$&$  0.438$&$-58.5$&$ 32.7$&$-88.7$&$ -0.7$&$15.5 $\\
 \hline
    \end{tabular}
  \end{center}
\end{table}

\newpage

\begin{table}[ht]
  \begin{center}
    \small
    \begin{tabular}{|c|c|c||l|c||c|c|c||c|}
      \hline
      $Q^2$  &$x$  & $y$ & \multicolumn{1}{c|}{$d^2\sigma_{CC}/dxdQ^2$} &  
      ${\phi}_{CC}$& 
      $\delta_{sta}$ & $\delta_{sys}$ &
      $\delta_{tot}$ & $\delta^{qed}_{CC}$ \htab \\
      \footnotesize $(\rm GeV^2)$ & &
      & \multicolumn{1}{c|} {\footnotesize$(\rm pb/GeV^2)$}
      & &\footnotesize $(\%)$ &
      \footnotesize $(\%)$ &\footnotesize $(\%)$&\footnotesize $(\%)$ \\[2mm]
\hline
$   300$&$  0.013$&$  0.255$&\phantom{c} $ 0.637\cdot 10^{0}$&$  1.075$&$ 27.4$&$ 16.0$&$ 31.8$&$\phantom{-}1.2   $ \\
$   300$&$  0.032$&$  0.104$&\phantom{c} $ 0.124\cdot 10^{0}$&$  0.514$&$ 28.1$&$ 10.3$&$ 30.0$&$\phantom{-}1.9   $ \\
$   300$&$  0.080$&$  0.041$&\phantom{c} $ 0.532\cdot 10^{-1}$&$  0.553$&$ 23.8$&$  7.5$&$ 25.5$&$\phantom{-}2.5   $ \\
 \hline
$   500$&$  0.013$&$  0.425$&\phantom{c} $ 0.468\cdot 10^{0}$&$  0.838$&$ 25.1$&$ 15.7$&$ 29.7$&$\phantom{-}0.3   $ \\
$   500$&$  0.032$&$  0.173$&\phantom{c} $ 0.177\cdot 10^{0}$&$  0.781$&$ 17.0$&$  8.7$&$ 19.2$&$\phantom{-}0.4   $ \\
$   500$&$  0.080$&$  0.069$&\phantom{c} $ 0.546\cdot 10^{-1}$&$  0.601$&$ 17.0$&$  6.5$&$ 18.9$&$\phantom{-}1.5   $ \\
$   500$&$  0.130$&$  0.043$&\phantom{c} $ 0.289\cdot 10^{-1}$&$  0.518$&$ 27.8$&$  8.0$&$ 29.4$&$\phantom{-}1.4   $ \\
 \hline
$  1000$&$  0.032$&$  0.345$&\phantom{c} $ 0.124\cdot 10^{0}$&$  0.630$&$ 15.0$&$  8.0$&$ 17.1$&$\phantom{-}0.2   $ \\
$  1000$&$  0.080$&$  0.138$&\phantom{c} $ 0.487\cdot 10^{-1}$&$  0.616$&$ 13.3$&$  6.1$&$ 14.8$&$-0.1   $ \\
$  1000$&$  0.130$&$  0.085$&\phantom{c} $ 0.199\cdot 10^{-1}$&$  0.410$&$ 20.9$&$  6.5$&$ 22.5$&$ 
\phantom{+}0.0   $ \\
$  1000$&$  0.250$&$  0.044$&\phantom{c} $ 0.105\cdot 10^{-1}$&$  0.415$&$ 31.7$&$ 11.7$&$ 34.1$&$-1.0   $ \\
 \hline
$  2000$&$  0.032$&$  0.690$&\phantom{c} $ 0.716\cdot 10^{-1}$&$  0.466$&$ 15.7$&$  8.8$&$ 18.1$&$-2.9$ \\
$  2000$&$  0.080$&$  0.276$&\phantom{c} $ 0.264\cdot 10^{-1}$&$  0.430$&$ 13.5$&$  5.8$&$ 14.8$&$-2.5$ \\
$  2000$&$  0.130$&$  0.170$&\phantom{c} $ 0.949\cdot 10^{-2}$&$  0.251$&$ 20.6$&$  5.7$&$ 21.4$&$\phantom{-}0.1$ \\
$  2000$&$  0.250$&$  0.088$&\phantom{c} $ 0.566\cdot 10^{-2}$&$  0.288$&$ 23.0$&$  7.3$&$ 24.6$&$-0.6$ \\
 \hline
$  3000$&$  0.080$&$  0.414$&\phantom{c} $ 0.156\cdot 10^{-1}$&$  0.317$&$ 15.2$&$  6.7$&$ 16.8$&$-2.6$ \\
$  3000$&$  0.130$&$  0.255$&\phantom{c} $ 0.872\cdot 10^{-2}$&$  0.288$&$ 17.0$&$  5.9$&$ 18.1$&$-4.1$ \\
$  3000$&$  0.250$&$  0.133$&\phantom{c} $ 0.283\cdot 10^{-2}$&$  0.180$&$ 23.6$&$  8.2$&$ 25.1$&$-1.6$ \\
 \hline
$  5000$&$  0.130$&$  0.425$&\phantom{c} $ 0.402\cdot 10^{-2}$&$  0.195$&$ 21.0$&$  7.4$&$ 22.3$&$-4.9$ \\
$  5000$&$  0.250$&$  0.221$&\phantom{c} $ 0.111\cdot 10^{-2}$&$  0.103$&$ 26.8$&$  6.5$&$ 27.6$&$-4.1$ \\
 \hline
$  8000$&$  0.130$&$  0.680$&\phantom{c} $ 0.125\cdot 10^{-2}$&$  0.097$&$ 35.7$&$ 14.3$&$ 38.5$&$-8.2$ \\
$  8000$&$  0.250$&$  0.354$&\phantom{c} $ 0.530\cdot 10^{-3}$&$  0.079$&$ 33.5$&$ 11.2$&$ 35.4$&$-5.3$ \\
$  8000$&$  0.400$&$  0.221$&\phantom{c} $ 0.235\cdot 10^{-3}$&$  0.056$&$ 50.0$&$ 15.6$&$ 52.4$&$-7.5$ \\
 \hline
$ 15000$&$  0.250$&$  0.663$&\phantom{c} $ 0.774\cdot 10^{-4}$&$  0.025$&$ 71.2$&$ 18.1$&$ 73.5$&$-10.1$\\
$ 15000$&$  0.400$&$  0.414$&\phantom{c} $ 0.114\cdot 10^{-3}$&$  0.059$&$ 40.9$&$ 17.4$&$ 44.5$&$-9.1$ \\
 \hline
    \end{tabular}
  \end{center}
\caption{\small \sl CC double differential cross-section
$d^2\sigma_{CC}/dx dQ^2$ and structure function term ${\phi}_{CC}$
(computed assuming $M_W=80.4$ \Gev )
with statistical error ($\delta_{sta}$), systematic error
($\delta_{sys}$), and total error ($\delta_{tot}$).  
The correction $\delta^{qed}_{CC}$ due to QED radiation effects, 
as defined in eq.~\ref{init}, is also given. 
The correction for weak 
radiative effects, $(1+\delta_{CC}^{weak})$, is given by
the ratio of $d^2\sigma_{CC}/dx dQ^2$ and ${\phi}_{CC}$, multiplied by
the factors $G_{F}^2/(2\pi x) M_W^4/(M_W^2+Q^2)^2$, see
eqs.~\ref{Rcc},\ref{Pcc}.
The normalization uncertainty, which
is not included in the systematic error, is 1.5\%.}
\label{tabf2cc}
\end{table}

\begin{table}[ht]
  \begin{center}
    \small
    \begin{tabular}{|c|l|l|*{4}{c|}|c|}
     \hline
      $Q^2$  & \multicolumn{1}{c|}{$d\sigma_{NC}/dQ^2$}&
       \multicolumn{1}{c|}{$ d\sigma_{NC}/dQ^2$} & 
      $\delta_{sta}$ & $\delta_{unc}$ &
      $\delta_{cor}$ &$\delta_{tot}$ & $\delta^{qed}_{NC}$ \\
      \footnotesize $(\rm GeV^2)$ & \multicolumn{1}{c|}{\footnotesize $(\rm pb/GeV^2)$}&
       \multicolumn{1}{c|}{\footnotesize $(\rm pb/GeV^2)$} 
      &\footnotesize $(\%)$ &\footnotesize $(\%)$ &\footnotesize $(\%)$ 
      &\footnotesize $(\%)$ &\footnotesize $(\%)$ \\[-1mm]
      &  \multicolumn{1}{c|}{\footnotesize $y < 0.9 $ }& 
       \multicolumn{1}{c|}{\footnotesize $y<0.9$} & &  &  & &  \\[-1mm]
       &  \multicolumn{1}{c|}{\footnotesize $ E_e^{\prime} > 11 
      {\rm GeV} $ } 
       & &  &  & & & \\
\hline
$   200$&$ 0.163\cdot 10^{2}$&$ 0.176\cdot 10^{2}$&$  0.9$&$  3.0$&$  1.0$&$  3.3$&$  7.1$\\
$   250$&$ 0.965\cdot 10^{1}$&$ 0.104\cdot 10^{2}$&$  0.8$&$  3.0$&$  0.8$&$  3.2$&$  7.9$\\
$   300$&$ 0.625\cdot 10^{1}$&$ 0.670\cdot 10^{1}$&$  0.9$&$  3.2$&$  1.0$&$  3.4$&$  7.3$\\
$   400$&$ 0.313\cdot 10^{1}$&$ 0.332\cdot 10^{1}$&$  1.1$&$  2.9$&$  1.0$&$  3.2$&$  8.8$\\
$   500$&$ 0.185\cdot 10^{1}$&$ 0.194\cdot 10^{1}$&$  1.2$&$  2.8$&$  1.0$&$  3.2$&$  8.5$\\
$   650$&$ 0.995\cdot 10^{0}$&$ 0.103\cdot 10^{1}$&$  1.5$&$  2.9$&$  1.2$&$  3.5$&$  9.6$\\
$   800$&$ 0.608\cdot 10^{0}$&$ 0.616\cdot 10^{0}$&$  1.7$&$  2.9$&$  1.1$&$  3.6$&$  8.9$\\
$  1000$&$ 0.347\cdot 10^{0}$&$ 0.347\cdot 10^{0}$&$  2.0$&$  2.8$&$  0.8$&$  3.6$&$ 10.7$\\
$  1200$&$ 0.211\cdot 10^{0}$&$ 0.211\cdot 10^{0}$&$  2.4$&$  2.8$&$  0.9$&$  3.8$&$ 10.5$\\
$  1500$&$ 0.112\cdot 10^{0}$&$ 0.112\cdot 10^{0}$&$  3.0$&$  2.8$&$  1.0$&$  4.2$&$  9.2$\\
$  2000$&$ 0.541\cdot 10^{-1}$&$ 0.541\cdot 10^{-1}$&$  3.5$&$  3.0$&$  1.1$&$  4.8$&$ 9.4 $\\
$  3000$&$ 0.188\cdot 10^{-1}$&$ 0.188\cdot 10^{-1}$&$  3.4$&$  2.8$&$  0.9$&$  4.5$&$ 9.1 $\\ 
$  5000$&$ 0.389\cdot 10^{-2}$&$ 0.389\cdot 10^{-2}$&$  5.0$&$  3.5$&$  0.9$&$  6.2$&$ 9.6 $\\
$  8000$&$ 0.987\cdot 10^{-3}$&$ 0.987\cdot 10^{-3}$&$  7.9$&$  4.9$&$  1.5$&$  9.4$&$11.0 $\\ 
$ 12000$&$ 0.158\cdot 10^{-3}$&$ 0.158\cdot 10^{-3}$&$ 18.3$&$  7.9$&$  1.7$&$ 20.0$&$11.3 $\\
$ 20000$&$ 0.386\cdot 10^{-4}$&$ 0.386\cdot 10^{-4}$&$ 28.1$&$ 12.7$&$  2.4$&$ 30.9$&$17.3 $\\ 
$ 30000$&$ 0.656\cdot 10^{-5}$&$ 0.656\cdot 10^{-5}$&$ 71.2$&$ 18.1$&$  3.3$&$ 73.6$&$25.9 $\\
\hline
    \end{tabular}
  \end{center}
\caption{\small \sl NC  cross-section $d\sigma_{NC}/ dQ^2$
measured for $y<0.9$ and $E^{\prime}_e > 11 \ {\rm{GeV}}$ 
and after correction according to SM expectations 
for the influence of the $E_e^{\prime}$ cut. 
 The   statistical error ($\delta_{sta}$),  the
correlated systematic error ($\delta_{cor}$), 
the uncorrelated  systematic error ($\delta_{unc}$)
and  the total error ($\delta_{tot}$) 
are given. 
The correction $\delta^{qed}_{NC}$ due to QED radiation effects, 
as defined in eq.~\ref{init}, is also given. 
The normalization 
uncertainty, which is not included in the systematic error, is $1.5\%$. }
\label{tabsnc}
\end{table}

\begin{table}[hb]
  \begin{center}
    \small
    \begin{tabular}{|c|c|*{5}{c|}|c|}
      \hline
      $Q^2$  &$ d\sigma_{CC}/dQ^2$  &
      $ d\sigma_{CC}/dQ^2$ &
      $\delta_{sta}$ & $\delta_{unc}$ &
      $\delta_{cor}$ &$\delta_{tot}$ & $\delta^{qed}_{CC}$ \htab \\
      \footnotesize $( \rm GeV^2)$ & \footnotesize  $(\rm pb/GeV^2)$    
      &\footnotesize $(\rm pb/GeV^2)$  
      &\footnotesize $(\%)$ &\footnotesize  $(\%)$ &\footnotesize  $(\%)$ 
      &\footnotesize $(\%)$ &\footnotesize $(\%)$ \\[-1mm]
       & \footnotesize $0.03 < y < 0.85$ 
       & \footnotesize $y<0.9$  &  &  &  &  & \\[-1mm]
       & \footnotesize $p_T^{\nu} > 12 \ {\rm{ GeV}} $ 
       &  &  &  &  &   & \\ 
\hline
$   300$&$ 0.164\cdot 10^{-1}$&$ 0.226\cdot 10^{-1}$&$ 14.5$&$  9.3$&$  7.3$&$ 18.8$&$  \phantom{-}3.5$\\
$   500$&$ 0.165\cdot 10^{-1}$&$ 0.193\cdot 10^{-1}$&$ 10.0$&$  7.7$&$  5.8$&$ 14.0$&$ -0.1$\\
$  1000$&$ 0.113\cdot 10^{-1}$&$ 0.118\cdot 10^{-1}$&$  8.2$&$  6.6$&$  3.7$&$ 11.4$&$ -2.3$\\
$  2000$&$ 0.472\cdot 10^{-2}$&$ 0.484\cdot 10^{-2}$&$  8.4$&$  6.2$&$  2.4$&$ 10.9$&$ -3.4$\\
$  3000$&$ 0.247\cdot 10^{-2}$&$ 0.255\cdot 10^{-2}$&$  9.6$&$  6.3$&$  2.2$&$ 11.8$&$ -6.6$\\
$  5000$&$ 0.794\cdot 10^{-3}$&$ 0.823\cdot 10^{-3}$&$ 13.1$&$  7.3$&$  2.5$&$ 15.3$&$ -9.0$\\
$  8000$&$ 0.220\cdot 10^{-3}$&$ 0.230\cdot 10^{-3}$&$ 21.4$&$ 10.8$&$  6.2$&$ 24.9$&$-11.6$\\
$ 15000$&$ 0.382\cdot 10^{-4}$&$ 0.405\cdot 10^{-4}$&$ 33.4$&$ 15.3$&$  7.9$&$ 37.7$&$-17.9$\\
\hline
    \end{tabular}
  \end{center}
\caption{\small \sl
 CC  cross-section $d\sigma_{CC}/ dQ^2$
measured for $0.03 < y<0.85 $ and  $p_T^{\nu} > 12 \ {\rm{GeV}} $,
and after correction according to SM expectations to $y < 0.9$ and 
for the influence of the $p_T^{\nu}$ cut. 
The  statistical error ($\delta_{sta}$),  the
correlated systematic error  ($\delta_{cor}$),  
the uncorrelated  systematic error ($\delta_{unc}$)
and  the total error ($\delta_{tot}$) 
are given.
The correction due to QED radiation effects, as defined in  eq.~\ref{init},
is also given.
The normalization 
uncertainty, which is not included in the systematic error, is $1.5\%$. }
\label{tabscc}
\end{table}                                                

%% file: longtab.tex
\begin{table}[ht]
  \begin{center}
    \tiny
    \begin{tabular}{|*{2}{c|}|*{3}{c|}|*{4}{c|}|*{1}{c|}*{5}{r|}}
      \hline
      $Q^2$  & $x$ & $\tilde{\sigma}_{NC}$ & 
      $\delta_{tot}$ & $\delta_{sta}$ & $\delta_{unc}$ &
      $\delta^{E}_{unc}$ & $\delta^{\theta}_{unc}$ & $\delta^{h}_{unc}$ &
      $\delta_{cor}$ & $\delta^{E^{+}}_{cor}$ & $\delta^{\theta^{+}}_{cor}$ &
      $\delta^{h^{+}}_{cor}$ & $\delta^{N^{+}}_{cor}$ & $\delta^{B^{+}}_{cor}$ \vtab \\
      \tiny $(\rm GeV^2)$ & & &
      \tiny $(\%)$ &\tiny $(\%)$ &
      \tiny $(\%)$ &\tiny $(\%)$ &
      \tiny $(\%)$ &\tiny $(\%)$ &
      \tiny $(\%)$ &\tiny $(\%)$ &
      \tiny $(\%)$ &\tiny $(\%)$ &
      \tiny $(\%)$ &\tiny  $(\%)$ \\[1mm]
      \hline
$  150$&$  .0032$&$  1.240$&$  5.5$&$  1.8$&$  4.9$&$  1.1$&$  0.8$&$  0.7$&$  1.7$&$ -1.0$&$ -0.4$&$  0.4$&$  1.1$&$ -0.4$\\
$  150$&$  0.005$&$  1.100$&$  3.8$&$  1.8$&$  3.3$&$  0.2$&$  0.7$&$  0.1$&$  0.6$&$ -0.2$&$  0.4$&$ -0.1$&$  0.3$&$ - 0.1$\\
$  150$&$  0.008$&$  0.920$&$  9.3$&$  2.9$&$  7.9$&$  4.0$&$  5.6$&$  0.2$&$  4.1$&$ -2.9$&$  2.8$&$ -0.1$&$ -0.5$&$  0.0$\\
 \hline
$  200$&$  0.005$&$  1.102$&$  5.3$&$  1.8$&$  4.7$&$  0.3$&$  0.6$&$  1.1$&$  1.7$&$ -0.4$&$ -0.3$&$  0.7$&$  1.4$&$  -0.3$\\
$  200$&$  0.008$&$  0.915$&$  4.0$&$  1.9$&$  3.3$&$  0.9$&$  0.9$&$  0.5$&$  1.2$&$  0.6$&$ -0.5$&$  0.2$&$  0.9$&$  0.0$\\
$  200$&$  0.013$&$  0.765$&$  4.3$&$  2.2$&$  3.5$&$  0.5$&$  1.7$&$  0.3$&$  1.0$&$ -0.4$&$  0.8$&$  0.0$&$  0.3$&$  0.0$\\
$  200$&$  0.020$&$  0.696$&$  5.5$&$  2.6$&$  4.6$&$  2.2$&$  2.3$&$  0.6$&$  1.7$&$ -1.0$&$  1.1$&$ -0.3$&$ -0.7$&$  0.0$\\
$  200$&$  0.032$&$  0.601$&$  8.1$&$  3.2$&$  6.6$&$  3.5$&$  4.3$&$  0.7$&$  3.6$&$ -2.3$&$  2.1$&$ -0.4$&$ -1.7$&$  0.0$\\
$  200$&$  0.050$&$  0.516$&$  9.0$&$  3.7$&$  7.3$&$  4.6$&$  4.3$&$  0.2$&$  3.7$&$ -2.9$&$  2.1$&$ -0.3$&$ -0.5$&$  0.0$\\
$  200$&$  0.080$&$  0.439$&$  9.9$&$  4.2$&$  7.7$&$  4.0$&$  5.2$&$  1.3$&$  4.5$&$ -3.2$&$  2.6$&$ -0.8$&$ -1.8$&$  0.0$\\
 \hline
$  250$&$  0.005$&$  1.113$&$  5.6$&$  2.3$&$  4.9$&$  0.3$&$  1.3$&$  0.7$&$  1.3$&$ -0.4$&$ -0.7$&$  0.3$&$  0.8$&$  -0.5$\\
$  250$&$  0.008$&$  1.018$&$  4.2$&$  2.0$&$  3.4$&$  0.2$&$  0.8$&$  1.0$&$  1.6$&$  0.2$&$ -0.4$&$  0.6$&$  1.4$&$  -0.1$\\
$  250$&$  0.013$&$  0.807$&$  4.4$&$  2.1$&$  3.5$&$  0.8$&$  1.3$&$  0.6$&$  1.7$&$  0.3$&$ -0.7$&$  0.3$&$  1.5$&$  0.0$\\
$  250$&$  0.020$&$  0.721$&$  4.1$&$  2.1$&$  3.5$&$  1.2$&$  0.7$&$  0.3$&$  0.9$&$  0.6$&$ -0.4$&$  0.2$&$  0.6$&$  0.0$\\
$  250$&$  0.032$&$  0.606$&$  4.3$&$  2.2$&$  3.5$&$  1.4$&$  0.7$&$  0.2$&$  0.9$&$  0.7$&$ -0.3$&$ -0.1$&$  0.3$&$  0.0$\\
$  250$&$  0.050$&$  0.529$&$  4.2$&$  2.4$&$  3.4$&$  0.8$&$  0.1$&$  0.1$&$  0.6$&$  0.2$&$ -0.1$&$  0.1$&$  0.6$&$  0.0$\\
$  250$&$  0.080$&$  0.430$&$  4.5$&$  2.7$&$  3.4$&$  0.3$&$  0.6$&$  0.4$&$  1.1$&$  0.5$&$ -0.3$&$ -0.5$&$  0.7$&$  0.0$\\
$  250$&$  0.130$&$  0.334$&$  5.5$&$  3.4$&$  4.1$&$  0.9$&$  0.3$&$  1.6$&$  1.5$&$  0.7$&$ -0.2$&$ -1.0$&$ -0.8$&$  0.0$\\
$  250$&$  0.250$&$  0.240$&$  8.1$&$  3.3$&$  4.8$&$  2.1$&$  0.9$&$  2.2$&$  5.7$&$  0.7$&$  0.5$&$ -1.3$&$ -5.4$&$  0.0$\\
$  250$&$  0.400$&$  0.122$&$ 13.4$&$  5.9$&$  5.6$&$  2.6$&$  1.2$&$  2.3$&$ 10.7$&$  1.8$&$ -0.6$&$ -1.3$&$-10.5$&$  0.0$\\
 \hline
$  300$&$  0.005$&$  1.139$&$  6.5$&$  3.4$&$  5.2$&$  1.8$&$  0.3$&$  0.2$&$  2.0$&$ -1.8$&$ -0.2$&$ -0.2$&$  0.0$&$  -0.6$\\
$  300$&$  0.008$&$  0.989$&$  5.7$&$  2.4$&$  4.9$&$  0.1$&$  1.1$&$  1.1$&$  1.5$&$  0.0$&$ -0.6$&$  0.6$&$  1.2$&$  -0.2$\\
$  300$&$  0.013$&$  0.846$&$  4.5$&$  2.4$&$  3.4$&$  0.6$&$  1.0$&$  0.7$&$  1.6$&$  0.5$&$ -0.5$&$  0.5$&$  1.4$&$  0.0$\\
$  300$&$  0.020$&$  0.740$&$  4.6$&$  2.4$&$  3.6$&$  1.1$&$  1.3$&$  0.6$&$  1.4$&$  0.8$&$ -0.7$&$  0.3$&$  1.0$&$  0.0$\\
$  300$&$  0.032$&$  0.629$&$  4.4$&$  2.4$&$  3.5$&$  1.3$&$  0.7$&$  0.4$&$  1.1$&$  0.7$&$ -0.4$&$  0.4$&$  0.7$&$  0.0$\\
$  300$&$  0.050$&$  0.499$&$  4.5$&$  2.6$&$  3.5$&$  1.2$&$  0.8$&$  0.3$&$  0.8$&$  0.4$&$ -0.4$&$  0.3$&$  0.6$&$  0.0$\\
$  300$&$  0.080$&$  0.456$&$  4.8$&$  2.7$&$  3.8$&$  1.6$&$  0.9$&$  0.4$&$  1.1$&$  0.6$&$ -0.5$&$ -0.2$&$  0.8$&$  0.0$\\
$  300$&$  0.130$&$  0.346$&$  6.8$&$  3.4$&$  5.3$&$  3.3$&$  1.5$&$  1.5$&$  2.5$&$  2.2$&$ -0.8$&$ -0.8$&$ -0.4$&$  0.0$\\
$  300$&$  0.250$&$  0.250$&$  8.7$&$  3.1$&$  6.2$&$  4.4$&$  1.9$&$  2.1$&$  5.2$&$  2.6$&$ -1.0$&$ -1.1$&$ -4.3$&$  0.0$\\
$  300$&$  0.400$&$  0.140$&$ 15.6$&$  5.7$&$  8.8$&$  6.8$&$  2.0$&$  3.6$&$ 11.6$&$  4.5$&$ -1.0$&$ -2.4$&$-10.3$&$  0.0$\\
 \hline
$  400$&$  0.008$&$  0.976$&$  6.0$&$  3.1$&$  4.9$&$  0.5$&$  0.8$&$  1.0$&$  1.4$&$ -0.5$&$ -0.4$&$  0.7$&$  1.0$&$ -0.5$\\
$  400$&$  0.013$&$  0.841$&$  4.8$&$  2.8$&$  3.6$&$  0.5$&$  1.1$&$  1.0$&$  1.6$&$  0.4$&$ -0.6$&$  0.7$&$  1.3$&$ -0.1$\\
$  400$&$  0.020$&$  0.739$&$  4.7$&$  2.8$&$  3.4$&$  0.2$&$  0.7$&$  0.9$&$  1.5$&$ -0.1$&$ -0.3$&$  0.5$&$  1.3$&$  0.0$\\
$  400$&$  0.032$&$  0.619$&$  4.6$&$  2.8$&$  3.4$&$  0.8$&$  0.8$&$  0.3$&$  0.9$&$  0.7$&$ -0.4$&$  0.1$&$  0.4$&$  0.0$\\
$  400$&$  0.050$&$  0.513$&$  4.8$&$  3.0$&$  3.6$&$  1.0$&$  0.9$&$  0.2$&$  1.2$&$  1.0$&$ -0.5$&$ -0.1$&$  0.3$&$  0.0$\\
$  400$&$  0.080$&$  0.455$&$  5.1$&$  3.1$&$  3.6$&$  0.5$&$  1.1$&$  0.5$&$  1.7$&$  0.3$&$ -0.5$&$  0.3$&$  1.6$&$  0.0$\\
$  400$&$  0.130$&$  0.373$&$  5.9$&$  3.8$&$  4.2$&$  1.5$&$  0.9$&$  1.0$&$  1.7$&$  1.4$&$ -0.5$&$ -0.9$&$ -0.2$&$  0.0$\\
$  400$&$  0.250$&$  0.241$&$  7.4$&$  3.5$&$  4.9$&$  2.6$&$  0.9$&$  2.2$&$  4.3$&$  2.4$&$ -0.5$&$ -1.1$&$ -3.4$&$  0.0$\\
$  400$&$  0.400$&$  0.155$&$ 13.2$&$  6.2$&$  5.8$&$  3.2$&$  0.5$&$  2.8$&$ 10.0$&$  2.9$&$ -0.2$&$ -1.8$&$ -9.4$&$  0.0$\\
 \hline
$  500$&$  0.008$&$  1.026$&$  6.6$&$  4.2$&$  5.0$&$  0.5$&$  0.6$&$  0.1$&$  0.9$&$ -0.5$&$ -0.3$&$ -0.1$&$  0.3$&$  -0.7$\\
$  500$&$  0.013$&$  0.906$&$  6.2$&$  3.3$&$  5.0$&$  0.3$&$  0.8$&$  1.2$&$  1.5$&$  0.4$&$ -0.4$&$  0.7$&$  1.2$&$  -0.3$\\
$  500$&$  0.020$&$  0.792$&$  5.2$&$  3.3$&$  3.7$&$  0.8$&$  1.2$&$  0.7$&$  1.3$&$  0.7$&$ -0.6$&$  0.3$&$  0.9$&$  0.0$\\
$  500$&$  0.032$&$  0.654$&$  5.2$&$  3.3$&$  3.6$&$  0.4$&$  0.4$&$  1.3$&$  1.7$&$ -0.4$&$ -0.2$&$  0.6$&$  1.5$&$  0.0$\\
$  500$&$  0.050$&$  0.508$&$  5.4$&$  3.5$&$  3.8$&$  1.3$&$  1.0$&$  0.4$&$  1.6$&$  1.4$&$ -0.5$&$  0.2$&$  0.4$&$  0.0$\\
$  500$&$  0.080$&$  0.445$&$  5.2$&$  3.6$&$  3.7$&$  0.2$&$  1.0$&$  0.3$&$  0.8$&$  0.3$&$ -0.5$&$  0.2$&$  0.4$&$  0.0$\\
$  500$&$  0.130$&$  0.368$&$  6.1$&$  4.3$&$  4.0$&$  1.1$&$  0.4$&$  0.7$&$  1.5$&$  1.0$&$  0.2$&$ -0.7$&$  0.8$&$  0.0$\\
$  500$&$  0.180$&$  0.287$&$  7.3$&$  4.9$&$  4.9$&$  1.8$&$  1.3$&$  1.9$&$  2.3$&$  1.9$&$ -0.7$&$ -1.2$&$  0.3$&$  0.0$\\
$  500$&$  0.250$&$  0.220$&$ 10.4$&$  5.9$&$  5.3$&$  2.5$&$  1.5$&$  1.9$&$  6.6$&$  2.5$&$ -0.8$&$ -0.9$&$ -6.0$&$  0.0$\\
$  500$&$  0.400$&$  0.143$&$ 17.5$&$  8.6$&$  8.6$&$  5.0$&$  1.5$&$  5.1$&$ 12.6$&$  5.2$&$ -0.8$&$ -2.8$&$-11.1$&$  0.0$\\
 \hline
$  650$&$  0.013$&$  0.903$&$  5.9$&$  4.0$&$  3.8$&$  1.0$&$  0.5$&$  1.4$&$  2.0$&$ -1.1$&$  0.2$&$  0.9$&$  1.3$&$  -0.5$\\
$  650$&$  0.020$&$  0.718$&$  5.7$&$  4.1$&$  3.7$&$  0.7$&$  0.6$&$  1.0$&$  1.4$&$  0.7$&$ -0.3$&$  0.4$&$  1.1$&$  -0.2$\\
$  650$&$  0.032$&$  0.633$&$  5.7$&$  4.0$&$  3.7$&$  0.7$&$  0.9$&$  0.7$&$  1.6$&$  0.8$&$ -0.5$&$  0.4$&$  1.2$&$  0.0$\\
$  650$&$  0.050$&$  0.521$&$  5.7$&$  4.1$&$  3.8$&$  0.9$&$  0.9$&$  0.5$&$  1.1$&$  0.8$&$ -0.5$&$  0.4$&$  0.4$&$  0.0$\\
$  650$&$  0.080$&$  0.436$&$  5.7$&$  4.0$&$  3.8$&$  1.1$&$  0.5$&$  0.2$&$  1.2$&$  1.1$&$ -0.3$&$  0.1$&$  0.6$&$  0.0$\\
$  650$&$  0.130$&$  0.413$&$  6.6$&$  4.6$&$  4.4$&$  1.1$&$  1.7$&$  0.1$&$  1.6$&$  1.1$&$ -0.9$&$ -0.2$&$  0.8$&$  0.0$\\
$  650$&$  0.180$&$  0.309$&$  7.9$&$  5.3$&$  5.2$&$  2.3$&$  1.3$&$  1.7$&$  2.7$&$  2.5$&$ -0.6$&$ -0.8$&$ -0.2$&$  0.0$\\
$  650$&$  0.250$&$  0.246$&$ 10.6$&$  6.2$&$  6.2$&$  3.5$&$  0.8$&$  2.9$&$  6.1$&$  3.6$&$ -0.4$&$ -1.6$&$ -4.7$&$  0.0$\\
$  650$&$  0.400$&$  0.125$&$ 15.2$&$  9.9$&$  7.3$&$  3.9$&$  0.7$&$  3.9$&$  9.0$&$  3.9$&$ -0.3$&$ -1.9$&$ -7.8$&$  0.0$\\
$  650$&$  0.650$&$  0.021$&$ 21.3$&$ 14.3$&$  7.8$&$  2.8$&$  2.8$&$  1.9$&$ 13.7$&$  2.9$&$ -1.4$&$ -0.9$&$-13.2$&$  0.0$\\
 \hline
$  800$&$  0.013$&$  1.000$&$  6.8$&$  5.0$&$  4.4$&$  0.4$&$  2.2$&$  0.4$&$  1.7$&$  0.8$&$ -1.1$&$  0.1$&$  0.7$&$  -0.6$\\
$  800$&$  0.020$&$  0.796$&$  6.3$&$  4.6$&$  4.0$&$  0.6$&$  0.9$&$  1.4$&$  1.6$&$  0.6$&$  0.4$&$  0.9$&$  1.1$&$  -0.3$\\
$  800$&$  0.032$&$  0.709$&$  6.0$&$  4.5$&$  3.8$&$  0.8$&$  0.8$&$  0.6$&$  1.4$&$  0.8$&$ -0.4$&$  0.6$&$  0.9$&$  0.0$\\
$  800$&$  0.050$&$  0.540$&$  6.0$&$  4.6$&$  3.8$&$  0.4$&$  0.9$&$  0.8$&$  1.0$&$  0.5$&$ -0.4$&$  0.3$&$  0.6$&$  0.0$\\
$  800$&$  0.080$&$  0.474$&$  6.2$&$  4.6$&$  3.9$&$  0.5$&$  1.1$&$  0.4$&$  1.4$&$  0.4$&$ -0.6$&$ -0.3$&$  1.2$&$  0.0$\\
$  800$&$  0.130$&$  0.370$&$  7.2$&$  5.4$&$  4.4$&$  1.5$&$  0.8$&$  0.7$&$  1.8$&$  1.7$&$ -0.4$&$ -0.4$&$ -0.2$&$  0.0$\\
$  800$&$  0.180$&$  0.333$&$  7.8$&$  6.0$&$  4.7$&$  1.4$&$  1.1$&$  1.2$&$  1.6$&$  1.3$&$ -0.5$&$ -0.5$&$ -0.6$&$  0.0$\\
$  800$&$  0.250$&$  0.208$&$  9.4$&$  7.5$&$  4.9$&$  2.1$&$  1.4$&$  1.1$&$  2.9$&$  1.9$&$ -0.7$&$ -0.6$&$ -2.0$&$  0.0$\\
$  800$&$  0.400$&$  0.150$&$ 14.2$&$  9.6$&$  6.9$&$  3.0$&$  0.8$&$  4.2$&$  8.0$&$  3.3$&$  0.4$&$ -2.6$&$ -6.8$&$  0.0$\\
$  800$&$  0.650$&$  0.018$&$ 26.9$&$ 19.6$&$  9.8$&$  4.8$&$  0.9$&$  4.8$&$ 15.5$&$  4.8$&$ -0.5$&$ -2.9$&$-14.5$&$  0.0$\\
 \hline

\end{tabular}
\caption{\small \sl NC reduced cross-section $\tilde{\sigma}_{NC}(x,Q^2)$
with total error ($\delta_{tot}$), statistical error ($\delta_{sta}$), 
uncorrelated systematic error ($\delta_{unc}$), and its contributions from
the positron energy error ($\delta^{E}_{unc}$),
the polar positron angle error($\delta^{\theta}_{unc}$) and
the hadronic energy error ($\delta^{h}_{unc}$). The effect of the other uncorrelated 
errors, as described in section~\ref{systemaa}, is included in $\delta_{unc}$.
Also given are the
correlated systematic error  ($\delta_{cor}$), and its contributions from 
a positive variation of one standard deviation
of the positron energy error ($\delta^{E^{+}}_{cor}$),
of the polar positron angle error ($\delta^{\theta^{+}}_{cor}$),
of the hadronic energy error ($\delta^{h^{+}}_{cor}$), 
of the error due to the noise subtraction ($\delta^{N^{+}}_{cor}$) and
of the error due to the background subtraction ($\delta^{B^{+}}_{cor}$).
The normalization uncertainty, 
which is not included in the systematic error, is 1.5\%. 
 The table continues on the next page.
\label{longtabnc}
}
\end{center}
\end{table}

\begin{table}[ht]
  \begin{center}
    \tiny
    \begin{tabular}{|*{2}{c|}|*{3}{c|}|*{4}{c|}|*{1}{c|}*{5}{r|}}
      \hline
      $Q^2$  & $x$ & $\tilde{\sigma}_{NC}$ & 
      $\delta_{tot}$ & $\delta_{sta}$ & $\delta_{unc}$ &
      $\delta^{E}_{unc}$ & $\delta^{\theta}_{unc}$ & $\delta^{h}_{unc}$ &
      $\delta_{cor}$ & $\delta^{E^{+}}_{cor}$ & $\delta^{\theta^{+}}_{cor}$ &
      $\delta^{h^{+}}_{cor}$ & $\delta^{N^{+}}_{cor}$ & $\delta^{B^{+}}_{cor}$ \vtab \\
      \tiny $(\rm GeV^2)$ & & & 
      \tiny $(\%)$ &\tiny $(\%)$ &
      \tiny $(\%)$ &\tiny $(\%)$ &
      \tiny $(\%)$ &\tiny $(\%)$ &
      \tiny $(\%)$ &\tiny $(\%)$ &
      \tiny $(\%)$ &\tiny $(\%)$ &
      \tiny $(\%)$ &\tiny  $(\%)$ \\[1mm]
 \hline
$ 1000$&$  0.020$&$  0.754$&$  6.6$&$  5.4$&$  3.7$&$  0.3$&$  1.1$&$  1.1$&$  1.2$&$ -0.2$&$ -0.6$&$  0.7$&$  0.6$&$  -0.5$\\
$ 1000$&$  0.032$&$  0.639$&$  6.9$&$  5.6$&$  3.6$&$  0.6$&$  0.5$&$  1.2$&$  2.0$&$  0.6$&$ -0.3$&$  0.7$&$  1.7$&$  -0.1$\\
$ 1000$&$  0.050$&$  0.566$&$  6.4$&$  5.1$&$  3.6$&$  0.7$&$  1.1$&$  0.9$&$  1.0$&$ -0.6$&$ -0.6$&$  0.3$&$  0.4$&$  0.0$\\
$ 1000$&$  0.080$&$  0.431$&$  6.5$&$  5.3$&$  3.5$&$  0.2$&$  0.7$&$  0.6$&$  1.3$&$  0.2$&$ -0.4$&$  0.7$&$  1.0$&$  0.0$\\
$ 1000$&$  0.130$&$  0.385$&$  7.7$&$  6.1$&$  4.2$&$  1.7$&$  1.0$&$  1.0$&$  2.1$&$  1.8$&$ -0.5$&$ -0.8$&$ -0.8$&$  0.0$\\
$ 1000$&$  0.180$&$  0.341$&$  7.9$&$  6.7$&$  4.0$&$  0.9$&$  0.4$&$  0.8$&$  1.5$&$  0.9$&$ -0.2$&$ -0.6$&$  1.1$&$  0.0$\\
$ 1000$&$  0.250$&$  0.244$&$  9.5$&$  7.8$&$  4.6$&$  1.5$&$  0.9$&$  1.7$&$  2.8$&$  1.8$&$ -0.5$&$ -1.1$&$ -1.8$&$  0.0$\\
$ 1000$&$  0.400$&$  0.111$&$ 18.1$&$ 12.1$&$  9.1$&$  5.4$&$  2.3$&$  5.3$&$  9.9$&$  5.3$&$ -1.2$&$ -2.3$&$ -7.9$&$  0.0$\\
$ 1000$&$  0.650$&$  0.013$&$ 29.2$&$ 25.0$&$  9.5$&$  4.1$&$  2.6$&$  4.6$&$ 11.7$&$  4.2$&$ -1.3$&$ -3.3$&$-10.3$&$  0.0$\\
 \hline
$ 1200$&$  0.020$&$  0.737$&$  8.1$&$  7.2$&$  3.5$&$  0.6$&$  0.6$&$  1.0$&$  1.1$&$ -0.4$&$ -0.3$&$  0.5$&$  0.6$&$  -0.6$\\
$ 1200$&$  0.032$&$  0.645$&$  7.4$&$  6.4$&$  3.5$&$  0.4$&$  0.3$&$  1.4$&$  1.5$&$ -0.4$&$  0.1$&$  0.8$&$  1.1$&$  -0.2$\\
$ 1200$&$  0.050$&$  0.531$&$  6.9$&$  6.0$&$  3.4$&$  0.2$&$  0.7$&$  1.0$&$  0.8$&$ -0.1$&$ -0.3$&$  0.3$&$  0.7$&$  0.0$\\
$ 1200$&$  0.080$&$  0.448$&$  6.9$&$  5.9$&$  3.4$&$  0.8$&$  0.5$&$  0.6$&$  1.3$&$  0.8$&$ -0.3$&$  0.6$&$  0.7$&$  0.0$\\
$ 1200$&$  0.130$&$  0.391$&$  7.8$&$  6.8$&$  3.5$&$  0.8$&$  0.5$&$  0.1$&$  1.1$&$  0.7$&$ -0.3$&$  0.3$&$  0.8$&$  0.0$\\
$ 1200$&$  0.180$&$  0.338$&$  8.9$&$  7.5$&$  4.3$&$  1.9$&$  1.3$&$  0.9$&$  2.1$&$  1.8$&$ -0.6$&$ -0.7$&$  0.4$&$  0.0$\\
$ 1200$&$  0.250$&$  0.250$&$ 10.9$&$  8.7$&$  5.3$&$  2.9$&$  0.4$&$  2.5$&$  4.1$&$  3.0$&$  0.2$&$ -1.2$&$ -2.5$&$  0.0$\\
$ 1200$&$  0.400$&$  0.129$&$ 14.8$&$ 12.1$&$  5.7$&$  3.4$&$  0.8$&$  2.1$&$  6.3$&$  3.6$&$ -0.4$&$ -1.8$&$ -4.8$&$  0.0$\\
$ 1200$&$  0.650$&$  0.017$&$ 29.9$&$ 24.2$&$ 11.1$&$  5.8$&$  1.6$&$  6.9$&$ 13.6$&$  5.9$&$ -0.8$&$ -3.4$&$-11.7$&$  0.0$\\
      \hline
$ 1500$&$  0.020$&$  0.789$&$ 10.5$&$  9.2$&$  4.8$&$  2.7$&$  0.6$&$  1.0$&$  1.7$&$ -0.7$&$ -0.3$&$ -1.0$&$ -0.6$&$  -0.9$\\
$ 1500$&$  0.032$&$  0.581$&$  9.2$&$  8.1$&$  4.0$&$  0.6$&$  1.2$&$  1.7$&$  1.7$&$ -0.2$&$ -0.6$&$  1.1$&$  1.0$&$  -0.4$\\
$ 1500$&$  0.050$&$  0.486$&$  8.1$&$  7.2$&$  3.6$&$  0.4$&$  0.7$&$  1.2$&$  1.4$&$  0.4$&$ -0.3$&$  0.7$&$  1.1$&$  -0.1$\\
$ 1500$&$  0.080$&$  0.457$&$  7.8$&$  6.8$&$  3.5$&$  0.9$&$  0.7$&$  0.3$&$  1.0$&$  0.9$&$ -0.4$&$  0.2$&$  0.2$&$  0.0$\\
$ 1500$&$  0.130$&$  0.376$&$  8.9$&$  8.0$&$  3.7$&$  0.7$&$  0.5$&$  0.6$&$  1.3$&$  0.6$&$ -0.2$&$  0.3$&$  1.0$&$  0.0$\\
$ 1500$&$  0.180$&$  0.345$&$  9.6$&$  8.6$&$  4.0$&$  0.9$&$  0.8$&$  0.8$&$  1.2$&$  1.1$&$  0.4$&$  0.2$&$ -0.4$&$  0.0$\\
$ 1500$&$  0.250$&$  0.268$&$ 11.0$&$  9.4$&$  4.9$&$  2.6$&$  0.8$&$  1.6$&$  3.1$&$  2.7$&$ -0.4$&$ -1.1$&$ -1.0$&$  0.0$\\
$ 1500$&$  0.400$&$  0.110$&$ 16.6$&$ 14.6$&$  5.9$&$  2.8$&$  0.3$&$  3.0$&$  5.1$&$  2.8$&$ -0.2$&$ -1.8$&$ -3.9$&$  0.0$\\
$ 1500$&$  0.650$&$  0.009$&$ 42.6$&$ 37.8$&$ 13.1$&$  8.0$&$  1.6$&$  7.4$&$ 14.6$&$  7.9$&$ -0.8$&$ -4.7$&$-11.3$&$  0.0$\\
 \hline
$ 2000$&$  0.032$&$  0.614$&$  9.9$&$  9.0$&$  4.0$&$  1.3$&$  0.5$&$  0.8$&$  1.0$&$ -0.4$&$ -0.3$&$  0.4$&$  0.5$&$  -0.7$\\
$ 2000$&$  0.050$&$  0.541$&$  9.7$&$  8.7$&$  4.0$&$  0.5$&$  0.3$&$  1.5$&$  1.5$&$  0.1$&$ -0.2$&$  1.1$&$  1.0$&$  -0.3$\\
$ 2000$&$  0.080$&$  0.428$&$  9.1$&$  8.3$&$  3.7$&$  0.1$&$  0.2$&$  1.0$&$  1.1$&$  0.4$&$  0.1$&$  0.4$&$  1.0$&$  0.0$\\
$ 2000$&$  0.130$&$  0.340$&$ 10.6$&$  9.6$&$  4.1$&$  1.4$&$  0.4$&$  0.4$&$  1.4$&$  1.3$&$ -0.2$&$ -0.2$&$ -0.3$&$  0.0$\\
$ 2000$&$  0.180$&$  0.331$&$ 11.1$&$ 10.1$&$  4.5$&$  1.5$&$  1.1$&$  0.7$&$  1.8$&$  1.4$&$ -0.6$&$  0.6$&$  0.7$&$  0.0$\\
$ 2000$&$  0.250$&$  0.249$&$ 12.2$&$ 10.7$&$  5.1$&$  2.5$&$  0.7$&$  1.8$&$  3.0$&$  2.5$&$ -0.3$&$ -1.2$&$ -1.1$&$  0.0$\\
$ 2000$&$  0.400$&$  0.114$&$ 17.2$&$ 15.1$&$  6.5$&$  3.7$&$  0.9$&$  2.9$&$  5.1$&$  3.8$&$  0.5$&$ -1.7$&$ -2.9$&$  0.0$\\
$ 2000$&$  0.650$&$  0.011$&$ 42.2$&$ 37.8$&$ 13.2$&$  7.2$&$  1.0$&$  7.9$&$ 13.3$&$  7.3$&$  0.5$&$ -4.4$&$-10.3$&$  0.0$\\
 \hline
$ 3000$&$  0.050$&$  0.513$&$  8.4$&$  7.3$&$  3.8$&$  0.8$&$  0.6$&$  1.5$&$  1.4$&$  0.7$&$ -0.3$&$  0.8$&$  0.6$&$  -0.6$\\
$ 3000$&$  0.080$&$  0.458$&$  8.7$&$  7.7$&$  4.0$&$  0.6$&$  0.5$&$  1.7$&$  1.3$&$ -0.5$&$ -0.2$&$  0.9$&$  0.7$&$  -0.2$\\
$ 3000$&$  0.130$&$  0.347$&$ 10.2$&$  9.1$&$  4.3$&$  2.1$&$  0.5$&$  0.1$&$  2.1$&$  2.0$&$ -0.2$&$  0.1$&$  0.6$&$  0.0$\\
$ 3000$&$  0.180$&$  0.324$&$ 10.0$&$  9.2$&$  4.0$&$  0.3$&$  0.4$&$  0.6$&$  1.0$&$  0.6$&$ -0.2$&$  0.3$&$  0.7$&$  0.0$\\
$ 3000$&$  0.250$&$  0.242$&$ 11.1$&$  9.9$&$  4.4$&$  2.1$&$  0.2$&$  0.7$&$  2.2$&$  2.1$&$ -0.1$&$ -0.4$&$ -0.5$&$  0.0$\\
$ 3000$&$  0.400$&$  0.127$&$ 15.4$&$ 12.5$&$  7.1$&$  4.6$&$  0.6$&$  3.5$&$  5.6$&$  4.7$&$ -0.3$&$ -2.2$&$ -2.1$&$  0.0$\\
$ 3000$&$  0.650$&$  0.012$&$ 33.6$&$ 30.1$&$ 10.9$&$  7.0$&$  0.5$&$  5.2$&$ 10.2$&$  6.2$&$ -0.3$&$ -3.2$&$ -7.4$&$  0.0$\\
 \hline
$ 5000$&$  0.080$&$  0.353$&$ 11.4$&$ 10.4$&$  4.4$&$  1.0$&$  0.6$&$  1.7$&$  1.6$&$  0.3$&$ -0.3$&$  1.1$&$  0.9$&$  -0.7$\\
$ 5000$&$  0.130$&$  0.392$&$ 11.6$&$ 10.4$&$  4.8$&$  2.0$&$  0.6$&$  0.6$&$  1.6$&$  1.4$&$ -0.3$&$  0.4$&$  0.5$&$  -0.3$\\
$ 5000$&$  0.180$&$  0.223$&$ 14.1$&$ 13.4$&$  4.5$&$  0.9$&$  0.2$&$  0.5$&$  0.6$&$  0.5$&$ -0.1$&$  0.2$&$  0.2$&$  -0.1$\\
$ 5000$&$  0.250$&$  0.217$&$ 15.4$&$ 13.9$&$  6.4$&$  4.5$&$  0.6$&$  0.3$&$  1.6$&$  1.4$&$ -0.3$&$ -0.1$&$  0.6$&$  0.0$\\
$ 5000$&$  0.400$&$  0.127$&$ 19.3$&$ 17.1$&$  7.8$&$  5.6$&$  0.4$&$  2.5$&$  4.0$&$  3.3$&$  0.2$&$ -1.4$&$ -1.9$&$  0.0$\\
$ 5000$&$  0.650$&$  0.012$&$ 40.6$&$ 37.8$&$ 13.8$&$ 10.1$&$  2.4$&$  4.4$&$  5.7$&$  4.2$&$  1.2$&$ -2.6$&$ -2.7$&$  0.0$\\
 \hline
$ 8000$&$  0.130$&$  0.283$&$ 17.2$&$ 16.5$&$  4.7$&$  0.5$&$  0.5$&$  1.5$&$  1.4$&$  0.2$&$ -0.2$&$  0.8$&$  0.9$&$  -0.6$\\
$ 8000$&$  0.180$&$  0.284$&$ 16.7$&$ 15.5$&$  6.1$&$  3.9$&$  0.3$&$  1.8$&$  2.0$&$  1.4$&$ -0.1$&$  0.9$&$  1.1$&$  -0.4$\\
$ 8000$&$  0.250$&$  0.273$&$ 16.6$&$ 15.1$&$  6.6$&$  4.8$&$  0.7$&$  0.5$&$  2.5$&$  2.3$&$  0.3$&$ -0.5$&$ -0.7$&$  -0.2$\\
$ 8000$&$  0.400$&$  0.093$&$ 26.2$&$ 24.2$&$  9.8$&$  8.5$&$  0.7$&$  0.4$&$  1.5$&$  1.4$&$ -0.4$&$  0.3$&$  0.5$&$  0.0$\\
$ 8000$&$  0.650$&$  0.013$&$ 48.9$&$ 44.7$&$ 18.3$&$ 15.6$&$  2.4$&$  4.8$&$  7.5$&$  6.3$&$  1.2$&$ -2.7$&$ -3.0$&$  0.0$\\
 \hline
$12000$&$  0.180$&$  0.153$&$ 34.6$&$ 34.4$&$  3.8$&$  1.9$&$  1.5$&$  0.2$&$  2.0$&$  1.6$&$ -0.8$&$ -0.4$&$  0.0$&$  -0.7$\\
$12000$&$  0.250$&$  0.127$&$ 32.7$&$ 32.1$&$  6.1$&$  5.1$&$  0.4$&$  1.6$&$  1.3$&$  0.5$&$  0.2$&$  1.1$&$  0.5$&$  -0.4$\\
$12000$&$  0.400$&$  0.085$&$ 35.2$&$ 33.3$&$ 11.2$&$ 10.8$&$  0.2$&$  0.3$&$  2.4$&$  2.3$&$  0.1$&$  0.2$&$  0.2$&$  -0.1$\\
$12000$&$  0.650$&$  0.015$&$ 62.6$&$ 57.7$&$ 23.4$&$ 22.8$&$  2.4$&$  2.8$&$  6.1$&$  5.7$&$  1.2$&$ -1.6$&$ -1.2$&$  0.0$\\
 \hline
$20000$&$  0.250$&$  0.090$&$ 62.2$&$ 61.9$&$  5.1$&$  3.1$&$  1.6$&$  1.6$&$  2.1$&$ -1.0$&$ -0.8$&$ -1.3$&$ -0.2$&$  -1.0$\\
$20000$&$  0.400$&$  0.142$&$ 37.0$&$ 35.7$&$  9.6$&$  8.8$&$  0.4$&$  2.3$&$  2.2$&$  1.3$&$ -0.2$&$  1.7$&$  0.6$&$  -0.5$\\
$20000$&$  0.650$&$  0.021$&$ 82.0$&$ 70.7$&$ 40.4$&$ 40.0$&$  3.9$&$  0.6$&$  9.7$&$  9.5$&$  1.9$&$ -0.5$&$ -0.3$&$  -0.1$\\
 \hline
$30000$&$  0.400$&$  0.182$&$ 72.6$&$ 71.9$&$  9.4$&$  6.8$&$  2.1$&$  1.6$&$  1.8$&$  0.8$&$ -1.0$&$ -0.7$&$ -0.4$&$  -0.9$\\
\hline
\end{tabular}
\end{center}
\end{table}

\begin{table}[ht]
  \begin{center}
    \tiny
    \begin{tabular}{|*{2}{c|}|l|*{2}{c|}|*{2}{c|}|*{1}{c|}*{4}{r|}}
      \hline 
      $Q^2$  & $x$ & \multicolumn{1}{c|}{$d^2 \sigma_{CC} /dx dQ^2$} & 
      $\delta_{tot}$ & $\delta_{sta}$ & $\delta_{unc}$ & $\delta^{h}_{unc}$ &
      $\delta_{cor}$ & $\delta^{V^{+}}_{cor}$ & 
      $\delta^{h^{+}}_{cor}$ & $\delta^{N^{+}}_{cor}$ & 
      $\delta^{B^{+}}_{cor}$ \vtab \\
      \tiny $(\rm GeV^2)$ & & \multicolumn{1}{c|}{\tiny $(\rm{pb/GeV^2})$} & 
      \tiny $(\%)$ &\tiny $(\%)$ &
      \tiny $(\%)$ &\tiny $(\%)$ &
      \tiny $(\%)$ &\tiny $(\%)$ &
      \tiny $(\%)$ &
      \tiny $(\%)$ &\tiny  $(\%)$ \\[1mm]
 \hline
$  300$&$  0.013$& \phantom{C} $ 0.637\cdot 10^{0}$&$ 31.8$&$ 27.4$&$  9.9$&$  2.9$&$ 12.6$&$ 11.9$&$ -1.6$&$  0.2$&$ -3.8$\\
$  300$&$  0.032$&\phantom{C}  $ 0.124\cdot 10^{0}$&$ 30.0$&$ 28.1$&$  7.9$&$  2.8$&$  6.6$&$  6.2$&$ -1.6$&$  0.9$&$ -0.7$\\
$  300$&$  0.080$&\phantom{C}  $ 0.532\cdot 10^{-1}$&$ 25.5$&$ 23.8$&$  7.0$&$  2.2$&$  2.8$&$  2.2$&$ -1.5$&$ -0.6$&$ -0.3$\\
 \hline
$  500$&$  0.013$&\phantom{C}  $ 0.468\cdot 10^{0}$&$ 29.7$&$ 25.1$&$  9.2$&$  1.9$&$ 12.8$&$ 12.1$&$ -1.5$&$  1.0$&$ -3.9$\\
$  500$&$  0.032$&\phantom{C}  $ 0.177\cdot 10^{0}$&$ 19.2$&$ 17.0$&$  6.7$&$  2.3$&$  5.5$&$  5.2$&$ -1.2$&$  0.4$&$ -0.3$\\
$  500$&$  0.080$&\phantom{C}  $ 0.546\cdot 10^{-1}$&$ 18.9$&$ 17.0$&$  6.1$&$  1.1$&$  2.3$&$  2.3$&$ -0.5$&$  0.2$&$ -0.2$\\
$  500$&$  0.130$&\phantom{C}  $ 0.289\cdot 10^{-1}$&$ 29.4$&$ 27.8$&$  7.8$&$  1.5$&$  1.6$&$  0.3$&$ -1.3$&$ -0.9$&$ -0.1$\\
 \hline
$ 1000$&$  0.032$&\phantom{C}  $ 0.124\cdot 10^{0}$&$ 17.1$&$ 15.0$&$  6.5$&$  1.8$&$  4.7$&$  4.6$&$ -1.3$&$  0.2$&$ -0.2$\\
$ 1000$&$  0.080$&\phantom{C} $ 0.487\cdot 10^{-1}$&$ 14.8$&$ 13.3$&$  5.7$&$  0.9$&$  2.3$&$  2.1$&$ -0.8$&$  0.6$&$ -0.1$\\
$ 1000$&$  0.130$&\phantom{C} $ 0.199\cdot 10^{-1}$&$ 22.5$&$ 20.9$&$  6.4$&$  0.9$&$  1.0$&$  0.6$&$ -0.7$&$ -0.4$&$ -0.1$\\
$ 1000$&$  0.250$&\phantom{C} $ 0.105\cdot 10^{-1}$&$ 34.1$&$ 31.7$&$ 10.4$&$  3.0$&$  5.2$&$  0.0$&$  3.1$&$ -4.2$&$ -0.1$\\
 \hline
$ 2000$&$  0.032$&\phantom{C} $ 0.716\cdot 10^{-1}$&$ 18.1$&$ 15.7$&$  6.9$&$  1.0$&$  5.5$&$  5.4$&$ -1.5$&$  1.7$&$ -0.4$\\
$ 2000$&$  0.080$&\phantom{C} $ 0.264\cdot 10^{-1}$&$ 14.8$&$ 13.5$&$  5.6$&$  0.1$&$  1.7$&$  1.6$&$  0.3$&$  0.3$&$ -0.2$\\
$ 2000$&$  0.130$&\phantom{C} $ 0.949\cdot 10^{-2}$&$ 21.4$&$ 20.6$&$  5.7$&$  0.6$&$  0.6$&$  0.4$&$  0.3$&$  0.4$&$ -0.2$\\
$ 2000$&$  0.250$&\phantom{C} $ 0.566\cdot 10^{-2}$&$ 24.6$&$ 23.0$&$  7.0$&$  0.4$&$  2.2$&$  0.1$&$ -0.2$&$ -2.2$&$ -0.1$\\
 \hline
$ 3000$&$  0.080$&\phantom{C} $ 0.156\cdot 10^{-1}$&$ 16.8$&$ 15.2$&$  6.2$&$  2.3$&$  2.7$&$  2.1$&$  1.5$&$  0.7$&$ -0.2$\\
$ 3000$&$  0.130$&\phantom{C} $ 0.872\cdot 10^{-2}$&$ 18.1$&$ 17.0$&$  5.8$&$  1.6$&$  1.1$&$  0.6$&$  0.8$&$  0.6$&$ -0.1$\\
$ 3000$&$  0.250$&\phantom{C} $ 0.283\cdot 10^{-2}$&$ 25.1$&$ 23.6$&$  7.8$&$  4.7$&$  2.6$&$  0.1$&$  2.5$&$ -0.6$&$ -0.1$\\
 \hline
$ 5000$&$  0.130$&\phantom{C} $ 0.402\cdot 10^{-2}$&$ 22.3$&$ 21.0$&$  7.0$&$  3.9$&$  2.6$&$  0.6$&$  2.5$&$  0.4$&$ -0.1$\\
$ 5000$&$  0.250$&\phantom{C} $ 0.111\cdot 10^{-2}$&$ 27.6$&$ 26.8$&$  6.5$&$  2.9$&$  0.8$&$  0.2$&$  1.2$&$  0.6$&$ -0.1$\\
 \hline
$ 8000$&$  0.130$&\phantom{C} $ 0.125\cdot 10^{-2}$&$ 38.5$&$ 35.7$&$ 12.6$&$ 10.8$&$  6.8$&$  1.2$&$  6.5$&$  1.5$&$ -0.3$\\
$ 8000$&$  0.250$&\phantom{C} $ 0.530\cdot 10^{-3}$&$ 35.4$&$ 33.5$&$  9.9$&$  7.9$&$  5.4$&$  0.2$&$  5.2$&$  1.2$&$ -0.1$\\
$ 8000$&$  0.400$&\phantom{C} $ 0.235\cdot 10^{-3}$&$ 52.4$&$ 50.0$&$ 13.1$&$ 10.9$&$  8.5$&$  0.0$&$  8.5$&$ -0.8$&$  0.0$\\
 \hline
$15000$&$  0.250$&\phantom{C} $ 0.774\cdot 10^{-4}$&$ 73.5$&$ 71.2$&$ 15.9$&$ 14.3$&$  8.5$&$  1.0$&$  8.4$&$  1.1$&$ -0.2$\\
$15000$&$  0.400$&\phantom{C} $ 0.114\cdot 10^{-3}$&$ 44.5$&$ 40.9$&$ 15.3$&$ 14.2$&$  8.2$&$  0.1$&$  8.1$&$  0.9$&$ -0.1$\\

\hline
\end{tabular}
\caption{\small \sl 
CC double differential  cross-section $d^2\sigma_{CC}/dx dQ^2$ 
with total error ($\delta_{tot}$),
statistical error ($\delta_{sta}$), 
uncorrelated systematic error ($\delta_{unc}$), and its contributions from
the hadronic energy error ($\delta^{h}_{unc}$).
The effect of the other uncorrelated 
errors, as described in section~\ref{systemaa}, is included in $\delta_{unc}$.
Also given are the
correlated systematic error ($\delta_{cor}$), and its contributions from
a positive variation of one standard deviation of the error coming from 
the cut on the $V_{ap}/V_p$ ratio ($\delta^{V^{+}}_{cor}$),
of the hadronic energy error ($\delta^{h^{+}}_{cor}$), 
of the noise contribution ($\delta^{N^{+}}_{cor}$) and of
the estimated background contribution ($\delta^{B^{+}}_{cor}$).
The normalization uncertainty, 
which is not included in the systematic error, is 1.5\%. 
\label{longtabcc}
}
\end{center}
\end{table}